\documentclass[11pt]{article}

\usepackage[nosort]{cite}
\usepackage[bulletsep]{collref}

\usepackage[british]{babel} 

\usepackage{epsfig,rotating}
\usepackage{amsmath,amssymb,amsthm}
\usepackage{amsfonts}
\usepackage{mathrsfs}
\usepackage{bbm}
\usepackage{bm}

\usepackage{color}
\usepackage[textwidth = 430 pt, textheight = 630 pt]{geometry}
\definecolor{MyDarkBlue}{rgb}{0.15,0.25,0.45}

\usepackage[linktocpage=true]{hyperref}
\hypersetup{
colorlinks=true,
citecolor=MyDarkBlue,
linkcolor=MyDarkBlue,
urlcolor=MyDarkBlue,
pdfauthor={Christian S\"amann and Martin Wolf},
pdftitle={On Twistors and Conformal Field Theories from Six Dimensions},
pdfsubject={hep-th}
breaklinks=true
}

\let\fn\footnote
\renewcommand{\footnote}[1]{\linespread{1.1}\fn{#1}\linespread{1.29}}

\makeatletter\renewcommand{\section}{\@startsection
{section}{1}{\z@}{-3.5ex plus -1ex minus
    -.2ex}{2.3ex plus .2ex}{\bf\mathversion{bold} }}
\makeatletter\renewcommand{\subsection}{\@startsection{subsection}{2}{\z@}{-3.25ex
plus -1ex minus
   -.2ex}{1.5ex plus .2ex}{\bf\mathversion{bold} }}
\makeatletter\renewcommand{\subsubsection}{\@startsection{subsubsection}{3}{-2.45ex}{-3.25ex
plus -1ex minus -.2ex}{1.5ex plus .2ex}{\it }}
\renewcommand{\thesection}{\arabic{section}}
\renewcommand{\thesubsection}{\arabic{section}.\arabic{subsection}}
\renewcommand{\@seccntformat}[1]{\@nameuse{the#1}.~~}

\renewcommand{\theequation}{\thesection.\arabic{equation}}
\makeatletter \@addtoreset{equation}{section}

\renewcommand*\l@section{\@dottedtocline{1}{0em}{2em}}
\renewcommand*\l@subsection{\@dottedtocline{2}{2em}{2.4em}}
\renewcommand*\l@subsubsection{\@dottedtocline{4}{3.8em}{3.7em}}

\renewcommand\tableofcontents{%
    \section*{\large\contentsname
        \@mkboth{%
          \MakeUppercase\contentsname}{\MakeUppercase\contentsname}}%
       {\baselineskip=15pt plus 2pt minus 1pt
    \@starttoc{toc}}%
}

\renewenvironment{thebibliography}[1]
     {\baselineskip=16pt plus 2pt minus 1pt
      \section*{\large\refname
        \@mkboth{\MakeUppercase\refname}{\MakeUppercase\refname}}%
     \list{\@biblabel{\@arabic\c@enumiv}}%
           {\settowidth\labelwidth{\@biblabel{#1}}%
            \leftmargin\labelwidth
            \advance\leftmargin\labelsep
            \@openbib@code
            \usecounter{enumiv}%
            \let\p@enumiv\@empty
            \renewcommand\theenumiv{\@arabic\c@enumiv}}%
      \sloppy
      \clubpenalty4000
      \@clubpenalty \clubpenalty
      \widowpenalty4000%
      \sfcode`\.\@m
 \catcode`\^^M=10%
}

\newcommand{\appendices}{
\section*{Appendix}\label{appendices}\setcounter{subsection}{0}
\addcontentsline{toc}{section}{Appendix}
\setcounter{equation}{0}
\makeatletter
\renewcommand{\theequation}{\Alph{subsection}.\arabic{equation}}
\renewcommand{\thesubsection}{\Alph{subsection}}
\@addtoreset{equation}{subsection}
\makeatother
}

\newtheorem{rem}{Remark}
\numberwithin{rem}{section}

\newtheorem{lemma}{Lemma}
\numberwithin{lemma}{section}

\numberwithin{definition}{section}

\newtheorem{theorem}{Theorem}
\numberwithin{theorem}{section}

\newtheorem{prop}{Proposition}
\numberwithin{prop}{section}

\numberwithin{cor}{section}

\def\periodb#1{\setbox0=\hbox{$#1$}#1\hskip-\wd0\hbox to\wd0{-}}

\newcommand{\unit}{\mathbbm{1}}   			
\newcommand{\im}{\mathrm{im}}   			

\newcommand{\CC}{\mathcal{C}}
\newcommand{\CCC}{\mathscr{C}}

\newcommand{\CF}{\mathcal{F}}

\newcommand{\CI}{\mathcal{I}}

\newcommand{\CL}{\mathcal{L}}

\newcommand{\CN}{\mathcal{N}}
\newcommand{\CO}{\mathcal{O}}

\newcommand{\CCP}{\mathscr{P}}

\newcommand{\CS}{\mathcal{S}}

\newcommand{\CZ}{\mathcal{Z}}

\newcommand{\CE}{\mathcal{E}}

\newcommand{\frU}{\mathfrak{U}}

\newcommand{\FR}{\mathbbm{R}}     			
\newcommand{\FC}{\mathbbm{C}}     			
\newcommand{\FQ}{\mathbbm{Q}}

\newcommand{\NN}{\mathbbm{N}}     			
\newcommand{\RZ}{\mathbbm{Z}}     			
\newcommand{\PP}{{\mathbbm{P}}}    			

\newcommand{\sh}{\hat{s}}
\newcommand{\Uh}{{\hat{U}}}
\newcommand{\thetah}{{\hat{\theta}}}
\newcommand{\Ah}{\hat{A}}
\newcommand{\Bh}{\hat{B}}

\newcommand{\fh}{\hat{f}}

\newcommand{\dd}{\mathrm{d}}     			
\newcommand{\dpar}{\partial}     			
\newcommand{\dparb}{{\bar{\partial}}}     		
\newcommand{\embd}{{\hookrightarrow}}     		
\newcommand{\diag}{{\mathrm{diag}}}     		
\newcommand{\de}{\mathrm{e}}     			
\newcommand{\di}{\mathrm{i}}     			
\newcommand{\eps}{{\varepsilon}}			

\newcommand{\ald}{{\dot{\alpha}}}     			
\newcommand{\bed}{{\dot{\beta}}}

\newcommand{\eand}{{~~~\mbox{and}~~~}}     		
\newcommand{\eon}{{~~\mbox{on}~~}}     		
\newcommand{\ewith}{{~~~\mbox{with}~~~}}
\newcommand{\efor}{{~~~\mbox{for}~~~}}

\newcommand{\der}[1]{\frac{\dpar}{\dpar #1}}   		

\newcommand{\asl}{\mathfrak{sl}}

\newcommand{\aso}{\mathfrak{so}}

\newcommand{\sU}{\mathsf{U}}     			

\newcommand{\sSU}{\mathsf{SU}}
\newcommand{\sSL}{\mathsf{SL}}

\newcommand{\sSO}{\mathsf{SO}}

\newcommand{\sSpin}{\mathsf{Spin}}

\newcommand{\remark}[1]{}     				
\def\tyng(#1){\hbox{\tiny$\yng(#1)$}}			
\def\tyoung(#1){\hbox{\tiny$\young(#1)$}}			

\begin{document}
\begin{titlepage}

\setcounter{page}{0}
\renewcommand{\thefootnote}{\fnsymbol{footnote}}

\begin{flushright}
 EMPG--11--26\\ HWM--11--24\\ DMUS--MP--11/01
\end{flushright}

\begin{center}

{\LARGE\textbf{On Twistors and Conformal Field Theories\\ from Six Dimensions}\par}

\vspace{1cm}

{\large
Christian S\"amann$^{a}$ and Martin Wolf$^{b}$
\footnote{{\it E-mail addresses:\/}
\href{mailto:c.saemann@hw.ac.uk}{\ttfamily c.saemann@hw.ac.uk},
\href{mailto:m.wolf@surrey.ac.uk}{\ttfamily m.wolf@surrey.ac.uk}
}}

\vspace{1cm}

{\it
$^a$ Maxwell Institute for Mathematical Sciences\\
Department of Mathematics,
Heriot--Watt University\\
Edinburgh EH14 4AS, United Kingdom\\[.5cm]

$^b$
Department of Mathematics,
University of Surrey\\
Guildford GU2 7XH, United Kingdom

}

\vspace{1cm}

{\bf Abstract}
\end{center}
\vspace{-.3cm}

\begin{quote}
We discuss chiral zero-rest-mass field equations on six-dimensional space-time from a twistorial point of view. Specifically, we present a detailed cohomological analysis, develop both Penrose and Penrose--Ward transforms, and analyse the corresponding contour integral formul{\ae}. We also give twistor space action principles. We then dimensionally reduce the twistor space of six-dimensional space-time to obtain twistor formulations of various theories in lower dimensions.  Besides well-known twistor spaces, we also find a novel twistor space amongst these reductions, which turns out to be suitable for a twistorial description of self-dual strings. For these reduced twistor spaces, we explain the Penrose and Penrose--Ward transforms as well as contour integral formul{\ae}.

\vfill
\noindent 24th August 2016\hfill{\it To the memory of Francis A.~Dolan}

\end{quote}

\setcounter{footnote}{0}\renewcommand{\thefootnote}{\arabic{thefootnote}}

\end{titlepage}

\tableofcontents

\bigskip
\bigskip
\hrule
\bigskip
\bigskip

\section{Introduction}

Since their discovery by Penrose \cite{Penrose:1967wn}, twistors have provided deep insights into various gauge and gravity theories, particularly into integrable ones. The cornerstone of twistor geometry is to replace space-time as a background for physical processes by an auxiliary space called twistor space. Differentially constrained data (such as solutions to field equations on space-time) are then encoded in differentially unconstrained complex analytic data (such as elements of cohomology groups) on twistor space. This allows for an elegant and complete description of solutions to certain classes of problems. The prime examples in this respect are all the solutions to the zero-rest-mass field equations on four-dimensional space-time \cite{Penrose:1967wn,Penrose:1968me,Penrose:1969aa,Penrose:1972ia}, instantons in Yang--Mills theory \cite{Ward:1977ta}, and self-dual Riemannian four-dimensional manifolds \cite{Penrose:1976js,Atiyah:1978wi,Ward:1980am}. Twistor geometry moreover underlies the well-known Atiyah--Drinfeld--Hitchin--Manin (ADHM) construction of Yang--Mills instanton solutions  \cite{Atiyah:1978ri}. Even solutions to non-linear second-order differential equations such as the full Yang--Mills and Einstein equations and their supersymmetric extensions can be captured in terms of holomorphic data using twistor methods \cite{Witten:1978xx,Isenberg:1978kk,Isenberg:1978qd,Eastwood:1986aa,Lebrun:1986it,Manin:1988ds,springerlink:10.1007/BF00419314,LeBrun:1991jh,0264-9381-2-4-020,Merkulov:1992qa}. In these latter cases, however, the power to explicitly construct solutions is limited.

Perhaps the simplest and oldest example of a twistor description of space-time is obtained by replacing Minkowski space $\FR^{1,3}$ by the space of projective light cones in this space, $\FR^{1,3}\times S^2$. This space can be shown to be diffeomorphic to the open subset $\PP^3_\circ:=\PP^3\setminus\PP^1$ of complex projective three-space $\PP^3$.\footnote{One may also conformally compactify space-time to obtain $\PP^3$ as twistor space. The line $\PP^1$ that is deleted from $\PP^3$ to obtain $\PP^3_\circ$ corresponds on space-time to the point infinity which is used for this conformal compactification.} In this paper, we will always work with complexified space-times. The twistor description of space-time $\FC^4$ is given by the space of totally null two-planes, which is again $\PP^3_\circ$. Moreover, holomorphic vector bundles over $\PP^3_\circ$, that are subject to a mild triviality condition, are in one-to-one correspondence with Yang--Mills instantons on $\FC^4$. There is a large variety of further examples of twistor spaces, on each of which there are such correspondences between cohomological data and solutions to field equations (via the so-called Penrose and Penrose--Ward transforms). In this paper, we shall encounter Penrose's twistor space \cite{Penrose:1967wn}, the ambitwistor space \cite{Witten:1978xx,Isenberg:1978kk}, and Hitchin's minitwistor space \cite{Hitchin:1982gh}. For detailed reviews of various aspects of twistor spaces, see for example \cite{Penrose:1985jw,Penrose:1986ca,Ward:1990vs}. See also \cite{Wolf:2010av,Adamo:2011pv} for recent reviews using conventions close to ours.

All the above-mentioned twistor spaces are suitable for capturing moduli spaces\footnote{These moduli spaces are obtained from the solution spaces by quotienting with respect to the group of gauge transformations.} of the four-dimensional Yang--Mills equations, their supersymmetric extensions, the BPS subsectors thereof and their dimensional reductions.\footnote{Examples of twistor spaces for higher-dimensional space-times including Penrose and Penrose--Ward transforms can be found e.g.\ in \cite{Ward:1983zm,Witten:1985nt,Eastwood:1985aa,Bailey:1991aa,springerlink:10.1007/BF00132253,Wolf:2009ep}.} In the light of the recent success of M2-brane models \cite{Bagger:2007jr,Gustavsson:2007vu,Aharony:2008ug}, it is natural to wonder about twistor spaces underlying the description of solution spaces of more general gauge theories. Staying within M-theory, there are essentially three theories one might be interested in. The three-dimensional M2-brane models, M5-brane models which should be given by some $\CN=(2,0)$ superconformal field theories in six dimension, and the self-dual string equation in four dimensions which describes a BPS subsector of the dimensionally reduced M5-brane model.

In this paper, we shall focus on the latter two. The self-dual string equation given by Howe, Lambert \& West \cite{Howe:1997ue} is a BPS equation that describes M2-branes ending on M5-branes. It can be seen as the M-theory analogue of the Bogomolny monopole equation describing D1-branes ending on D3-branes. Moreover, when re-phrased on loop space, all solutions to this equation have been shown to arise via an ADHM-like construction \cite{Gustavsson:2008dy,Saemann:2010cp,Palmer:2011vx}. This is a strong hint that a twistor description of the solutions and the moduli space of this equation should be possible. Since the self-dual string equation arises from a reduction of the six-dimensional theory of self-dual three-forms, one should first consider a twistor space describing such three-forms and then further reduce it. Fortunately, a candidate twistor space that is suitable for a twistorial description of chiral theories has already appeared  \cite{Penrose:1985jw,Penrose:1986ca,springerlink:10.1007/BF00132253,Hughston:1986hb,Hughston:1979TN,Hughston:1982TN,Hughston:1984TN,Hughston:1987aa,Hughston:1987he,Hughston:1988nz,Baston:1989,Inoue:1992aa,Berkovits:2004bw,Chern:2009nt}. We shall denote this twistor space by $P^6$. Since the non-Abelian extensions of both the self-dual string equation and the self-dual three-forms are still essentially unknown\footnote{There is a non-Abelian extension of the self-dual string equation on loop space \cite{Gustavsson:2008dy,Saemann:2010cp,Palmer:2011vx}. Also there have been some recent proposals for non-Abelian M5-brane models, see e.g.\ \cite{Lambert:2010wm,Papageorgakis:2011xg,Samtleben:2011fj,Chu:2011fd}.}, we shall restrict ourselves to the Abelian cases. The hope is certainly that the twistor descriptions presented below might shed light on the issue of non-Abelian extensions.

Our first aim is to establish both Penrose and Penrose--Ward transforms for the construction of chiral zero-rest-mass spinor fields on six-dimensional space-time using the twistor space $P^6$. In particular, we shall give a detailed proof of a Penrose transform to establish an isomorphism between certain cohomology groups on $P^6$ and chiral zero-rest-mass fields on space-time. Our discussion follows the corresponding one in the four-dimensional case as given, e.g.~in \cite{Eastwood:1981jy,Ward:1990vs}.\footnote{An alternative proof can be found in \cite{springerlink:10.1007/BF00132253}; see also \cite{Baston:1989}.} Moreover, we show how to generalise the Penrose--Ward transform to $P^6$ and how to obtain spinor fields via this transform. We shall also introduce twistor space action principles for chiral fields which might be the twistor analog of the space-time actions of Pasti, Sorokin \& Tonin \cite{Pasti:1995ii,Pasti:1995tn,Pasti:1996vs,Pasti:1997gx}.

Our second aim is to demonstrate how the dimensional reductions of the six-dimensional spinor fields to four and three space-time dimensions is reflected in certain reductions of $P^6$. In particular, we find that the twistor space $P^6$ contains naturally the ambitwistor space, which provides a twistor description of the Maxwell and Yang--Mills equations, a twistor space we shall refer to as the hyperplane twistor space and which turns out to be suitable for a twistor description of the self-dual string equation, and the minitwistor space underlying a twistor description of monopoles. To our knowledge, the hyperplane twistor space has not been discussed in the literature before. Therefore, we shall be explicit in constructing both Penrose and Penrose--Ward transforms over this twistor space.

This paper is structured as follows. We begin our considerations with a brief review of spinors and free fields in six dimensions. We then present the construction of the twistor space for six-dimensional space-time from various perspectives in Section \ref{sec:TwistorSpace}. In Section \ref{sec:6DPenrose}, we  lay down the cohomological foundation on which all of our later analysis is based. This section also contains a detailed proof of the Penrose transform and explicit integral formul\ae{} yielding zero-rest-mass fields. The Penrose--Ward transform is presented in Section \ref{sec:PWTransform}, where we also comment on the aforementioned action principle. We then continue with discussing various dimensional reductions in Section \ref{sec:RedToLowerDim}. In particular, we show how the six-dimensional picture reduces to the ambitwistor space describing Maxwell fields in four dimensions, the twistor description of self-dual strings and the twistor description of monopoles. We summarise our conclusions in Section \ref{sec:ConclusionsAndOutlook}, where we also present an outlook. Three appendices collect some technical background material.

\paragraph{Remark.} Whilst finalising the draft, we became aware of the results of Mason, Reid-Edwards \& Taghavi-Chabert \cite{Mason:2011nw}, which partially overlap with the results presented in this work.

\paragraph{Dedication.} We would like to dedicate this work to our friend and colleague Francis A.~Dolan, who passed  away very unexpectedly in September 2011.

\section{Spinors and free fields in six dimensions}\label{sec:6dZRMfields}

\subsection{Spinors in six dimensions}

In the following, we shall be working with the complexification of flat six-dimensional space-time $M^6:=\FC^6$. Notice that reality conditions leading to real slices of $M^6$ with Minkowski or split signature can be imposed if desired. These are briefly discussed in Appendix \ref{app:spinors}.

The spin bundle on $M^6$  is of rank eight and decomposes into the direct sum $S\oplus \tilde{S}$ of the two rank-4 subbundles  of anti-chiral spinors, $S$, and chiral spinors, $\tilde{S}$. There is a natural isomorphism identifying $S$ and $\tilde{S}$ with the duals\footnote{Given a linear space $V$, we denote its dual by $V^\vee$.} $\tilde{S}^\vee$ and $S^\vee$ (see e.g.~Penrose \& Rindler \cite{Penrose:1986ca} for details; this identification basically works via an automorphism of the Clifford algebra corresponding to charge conjugation). Therefore, we may exclusively work with, say, $S$ and $S^\vee$. We shall label the corresponding spinors by upper and lower capital Latin letters from the beginning of the alphabet, e.g.\ $\psi^A$ for a section of $S$ and $\psi_A$ for a section of $S^\vee$, with $A, B,\ldots=1,\ldots,4$.

We may identify the tangent bundle $T_{M^6}$ with the anti-symmetric tensor product of the chiral spinor bundle with itself via
\begin{equation}\label{eq:TangentBundle}
\begin{aligned}
 T_{M^6}\ \cong\ S\wedge S~,\kern1.8cm\\
 \partial_M\ :=\ \frac{\partial}{\partial x^M} \ \overset{\tilde\sigma_*}{\longleftrightarrow}\ \partial_{AB}\ :=\ \frac{\partial}{\partial x^{AB}}~.
  \end{aligned}
\end{equation}
Here, we coordinatised $M^6$ by $x^M$, for $M, N,\ldots=1,\ldots,6$ and used the identification $\tilde\sigma\,:\,x=(x^M)\mapsto \tilde\sigma(x)=(x^{AB})$ with $x^{AB}=\tilde\sigma_M^{AB}x^M$ $\Leftrightarrow$ $x^M=\frac14\sigma^M_{AB}x^{AB}$, where $\tilde\sigma_M^{AB}$, $\sigma^M_{AB}$ are the six-dimensional sigma-matrices, cf.\ Appendix \ref{app:spinors}. The induced linear mapping $\tilde\sigma_*$ is explicitly given as $\partial_{AB}=\frac14\sigma^M_{AB}\partial_M$ and the (flat) metric $\eta_{MN}$ on $M^6$ can be identified with the Levi-Civita symbol $\frac12\varepsilon_{ABCD}$ in spinor notation. Hence, $\sigma^M_{AB}=\frac12\varepsilon_{ABCD}\tilde\sigma^{MCD}$ and we can raise and lower indices according to
\begin{equation}
 \partial_{AB}\ =\ \tfrac12\varepsilon_{ABCD}\partial^{CD}\qquad\Longleftrightarrow\qquad
 \partial^{AB}\ =\ \tfrac12\varepsilon^{ABCD}\partial_{CD}~.
\end{equation}
For any two six-vectors $p=(p^M)$ and $q=(q^M)$, we shall write
\begin{equation}
 p\cdot q\ :=\ p_M q^M\ =\ \tfrac14 p_{AB} q^{AB}\ =\ \tfrac18 \varepsilon_{ABCD}p^{AB}q^{CD}~,
\end{equation}
and we have $p^2:=p\cdot p=\sqrt{\det p^{AB}}$.

\subsection{Zero-rest-mass fields in six dimensions}\label{ssec:zrmfields}

Next we wish to discuss zero-rest-mass fields in the six-dimensional spinor-helicity formalism, borrowing some of the ideas of \cite{Cheung:2009dc,Dennen:2009vk,Huang:2010rn,Czech:2011dk}. Let us start by considering a momentum six-vector $p=(p_M)$. If we impose the null-condition $p^2=0$, then  we have $\det p_{AB}=0=\det p^{AB}$. These equations are solved most generally by
\begin{equation}
 p_{AB}\ =\  k_{Aa} k_{Bb}\varepsilon^{ab}\eand
 p^{AB}\ =\  \tilde{k}^{A\dot a} \tilde{k}^{B\dot b}\varepsilon_{\dot a\dot b}
\end{equation}
with $a,b,\ldots,\dot a,\dot b,\ldots=1,2$ and $\varepsilon^{ab}=-\varepsilon^{ba}$ and $\varepsilon_{\dot a\dot b}=-\varepsilon_{\dot b\dot a}$. We shall refer to such a momentum as null-momentum. Moreover, transformations of the form $k_{Aa}\mapsto M\ \!\!_a^b  k_{Ab}$ and $\tilde{k}^{A\dot a}\mapsto \tilde{M}\ \!\!^{\dot a}_{\dot b} \tilde{k}^{A\dot b}$ with $\det M=1=\det \tilde{M}$ will leave $p$ invariant, which shows that the indices $a,\dot a,\ldots$ are little group indices. The little group of (complex) null-vectors in six dimensions is therefore $\sSL(2,\FC)\times \widetilde{\sSL(2,\FC)}$. Notice that $k_{Aa} \tilde{k}^{A\dot b}=0$ since $p_{AB}=\frac12\varepsilon_{ABCD}p^{CD}$, which, in turn, shows that  $k_{Aa}$ and $\tilde{k}^{A\dot a}$ are not independent. Notice also that $k_{Aa}$ has $4\times 2=8$ components, but three of them can be fixed by little group transformations. Thus, $k_{Aa}$ has indeed exactly the five independent components needed to describe the (five-dimensional) null-cone in six dimensions.

Fields form irreducible representations of the Lorentz group which are induced from representations of the little group. In six dimensions, the spin label of fields therefore has to be generalised to a pair of integers, labelling the irreducible representations of the little group $\sSL(2,\FC)\times \widetilde{\sSL(2,\FC)}$. As an example of zero-rest-mass fields, let us consider the fields in the $\CN=(2,0)$ tensor multiplet \cite{Howe:1983fr}. This multiplet is a chiral multiplet and hence the fields transform trivially under the $\widetilde{\sSL(2,\FC)}$ subgroup. Amongst these fields, there is a self-dual three-form $H=\dd B$, which transforms as the $(\mathbf{3},\mathbf{1})$ of the little group. In spinor notation, $H$ has components\footnote{\label{foot:SD} A general three-form $H=\dd B$ in six dimensions is described by a pair of  symmetric bi-spinors $H=(H_{AB},H^{AB})=(\partial_{C(A}B_{B)}{}^C,\partial^{C(A}B_C{}^{B)})$ and transforms as the $(\mathbf{3},\mathbf{1})\oplus(\mathbf{1},\mathbf{3})$ of the little group. We use parentheses and square brackets to denote normalised symmetrisation and normalised anti-symmetrisation, respectively. By imposing either self-duality or anti-self-duality, one of the bi-spinors is put to zero. In our conventions, self-duality implies $H^{AB}=0$ while anti-self-duality amounts to $H_{AB}=0$.}  $H_{AB}=\partial_{C(A}B_{B)}{}^C$, where $B_{B}{}^C$ is trace-less and denotes the components of a two-form potential $B$ in spinor notation. In addition, we have four Weyl spinors $\psi^I_A$ in the $(\mathbf{2},\mathbf{1})$ and five scalars $\phi^{IJ}$ in the trivial representation $(\mathbf{1},\mathbf{1})$ of the little group. Notice that the a priori six components of $\phi^{IJ}=-\phi^{JI}$ are reduced to five by the condition $\phi^{IJ}\Omega_{IJ}=0$, where $I,J,\ldots=1,\ldots,4$ and $\Omega_{IJ}$ is an invariant form of the underlying $R$-symmetry; see e.g.\ \cite{Claus:1997cq} for more details. In the following, we shall work with complex fields. However, one may impose reality conditions on all fields as briefly discussed in Appendix \ref{app:spinors}. The zero-rest-mass field equations  (i.e.~the free equations of motion) for the fields in the tensor multiplet read as
\begin{equation}\label{eq:PSWE}
 H^{AB}\ =\ 0\ewith \partial^{AC}H_{CB}\ =\ 0~,\quad \partial^{AB}\psi_B\ =\ 0~,\eand\ \Box\phi\ =\ 0~,
\end{equation}
where we suppressed the $R$-symmetry indices.  Notice that the second equation is the Bianchi identity (which, of course, is equivalent to the field equation for self-dual three-forms). The corresponding plane waves are given by the expressions (${\rm i}:=\sqrt{-1}$)
\begin{equation}\label{eq:FExp6d}
 H_{AB\,ab}\ =\ k_{A(a} k_{Bb)}\, \de^{\di x\cdot p}~,\quad \psi_{Aa}\ =\  k_{Aa}\,\de^{\di x\cdot p}~,\eand \phi\ =\   \de^{\di  x\cdot p}~.
\end{equation}
This follows from straightforward differentiation. Here, the representations of the little group formed by the fields become explicit.  Furthermore, since $H_{AB}=\partial_{C(A}B_{B)}{}^C$, we can express the plane waves of $H_{AB}$ in terms of the plane waves of the potential two-form $B_B{}^A$. To this end, we note that in spinor notation, gauge transformations of $B_B{}^A$ are mediated by gauge parameters $\Lambda_{AB}=\Lambda_{[AB]}$ via  $B_B{}^A\mapsto B_B{}^A+\partial^{AC}\Lambda_{CB}-\partial_{BC}\Lambda^{CA}$. We shall choose Lorenz gauge, which in spinor notation reads as $\partial_{C[A}B_{B]}{}^C=0=\partial^{C[A}B_C{}^{B]}$. The residual gauge transformations are given by gauge parameters that obey $\partial\cdot\Lambda=0$. Let us now choose reference spinors $\mu_{Aa}$ and define the null-momentum  $q_{AB}:=\mu_{Aa}\mu_{Bb}\eps^{ab}$ so that $p\cdot q\neq0$. Then the plane waves of the potential two-form $B_B{}^A$ in Lorenz gauge are given by
\begin{equation}\label{eq:SolOfB}
 B_B{}^A{}_{ab}\ =\  \kappa^A_{(a}k_{Bb)}\,\de^{\di x\cdot p}\ewith \kappa^A_a\ :=\ -2\di\frac{q^{AB} k_{Ba}}{p\cdot q}~.
\end{equation}
Clearly,  $B_B{}^A$ is trace-less and one can check that $\partial_{C(A}B_{B)}{}^C$ yields the components for $H_{AB}$ given in \eqref{eq:FExp6d}. Since $\partial^{CA}B_C{}^B=0$, we also have $H^{AB}=\partial^{C(A}B_C{}^{B)}=0$, which implies that $B_B{}^A$ does indeed yield a self-dual field strength.  Furthermore, the choice of $\mu_{Aa}$ is irrelevant since changes in  $\mu_{Aa}$ merely correspond to (residual) gauge transformations of $B_B{}^A$, a fact that is already familiar from four dimensions \cite{Witten:2003nn}.\footnote{Since the $\mu_{Aa}$ appear only in the combination $q_{AB}=\mu_{Aa}\mu_{Bb}\eps^{ab}$, we may focus on the induced changes in $q$. The space of the $q$ is five-dimensional, so the most general change is of the form $q_{AB}\mapsto q_{AB}+\alpha\, q_{AB} +\beta_{AB}$ with $\beta^2=0$ and $\beta\cdot p=0$. Therefore, $B_B{}^A{}_{ab}\mapsto B_B{}^A{}_{ab}+\Lambda^A_{(a}k_{Bb)}\,\de^{\di x\cdot p}$, with $\Lambda^A_a:=-2\di\frac{\beta^{AB} k_{Ba}}{(1+\alpha)p\cdot q}$. This is just a (residual) gauge transformation as a consequence of $\beta\cdot p=0$.} One may analyse other spin fields in a very similar way and we shall present a few more comments in Remark \ref{rem:HigherSpinFields} below.

Let us now introduce some notions and notation that we shall make use of in this paper. We shall mostly be interested in {\em chiral zero-rest-mass fields}, i.e.\ fields forming representations $(\mathbf{2h+1},\mathbf{1})$, $h\in \tfrac{1}{2}\NN_0$, of the little group $\sSL(2,\FC)\times \widetilde{\sSL(2,\FC)}$. These fields will carry $2h$ symmetrised spinor indices. Specifically, using the conventions
\begin{equation}\label{eq:ConformalWeights6d}
\begin{aligned}
 \quad [k]\ :=\ \otimes^k\det S^\vee~,\quad [-k]\ :=\ [k]^\vee\eand [0]\ :=\ [k]\otimes[-k]\efor k\ \in\ \NN~,\\
 \CS[\pm k]\ :=\ \CS\otimes_{\CO_{M^6}}[\pm k]~~\mbox{for some Abelian sheaf $\CS$ on $M^6$}~,\kern1.6cm
 \end{aligned}
\end{equation}
we shall denote the sheaf of  chiral zero-rest-mass fields on $M^6$ by $\CZ_h$,
\begin{equation}\label{eq:DefZRM}
 \CZ_h\ :=\ \begin{cases}
\ker\left\{\partial^{AB}\,:\,(\odot^{2h}S^\vee)[1]\ \to\
      (\odot^{2h-1}S^\vee\otimes_{\CO_{M^6}} S)_0[2]    \right\} &\efor h\ \geq\ \frac12~,\\
\ker\left\{\square:=\tfrac14 \partial^{AB}\partial_{AB}\,:\,[1]\ \to\
      [2]    \right\} &\efor h\ =\ 0~.\\
               \end{cases}
\end{equation}
Here, the subscript zero refers to the totally trace-less part. The factors $[\pm k]$ are referred to as {\em conformal weights}, as they render the zero-rest-mass field equations conformally invariant. See also Penrose \& Rindler \cite{Penrose:1985jw,Penrose:1986ca} or Ward \& Wells  \cite{Ward:1990vs} for the discussion of conformal weights in the four-dimensional setting.

{\rem\label{rem:HigherSpinFields}
Recall that there is a potential formulation of zero-rest-mass fields in four dimensions, cf.\ e.g.~Ward \& Wells \cite{Ward:1990vs}. This formulation generalises to six dimensions, as we shall demonstrate now. Consider an $h\in \tfrac12\NN^*$. From the potential fields
\begin{equation}
B_A{}^{A_1\cdots A_{2h-1}}\ =\ B_A{}^{(A_1\cdots A_{2h-1})}\ \in\ H^0(M^6,(\odot^{2h-1}S\otimes_{\CO_{M^6}}S^\vee)_0[1])~,
\end{equation}
we derive a field strength $H_{A_1\cdots A_{2s}}\in H^0(U,(\odot^{2h}S^\vee)[1])$ according to
\begin{equation}
 H_{A_1\cdots A_{2h}}\ :=\ \partial_{(A_1B_1}\cdots\partial_{A_{2h-1}B_{2h-1}} B_{A_{2h})}{}^{B_1\cdots B_{2h-1}}~.
\end{equation}
The equations
\begin{equation}\label{eq:PotEqB-Intro}
 H^{A_1\cdots A_{2h}}\ :=\ \partial^{A (A_1}B_A{}^{A_2\cdots A_{2h})}\ =\ 0~
\end{equation}
then imply that
\begin{equation}
 \partial^{AA_1}H_{A_1\cdots A_{2h}}\ =\ 0~.
\end{equation}
Furthermore, the pair of spinors $(H_{A_1\cdots A_{2h}},H^{A_1\cdots A_{2h}})$ is invariant under gauge transformations of the form
\begin{equation}\label{eq:GaugeTrafePotB-Intro}
 B_B{}^{A A_1\cdots A_{2h-2}}\ \mapsto\  B_B{}^{A A_1\cdots A_{2h-2}}+\left[\partial_{CB}\Lambda^{C(AA _1\cdots A_{2h-2})}-\partial^{C(A}\Lambda_{CB}{}^{A_1\cdots A_{2h-2})}\right]_0~,
\end{equation}
where the subscript zero refers again to the totally trace-less part and $\Lambda_{AB}{}^{A_1\cdots A_{2h-2}}=\Lambda_{[AB]}{}^{(A_1\cdots A_{2h-2})}$ is totally trace-less itself. Note that the traces of $\big[\partial_{CB}\Lambda^{C(AA _1\cdots A_{2h-2})}-\partial^{C(A}\Lambda_{CB}{}^{A_1\cdots A_{2h-2})}\big]$ always drop out of the above definition of $(H_{A_1\cdots A_{2h}},H^{A_1\cdots A_{2h}})$. Altogether, the spinor field $H_{A_1\cdots A_{2h}}$ can therefore be regarded as a section of the sheaf $\CZ_h$. We shall make use of this formulation in Section \ref{sec:PWTransform} when dealing with the Penrose--Ward transform.
}

\section{Twistor space of six-dimensional space-time}\label{sec:TwistorSpace}

In this section, we shall review a particular twistor space associated with $M^6$ that is a very natural generalisation of known twistor spaces and suitable for the description of chiral theories in six dimensions. This twistor space has appeared earlier e.g.\ in
\cite{Penrose:1985jw,Penrose:1986ca,springerlink:10.1007/BF00132253,Hughston:1986hb,Hughston:1979TN,Hughston:1982TN,Hughston:1984TN,Hughston:1987aa,Hughston:1987he,Hughston:1988nz,Baston:1989,Inoue:1992aa,Berkovits:2004bw,Chern:2009nt}. Here we shall present a detailed discussion of its constructions from an alternative point of view.

{\rem We shall always be working with locally free sheaves and therefore we shall not make a notational distinction between vector bundles and their corresponding sheaves of sections. We shall switch between the two notions freely depending on context.}

\subsection{Twistor space from space-time}

Let us consider the projectivisation  $\PP(S^\vee)$ of the dual anti-chiral spin bundle $S^\vee$. Since $S^\vee$ is of rank four, $\PP(S^\vee)\to M^6$ is a $\PP^3$-bundle over $M^6$. Hence,  the projectivisation $\PP(S^\vee)$ is a nine-dimensional complex manifold $F^9\cong\FC^6\times\PP^3$, the {\em correspondence space}. We take $(x,\lambda)=(x^{AB},\lambda_A)$ as coordinates on $F^9$, where $\lambda_A$ are homogeneous coordinates on $\PP^3$.

Consider now the following vector fields on $F^9$:
\begin{equation}\label{eq:TwistorDistribution}
V^A\ :=\ \lambda_B\frac{\partial}{\partial x_{AB}}~.
\end{equation}
Note that $\lambda_A V^A=0$ because of the anti-symmetry of the spinor indices in the partial derivative. These vector fields define an integrable rank-3 distribution on $F^9$, which we call {\em twistor distribution}. Therefore, we have a foliation of $F^9$ by three-dimensional complex manifolds. The resulting quotient will be {\em twistor space}, a six-dimensional manifold denoted by $P^6$. We have thus established the following double fibration:
\begin{equation}\label{eq:DoubleFibration}
 \begin{picture}(50,40)
  \put(0.0,0.0){\makebox(0,0)[c]{$P^6$}}
  \put(64.0,0.0){\makebox(0,0)[c]{$M^6$}}
  \put(34.0,33.0){\makebox(0,0)[c]{$F^9$}}
  \put(7.0,18.0){\makebox(0,0)[c]{$\pi_1$}}
  \put(55.0,18.0){\makebox(0,0)[c]{$\pi_2$}}
  \put(25.0,25.0){\vector(-1,-1){18}}
  \put(37.0,25.0){\vector(1,-1){18}}
 \end{picture}
\end{equation}
Let  $(z,\lambda)=(z^A,\lambda_A)$ be homogeneous coordinates on $\PP^7$ and assume that $\lambda_A\neq0$. This effectively means that we are working on the open subset
\begin{equation}
\PP^7_\circ\ :=\ \PP^7\setminus\PP^3
\end{equation}
of $\PP^7$, where the removed $\PP^3$ is given by $z^A\neq0$ and $\lambda_A=0$. In the double fibration \eqref{eq:DoubleFibration}, the projection $\pi_2$ is the trivial projection and $\pi_1\,:\,(x^{AB},\lambda_A)\mapsto (z^A,\lambda_A)=(x^{AB}\lambda_B,\lambda_A)$. Thus, $P^6$ forms a quadric hypersurface inside $\PP^7_\circ$, which is given by the equation
\begin{equation}\label{eq:quadric}
 z^A\lambda_A\ =\ 0~.
\end{equation}
We shall refer to the relation
\begin{equation}\label{eq:incidence}
 z^A\ =\ x^{AB}\lambda_B
\end{equation}
as {\em incidence relation}, because it is a direct generalisation of Penrose's incidence relation in four dimensions.

\paragraph{Geometric twistor correspondence.}
The double fibration \eqref{eq:DoubleFibration} shows that points in either of the base spaces $M^6$ and $P^6$ correspond to subspaces of the other base space: For any point $x\in M^6$, the corresponding manifold $\hat x:=\pi_1(\pi_2^{-1}(x))\hookrightarrow P^6$ is a three-dimensional complex manifold bi-holomorphic to $\PP^3$ as follows from \eqref{eq:incidence}. Conversely, for any fixed $p=(z,\lambda)\in P^6$, the most general solution to the incidence relation \eqref{eq:incidence} is given by
\begin{equation}\label{eq:IncidenceSolution}
 x^{AB}\ =\ x_0^{AB}+\varepsilon^{ABCD}\mu_C\lambda_D~,
\end{equation}
where $x_0^{AB}$ is a particular solution and $\mu_A$ is arbitrary. This defines a totally null-plane $\pi_2(\pi_1^{-1}(p))$ in $M^6$. This plane is three-dimensional because of the freedom in the choice of $\mu_A$ given by the shifts $\mu_A\mapsto\mu_A+\varrho\,\lambda_A$ for $\varrho\in\FC$ which do not alter the solution \eqref{eq:IncidenceSolution}.

Altogether, points in space-time correspond to complex projective three-spaces in twistor space while points in twistor space correspond to totally null three-planes in space-time. Thus, twistor space parametrises all totally null three-planes of space-time.

\paragraph{Twistor space as a normal bundle.}
The above considerations imply that $P^6$ can be viewed as a holomorphic vector bundle over $\PP^3$, where the global holomorphic sections are given by the incidence relation \eqref{eq:incidence}. In fact, \eqref{eq:incidence} shows that $P^6$ is a rank-3 subbundle of the bundle $\CO_{\PP^3}(1)\otimes\FC^4\to\PP^3$, whose total space is $\PP^7_\circ$. Here and in the following, $\CO_{\PP^3}(1)$ denotes the dual tautological bundle\footnote{or hyperplane bundle} over $\PP^3$.

To identify the subbundle $P^6$, let us denote by $N_{Y|X}$ the normal bundle of some complex submanifold $Y$ of a complex manifold $X$, $i\,:\,Y\hookrightarrow X$.  This bundle is defined by the following short exact sequence:
\begin{equation}\label{eq:NormalSequence0}
0\ \longrightarrow\ T_Y\ \longrightarrow\ i^*T_X\ \longrightarrow\  N_{Y|X}\ \longrightarrow\ 0~.
\end{equation}
Let us now specialise to $Y=\PP^3$ and $X=\PP^7$ with coordinates $(z^A,\lambda_A)$ on $\PP^7$ as before. If $\PP^3\embd \PP^7$ is given by $z^A=0$ and $\lambda_A\neq0$, then $T_{\PP^3}=\langle \frac{\partial}{\partial\lambda_A} \rangle$ and $T_{\PP^7}=\langle \frac{\partial}{\partial z^A}, \frac{\partial}{\partial\lambda_A}\rangle$. The normal bundle of $N_{\PP^3|\PP^7}$ of $\PP^3$ inside $\PP^7$ is given by
\begin{subequations}
\begin{equation}\label{eq:NormalSequence1}
0\ \longrightarrow\ T_{\PP^3}\ \longrightarrow\ i^*T_{\PP^7}\ \longrightarrow\  N_{\PP^3|\PP^7}\ \longrightarrow\ 0~,
\end{equation}
which implies that
\begin{equation}
 N_{\PP^3|\PP^7}\ \cong\ \CO_{\PP^3}(1)\otimes\FC^4~,
\end{equation}
\end{subequations}
since the coefficient functions of the basis vector fields $\frac{\partial}{\partial z^A}$ and $\frac{\partial}{\partial\lambda_A}$ are linear in the coordinates. Hence, the $z^A$ can be regarded as fibre coordinates of $N_{\PP^3|\PP^7}$, while the $\lambda_A$ are base coordinates.

Using these results, we find that our twistor space $P^6$ fits into the short exact sequence\footnote{We use the notation $\CO_{\PP^3}(k):=\otimes^k\CO_{\PP^3}(1)$ and $\CO_{\PP^3}(-k):=\CO^\vee_{\PP^3}(k)$, $k>0$, as well as $\CO_{\PP^3}(0)=\CO_{\PP^3}$.}
\begin{subequations}
\begin{equation}\label{eq:NormalSequence2}
 0\ \longrightarrow\ P^6\ \longrightarrow\ N_{\PP^3|\PP^7}\ \overset{\kappa}{\longrightarrow}\ \CO_{\PP^3}(2)\ \longrightarrow\ 0~,
\end{equation}
where
\begin{equation}
 \kappa\,:\, (z^A,\lambda_A)\ \mapsto\ z^A\lambda_A~.
\end{equation}
\end{subequations}
Note that the sequence \eqref{eq:NormalSequence2} can be regarded as an alternative definition of twistor space. Again, we see that $P^6$ is a rank-3 subbundle of $\CO_{\PP^3}(1)\otimes\FC^4\to\PP^3$ as stated earlier. It also shows that $P^6$ is the normal bundle of $\PP^3$ inside the quadric hypersurface $\FQ^6\hookrightarrow\PP^7$ given by the zero locus $z^A\lambda_A=0$. Moreover, notice that the open subset $\FQ^6\cap\PP^7_\circ$ can be identified with $P^6$.

\subsection{Space-time from twistor space}

Next we wish to address the problem of how to obtain space-time $M^6$, and in particular the factorisation \eqref{eq:TangentBundle} of the tangent bundle, from twistor space using \eqref{eq:NormalSequence2}. To this end, consider the long exact sequence of cohomology groups induced by the short exact sequence \eqref{eq:NormalSequence2},
\begin{equation}\label{eq:CohomologySequence1}
\begin{aligned}
 &0\ \longrightarrow\ H^0(\PP^3,P^6)\ \longrightarrow\  H^0(\PP^3,N_{\PP^3|\PP^7})\ \overset{\kappa}{\longrightarrow}\  H^0(\PP^3,\CO_{\PP^3}(2))\ \longrightarrow\\
&\kern1cm \longrightarrow\ H^1(\PP^3,P^6)\ \longrightarrow\  H^1(\PP^3,N_{\PP^3|\PP^7})\ \longrightarrow\  H^1(\PP^3,\CO_{\PP^3}(2))\ \longrightarrow\ \cdots~,
\end{aligned}
\end{equation}
where we have slightly abused notation by again using the letter $\kappa$. To compute these cohomology groups, we recall a special case of the Borel--Weil--Bott theorem:

{\lemma (Bott's rule \cite{Bott57homogeneousvector})\label{lem:Bott}
Let $V$ be an $n$-dimensional complex vector space. Consider its projectivisation $\PP(V)$ together with the hyperplane bundle $\CO_{\PP(V)}(1)$.  Furthermore, set $\CO_{\PP(V)}(k):=\otimes^k\CO_{\PP(V)}(1)$,
 $\CO_{\PP(V)}(-k):=\CO^\vee_{\PP(V)}(k)$ and $\CO_{\PP(V)}(0):=\CO_{\PP(V)}$ for $k\in\NN$. Then
\begin{equation}
 H^q(\PP(V),\CO_{\PP(V)}(k))\ \cong\ \begin{cases}
                                               \odot^k V^\vee &\efor q=0\mbox{ \& } k\geq0~,\\
                                               \odot^{-k-n}V\otimes\det V &\efor q=n-1\mbox{ \& } k\leq -n~,\\
                                               0 &\quad\mbox{otherwise}~,
                                          \end{cases}
\end{equation}
where $\det V\equiv\Lambda^n V$.
}

\vspace{10pt}
\noindent
From Bott's rule for $V=\FC^4$, we find that $H^1(\PP^3,N_{\PP^3|\PP^7})=0=H^1(\PP^3,\CO_{\PP^3}(2))$ and furthermore
\begin{equation}\label{eq:Kodaira1}
 H^1(\PP^3,P^6)\ =\ 0~,
\end{equation}
since $\kappa$ is surjective. Therefore, the long exact sequence of cohomology groups \eqref{eq:CohomologySequence1} reduces to
\begin{equation}
 0\ \longrightarrow\ H^0(\PP^3,P^6)\ \longrightarrow\  H^0(\PP^3,N_{\PP^3|\PP^7})\ \overset{\kappa}{\longrightarrow}\  H^0(\PP^3,\CO_{\PP^3}(2))\ \longrightarrow\ 0~.
\end{equation}
By applying Bott's rule again, we deduce from the latter sequence that
\begin{equation}\label{eq:Kodaira2}
 \dim_\FC H^0(\PP^3,P^6)\ =\ 6~.
\end{equation}
Because of \eqref{eq:Kodaira1} and \eqref{eq:Kodaira2}, we may now apply Kodaira's theorem of relative deformation theory\footnote{Recall that Kodaira's theorem states that if $Y$ is a compact complex submanifold of a not necessarily compact complex manifold $X$ with $H^1(Y,N_{Y|X})=0$, then there exists a $\dim_\FC H^0(Y,N_{Y|X})$-dimensional family of deformations of $Y$ inside $X$. For more details, see e.g.~\cite{Burns:1979,Kodaira:1986}.} to conclude that there is a six-dimensional family of deformations of  $\PP^3$ inside the quadric hypersurface $\FQ^6\hookrightarrow\PP^7$. We shall denote this family by $M^6$ and the individual deformation of $\PP^3$ labelled by $x\in M^6$ as $\hat{x}$.

Next we define the correspondence space $F^9$ according to
\begin{equation}
 F^9\ :=\ \{(p,x)\in P^6\times M^6\,|\,p\in \hat x\}~,
\end{equation}
Notice that $F^9$ is fibred over both $P^6$ and $M^6$.  The typical fibres of $\pi_2\,:\,F^9\to M^6$ are complex projective three-spaces $\PP^3$. Hence, we have again established a double fibration of the form \eqref{eq:DoubleFibration}, where the fibres of $F^9\to P^6$ are three-dimensional complex submanifolds of $M^6$.

On $F^9$, we may consider the relative tangent bundle, denoted by $T_{\pi_1}$, along the fibration $\pi_1\,:\,F^9\to P^6$. It is of rank three and defined by
\begin{equation}\label{eq:RelTanBun}
 0\ \longrightarrow\ T_{\pi_1}\ \longrightarrow\ T_{F^9}\ \longrightarrow\ \pi_1^*T_{P^6}\ \longrightarrow\ 0~.
\end{equation}
By construction, the  vector fields $V^A$ given in \eqref{eq:TwistorDistribution} annihilate $z^A=x^{AB}\lambda_B$ and therefore, $T_{\pi_1}$ can be identified with the twistor distribution generated by $V^A$, cf.\ \eqref{eq:TwistorDistribution}. Hence, sections $\mu_A$ of $T_{\pi_1}$ are defined up to shifts by terms proportional to $\lambda_A$ (recall that $\lambda_AV^A=0$). Then we define a bundle $N$ on $F^9$ by
\begin{equation}\label{eq:DefOfN}
\begin{aligned}
0\kern15pt\longrightarrow\kern10pt &T_{\pi_1}\kern15pt \longrightarrow\kern15pt &&\pi_2^*T_{M^6}\kern15pt \longrightarrow\kern15pt N\kern15pt \longrightarrow \kern15pt 0~,\\
                &\kern2pt \mu_A\kern20pt \mapsto     &&\kern-18pt\ \varepsilon^{ABCD}\mu_C\lambda_D~,\\
                &                                                                   &&\kern10pt \xi^{AB}\kern20pt \mapsto\kern10pt  \xi^{AB}\lambda_B~.
\end{aligned}
\end{equation}
Clearly, the rank of $N$ is three and the restriction of $N$ to the fibre $\pi_2^{-1}(x)$ of $F^9\to M^6$ for $x\in M^6$ is isomorphic  to the pull-back $\pi_1^*N_{\hat x|P^6}$ of the normal bundle $N_{\hat x|P^6}$ of $\hat x\hookrightarrow P^6$. Thus, $N$ can be identified with  $\pi_1^*N_{\hat x|P^6}$.\footnote{Note that $N_{\hat x|P^6}$ is bi-holomorphic to $P^6\to\PP^3$.}

These considerations allow us to reconstruct the tangent bundle $T_{M^6}$ from twistor space. In fact, we may apply the direct image functor (with regard to $\pi_2$) to the short exact sequence \eqref{eq:DefOfN}. Since both direct images\footnote{\label{foot:DirectImage}Remember that the $q$-th direct image  sheaf $\pi_*^q\CS$ of some Abelian sheaf $\CS$ on $X$ for some map $\pi\,:\,X\to Y$ is defined by the pre-sheaf $Y\supset U\ \mbox{open}\mapsto H^q(\pi^{-1}(U),\CS)$. We abbreviate $\pi_*\CS:=\pi^0_*\CS$. See also Section \ref{sec:6DPenrose} for details of the computation of direct image sheaves, in particular Proposition \ref{prop:DirectImage}.} $\pi_{2*}T_{\pi_1}$ and $\pi_{2*}^1T_{\pi_1}$ vanish, we obtain
\begin{equation}\label{eq:TanBunTwis}
 T_{M^6}\ \cong\ \pi_{2*}\,\pi^*_1N_{\hat x|P^6}
 \qquad\Longleftrightarrow\qquad
 (T_{M^6})_x\ \cong\ H^0(\hat x,N_{\hat x|P^6})~.
\end{equation}
Elements of $H^0(\hat x,N_{\hat x|P^6})$ are given in terms of elements of $H^0(\PP^3,P^6)$ by allowing the latter to depend on $x$. One can check that this dependence is holomorphic in an open neighbourhood of $x$.

What remains to be understood is how the explicit factorisation \eqref{eq:TangentBundle} of the tangent bundle emerges from the above construction and in particular from  $H^0(\PP^3,P^6)$. To show this, we consider the Euler sequence for $\PP^3$,
\begin{equation}\label{eq:EulerSequence}
 0\ \longrightarrow\ \CO_{\PP^3}\ \longrightarrow\  \CO_{\PP^3}(1)\otimes\FC^4\ \longrightarrow\ T_{\PP^3}\ \longrightarrow\ 0~.
\end{equation}
Upon dualising this sequence and twisting by $\CO_{\PP^3}(2)$, we find\footnote{Here and in the following, we shall denote the sheaf of one-forms on some manifold $X$ by $\Omega^1_X$.}
\begin{equation}\label{eq:EulerSequence2}
 0\ \longrightarrow\  \Omega^1_{\PP^3}\otimes\CO_{\PP^3}(2)\ \longrightarrow\  \CO_{\PP^3}(1)\otimes\FC^4\ \longrightarrow\ \CO_{\PP^3}(2)\ \longrightarrow\ 0~.
\end{equation}
By comparing with \eqref{eq:NormalSequence2}, we may conclude that
\begin{equation}
 P^6\ \cong\ \Omega^1(2) \ewith \Omega^p(k)\ :=\ \Omega^p_{\PP^3}\otimes\CO_{\PP^3}(k)~.
\end{equation}
Thus, elements of $H^0(\PP^3,P^6)$ can also be viewed as elements of $H^0(\PP^3,\Omega^1(2))$. The latter are of the form $\omega=\omega^{AB}\lambda_A\dd\lambda_B$ with $\omega^{AB}=-\omega^{BA}$. Since
\begin{equation}
 S_x\ \cong\ H^0(\hat x,\CO_{\hat x}(1))
\end{equation}
via $s^A\mapsto s^A\lambda_A$ for $s^A\in S_x$, we indeed find the factorisation $(T_{M^6})_x\cong S_x\wedge S_x$. This concludes our construction of space-time from twistor space.

{\rem
Notice that an identification of the form \eqref{eq:TangentBundle} amounts to choosing a (holomorphic) conformal structure. This can be seen as follows: Let $X$ be a six-dimensional complex spin manifold. The first definition of a conformal structure on $X$ (and perhaps the standard one) assumes an equivalence class $[g]$, the conformal class, of holomorphic metrics $g$ on $X$. Two given metrics $g$ and $g'$ are called equivalent if $g'=\gamma^2 g$ for some nowhere vanishing holomorphic function $\gamma$. Thus, a conformal structure is a line subbundle $L$ in $T^\vee_X\odot T^\vee_X$.

An alternative definition of a conformal structure assumes a factorisation of the form $T_X\cong S\wedge S$, where $S$  is the rank-4 chiral spin bundle. This isomorphism in turn gives (canonically) the line subbundle $\det S^\vee\equiv\Lambda^4S^\vee$ in $T^\vee_X\odot T^\vee_X$ since upon using splitting principle arguments (see~e.g.~\cite{Griffiths:1978}), one finds the identification $K_X:=\det T^\vee_X\cong\otimes^3\det S^\vee$ for the canonical bundle $K_X$. Hence, $\det S^\vee$ can be identified with the line bundle $L$ from above, and the metric $g$ is then of the form $\gamma^2\,\varepsilon_{ABCD}$.
}

\section{Penrose transform in six dimensions}\label{sec:6DPenrose}

Having defined twistor space, we would like to understand differentially constrained data on space-time in terms of differentially unconstrained data on twistor space. Specifically, we are interested in the chiral fields introduced in Section \ref{ssec:zrmfields} and prove the following theorem:\footnote{We shall use the notation $\CO_{P^6}(k):={\rm pr}^*\CO_{\PP^3}(k)$ where ${\rm pr}:P^6\to\PP^3$ is the bundle projection and likewise for the open sets.}

{\theorem\label{thm:6DPenrose}
Consider the double fibration \eqref{eq:DoubleFibration}. Let $U\subset M^6$ be open and convex and set  $U':=\pi_2^{-1}(U)\subset F^9$ and $\hat U:=\pi_1(\pi_2^{-1}(U))\subset P^6$, respectively. For $h\in\frac12\NN_0$, there is a canonical isomorphism
\begin{equation}
\CCP\,:\,H^3(\hat U,\CO_{\hat U}(-2h-4))\ \to\ H^0(U,\CZ_h)~,
\end{equation}
where $\CZ_h$ is the sheaf of chiral zero-rest-mass fields defined in \eqref{eq:DefZRM}. This transformation is called the Penrose transform.
}

\vspace{10pt}
\noindent
Note that contrary to the corresponding theorem in four dimensions, $h$ is restricted to be non-negative. The case of negative $h$ has to be treated differently as the cohomology group $H^3(\hat U,\CO_{\hat U}(-2h-4))$ in this case yields nothing non-trivial on space-time. We shall come to this issue in Section \ref{sec:PWTransform}.

There are various steps involved in proving Theorem \ref{thm:6DPenrose} and the structure of the proof we shall present is similar to the one given in the four-dimensional setting. Therefore, the reader might find it useful to consult additionally e.g.\ the article by Eastwood, Penrose \& Wells \cite{Eastwood:1981jy} or the text book by Ward \& Wells \cite{Ward:1990vs} and references therein, where the four-dimensional case is presented in detail. We also refer to Buchdahl \cite{Buchdahl:1985aa} and Pool \cite{Pool:1981aa} for very good accounts on various cohomological constructions which we shall make use of below. Murray \cite{springerlink:10.1007/BF00132253} gave a proof for Penrose transforms on twistor spaces of certain even-dimensional Riemannian manifolds and Baston \& Eastwood \cite{Baston:1989} provided an abstract discussion. Here, we shall present a very detailed proof which resembles the one in \cite{Eastwood:1981jy} or \cite{Ward:1990vs} in four dimensions and which, in addition, can be transferred relatively straightforwardly to our discussion presented in Section \ref{sec:HPT-SDS}. To simplify our discussion, we shall restrict ourselves to complex holomorphic fields which correspond to real-analytic solutions upon imposing reality conditions as briefly discussed in Appendix \ref{app:spinors}. Of course, it would be interesting to extend the discussion to hyperfunction solutions as well.

We first present the cohomological foundations needed for proving the above Penrose transform; these considerations are also needed in Section \ref{sec:PWTransform}. In particular, we shall introduce the so-called relative de Rham complex on the correspondence space of the double fibration  \eqref{eq:DoubleFibration} and compute its cohomology by applying the direct image functor with respect to the projection $\pi_2$. We shall also recall a result of Buchdahl which allows us to pull-back cohomological data from the twistor space to the correspondence space. Once we have presented the setup, we give the proof of Theorem \ref{thm:6DPenrose}.

\subsection{Cohomological considerations}

\paragraph{Relative de Rham complex.}
The starting point of our considerations is the double fibration \eqref{eq:DoubleFibration}. As a first tool in proving the Penrose transform, we introduce the relative differential forms $\Omega^p_{\pi_1}$, i.e.\ the differential $p$-forms along the fibres of the fibration $\pi_1:F^9\to P^6$. We have already introduced the corresponding relative tangent bundle in \eqref{eq:RelTanBun}. Simply dualising this sequence, we obtain the definition of the sheaf of relative one-forms from
\begin{equation}\label{eq:RelOneForms}
 0\ \longrightarrow\ \pi_1^*\Omega^1_{P^6}\ \longrightarrow\ \Omega^1_{F^9}\ \longrightarrow\ \Omega^1_{\pi_1}\ \longrightarrow\ 0~.
\end{equation}
Recall from our previous discussion that in our parametrisation, sections $\mu_A$ of the relative tangent bundle $T_{\pi_1}$ are defined up to shifts by terms proportional to $\lambda_A$. This, in turn, induces the condition $\omega^A\lambda_A=0$ on sections $\omega^A$ of $\Omega^1_{\pi_1}$. We shall come back to this point when discussing the direct images of $\Omega^1_{\pi_1}$.

In general, we introduce the relative $p$-forms $\Omega^p_{\pi_1}$ on $F^9$ with respect to the fibration $\pi_1:F^9\to P^6$ according to
\begin{equation}\label{eq:RelOneForms2}
 0\ \longrightarrow\ \pi_1^*\Omega^1_{P^6}\wedge\Omega^{p-1}_{F^9}\ \longrightarrow\ \Omega^p_{F^9}\ \longrightarrow\ \Omega^p_{\pi_1}\ \longrightarrow\ 0~.
\end{equation}
Thus, relative $p$-forms have components only along the fibres of $\pi_1:F^9\to P^6$ (i.e.\ any contraction with a vector field which is a section of $\pi_1^* T P^6$ vanishes). The coefficient functions in local coordinates, however, depend on both the base and the fibre coordinates. Note that the maximum value of $p$ here is three. If we let $\mbox{pr}_{\pi_1}\!:\Omega_{F^9}^p\to \Omega^p_{\pi_1}$ be the quotient mapping, we can define the relative exterior derivative $\dd_{\pi_1}$ by setting
\begin{equation}\label{eq:RED1}
 \dd_{\pi_1}\ :=\ \mbox{pr}_{\pi_1}\circ\dd\,:\, \Omega^p_{\pi_1}\ \to\ \Omega^{p+1}_{\pi_1}~,
\end{equation}
where $\dd$ is the usual exterior derivative on $F^9$. In local coordinates $(x^{AB},\lambda_A)$ on $F^9$, the relative exterior derivative can be presented in terms of the vector fields \eqref{eq:TwistorDistribution}.

Next, observe that the relative differential $\dd_{\pi_1}$ induces the {\em relative de Rham complex}. This complex is given in terms of an injective  resolution of the topological inverse\footnote{Remember that the topological inverse $\pi^{-1}\CS$ of some Abelian sheaf $\CS$ on $Y$ for  $\pi\,:\,X\to Y$ is defined by the pre-sheaf $X\supset U\ \mbox{open} \mapsto H^0(\pi(U),\CS)$.} $\pi_1^{-1}\CO_{P^6}$ of $\CO_{P^6}$ on the correspondence space $F^9$:
\begin{equation}\label{eq:RelDeRhamCom}
 0\ \longrightarrow\ \pi_1^{-1}\CO_{P^6}\ \longrightarrow\ \CO_{F^9}\ \xrightarrow{\dd_{\pi_1}}\  \Omega^1_{\pi_1}\ \xrightarrow{\dd_{\pi_1}}\  \Omega^2_{\pi_1}\ \xrightarrow{\dd_{\pi_1}}\  \Omega^3_{\pi_1}\ \longrightarrow\ 0~.
\end{equation}
We shall not explain here why this sequence is exact but instead refer the interested reader to Ward \& Wells \cite{Ward:1990vs} or Buchdahl \cite{Buchdahl:1985aa}, where more general discussions are given.

A natural question is now if the sheaves  $\Omega^p_{\pi_1}$ have an interpretation in terms of certain pull-back sheaves from space-time and twistor space. Notice that the vectors fields \eqref{eq:TwistorDistribution} are given by $V^A=\frac12\varepsilon^{ABCD}\lambda_B\partial_{CD}$, where $\dpar_{AB}$ are the vector fields spanning $T_{M^6}$. In terms of the $V^A$, the map $\dd_{\pi_1}:\CO_{F^9}\to \Omega^1_{\pi_1}$ reads explicitly as
\begin{equation}
 V^A\,:f\ \mapsto\ \omega^A\ =\ V^Af\ =\ \tfrac12\varepsilon^{ABCD}\lambda_B\partial_{CD} f~,\quad f\ \in\ \CO_{F^9}~.
\end{equation}
This  shows that $\omega^A=V^Af$ is a section of $\pi_2^*(\det S^\vee\otimes_{\CO_{M^6}} S)\otimes_{\CO_{F^9}}\pi_1^*\CO_{P^6}(1)$. Clearly, it is not the most general section of this sheaf, since we have $\lambda_A\omega^A=\lambda_AV^Af=0$; see also our comments given below \eqref{eq:RelOneForms}. For a general section $s^A$ of $\pi_2^*(\det S^\vee\otimes_{\CO_{M^6}} S)\otimes_{\CO_{F^9}}\pi_1^*\CO_{P^6}(1)$,  the map $\lambda_A:s^A\mapsto s^A\lambda_A$ gives a section of $\pi_2^*\det S^\vee\otimes_{\CO_{F^9}}\pi_1^*\CO_{P^6}(2)$ and its kernel gives  $\Omega^1_{\pi_1}$.
Altogether, we conclude that $\Omega^1_{\pi_1}$ fits into the following short exact sequence:
\begin{equation}\label{eq:Rel1FSES}
\begin{aligned}
 0\ \longrightarrow\ \Omega^1_{\pi_1}\ \longrightarrow\ \pi_2^*(\det S^\vee\otimes_{\CO_{M^6}} S)\otimes_{\CO_{F^9}}\pi_1^*\CO_{P^6}(1)\ \longrightarrow\kern2cm \\
\longrightarrow\  \pi_2^*\det S^\vee\otimes_{\CO_{F^9}}\pi_1^*\CO_{P^6}(2)\ \longrightarrow\ 0~.
 \end{aligned}
\end{equation}
Using the notation \eqref{eq:ConformalWeights6d}, we then obtain the following proposition:

{\prop\label{prop:RelFormsPullBackSheaves}
The sheaves appearing in the relative de Rham sequence \eqref{eq:RelDeRhamCom} can be canonically identified as follows. With $\Omega^{p}_{\pi_1}(k):=\Omega^{p}_{\pi_1}\otimes_{\CO_{F^9}}\pi_1^*\CO_{P^6}(k)$, we have
\begin{equation}\label{eq:Rel1FormsPullBackSheaves}
\begin{aligned}
 0\ \longrightarrow\ \Omega^p_{\pi_1}\ \longrightarrow\ \pi_2^*(\Lambda^pS)[p]\otimes_{\CO_{F^9}}\pi_1^*\CO_{P^6}(p)\
\longrightarrow\  \pi_2^*[1]\otimes_{\CO_{F^9}}\Omega^{p-1}_{\pi_1}(2)\ \longrightarrow\ 0~.
 \end{aligned}
\end{equation}
}

\noindent
{\em Proof:} Using the fact that short exact sequences of the form $0\to\CE\to\CF\to\CL\to0$, where $\CL$ is the sheaf of sections of some line bundle, always induce $0\to\Lambda^p\CE\to \Lambda^p\CF\to\Lambda^{p-1}\CE\otimes\CL\to0$, the sequence \eqref{eq:Rel1FSES} immediately leads to  \eqref{eq:Rel1FormsPullBackSheaves}. \hfill $\Box$

\vspace{10pt}
Finally, we point out that the relative de Rham sequence \eqref{eq:RelDeRhamCom} has a natural extension via twisting by a holomorphic vector bundle. Specifically, let $E\to P^6$ be a holomorphic vector bundle over $P^6$ and consider the pull-back bundle $\pi_1^*E$ over the correspondence space $F^9$. We may tensor \eqref{eq:RelDeRhamCom} by $\pi_1^{-1}\CO_{P^6}(E)$, which is the sheaf of sections of $\pi_1^*E$ that are constant along $\pi_1:F^9\to P^6$. Because $\CO_{F^9}(\pi_1^*E)\cong\pi_1^*\CO_{P^6}(E)$ and $\CO_{F^9}\otimes_{\pi_1^{-1}\CO_{P^6}}\pi_1^{-1}\CO_{P^6}(E)$ are canonically isomorphic, we find
\begin{subequations}
\begin{equation}\label{eq:RelDeRhamComVect}
 0\ \longrightarrow\ \pi_1^{-1}\CO_{P^6}(E)\ \longrightarrow\ \Omega^0_{\pi_1}(E)\ \xrightarrow{\dd_{\pi_1}}\   \cdots\ \xrightarrow{\dd_{\pi_1}}\  \Omega^3_{\pi_1}(E)\ \longrightarrow\ 0~,
\end{equation}
where we have defined
\begin{equation}
  \Omega^0_{\pi_1}(E)\ :=\ \CO_{F^9}(\pi_1^*E)
  \eand
   \Omega^p_{\pi_1}(E)\ :=\   \Omega^p_{\pi_1}\otimes_{\CO_{F^9}}\CO_{F^9}(\pi_1^*E)~.
\end{equation}
\end{subequations}

\paragraph{Direct image sheaves.}
The next important ingredient for our subsequent discussion is the direct images of $\Omega^p_{\pi_1}(E)$ with respect to the fibration $\pi_2:F^9\to M^6$ for the special case $E=\CO_{P^6}(k)$, $k\in\RZ$.  To  compute those, we shall make use of the following lemma:

{\lemma\label{lem:Dolbeault}
Let $V$ be a four-dimensional complex vector space together with its projectivisation $\PP(V)$. Using the shorthand notations
$\Omega^p(k):=\Omega^p_{\PP(V)}\otimes\CO_{\PP(V)}(k)$ and $\Omega^0(k):=\CO_{\PP(V)}(k)$, we have the following list of sheaf cohomology groups:
\begin{subequations}
\begin{equation}\label{eq:Cohom1}
 H^q(\PP(V),\Omega^0(k))\ \cong\ \begin{cases}
                                               \odot^k V^\vee &\efor q\ =\ 0\ \mbox{ \& }\ k\ \geq\ 0~,\\[8pt]
                                               \odot^{-k-4}V\otimes\det V &\efor q\ =\ 3\ \mbox{ \& }\ k\ \leq\ -4~,\\[8pt]
                                               0 &\quad\mbox{otherwise}~,
                                          \end{cases}
\end{equation}
\begin{equation}\label{eq:Cohom2}
 H^q(\PP(V),\Omega^1(k))\ \cong\ \begin{cases}
                                               \left[ \displaystyle{\frac{\odot^{k-1} V\otimes V}{\odot^k V}}\right]^\vee &\efor q\ =\ 0\ \mbox{ \& }\ k\ \geq\ 2~,\\[8pt]
                                               \FC &\efor q\ =\ 1\ \mbox{ \& }\ k\ =\ 0~,\\[8pt]
                                                V^\vee\otimes\det V &\efor q\ =\ 3\ \mbox{ \& }\ k\ =\ -3~,\\[8pt]
                                               \left[ \displaystyle{\frac{\odot^{-k-3} V^\vee\otimes V}{\odot^{-k-4} V^\vee}}\right]^\vee\otimes\det V &\efor q\ =\ 3\ \mbox{ \& }\ k\ <\ -3~,\\[8pt]
                                               0 &\quad\mbox{otherwise}~,
                                          \end{cases}
\end{equation}
\begin{equation}\label{eq:Cohom3}
H^q(\PP(V),\Omega^2(k))\ \cong\ \begin{cases}
                                              V\otimes\det V^\vee &\efor q\ =\ 0\ \mbox{ \& }\ k\ =\ 3~,\\[8pt]
                                                \displaystyle{\frac{\odot^{k-3} V^\vee\otimes V}{\odot^{k-4} V^\vee}}\otimes\det V^\vee &\efor q\ =\ 0\ \mbox{ \& }\ k\ >\ 3~,\\[8pt]
                                               \FC &\efor q\ =\ 2\ \mbox{ \& }\ k\ =\ 0~,\\[8pt]
                                               \displaystyle{\frac{\odot^{-k-1} V\otimes V}{\odot^{-k} V}} &\efor q\ =\ 3\ \mbox{ \& }\ k\leq\ -2~,\\[8pt]
                                               0 &\quad\mbox{otherwise}~,
                                          \end{cases}
\end{equation}
\begin{equation}\label{eq:Cohom4}
 H^q(\PP(V),\Omega^3(k))\ \cong\ \begin{cases}
                                              \odot^{k-4}V^\vee\otimes\det V^\vee &\efor q\ =\ 0\ \mbox{ \& }\ k\ \geq\ 4~,\\[8pt]
                                               \odot^{-k} V &\efor q\ =\ 3\ \mbox{ \& }\ k\ \leq\ 0~,\\[8pt]
                                               0 &\quad\mbox{otherwise}~.
                                          \end{cases}
\end{equation}
\end{subequations}
}

\vspace{10pt}
\noindent
Notice that here, we are essentially computing the Dolbeault cohomology groups $H^{p,q}_{\bar\partial}(\PP^3,$ $\CO_{\PP^3}(k))$ of the complex projective three-space $\PP^3$ with values in $\CO_{\PP^3}(k)$ via the Dolbeault isomorphism.

\vspace{10pt}
\noindent
{\it Proof:} We already know the cohomology groups \eqref{eq:Cohom1} from Bott's rule given in Lemma \ref{lem:Bott}. Moreover, after computing \eqref{eq:Cohom2}, all remaining cases follow directly from \eqref{eq:Cohom1} and \eqref{eq:Cohom2} via Serre duality.\footnote{Serre duality (cf.\ e.g.~Griffiths \& Harris \cite{Griffiths:1978}) asserts that if $X$ is a compact $n$-dimensional complex manifold and $\CS$ an Abelian sheaf on $X$, then $ H^q(X,\Omega^p_X(\CS))\cong\left[H^{n-q}(X,\Omega^{n-p}_X(\CS^\vee))\right]^\vee~$.} In fact, we find the cohomology groups \eqref{eq:Cohom3} and \eqref{eq:Cohom4} from
\begin{equation}
\begin{aligned}
  H^q(\PP(V),\Omega^2(k))\ &\cong\ \left[H^{3-q}(\PP(V),\Omega^{1}(-k))\right]^\vee~,\\
  H^q(\PP(V),\Omega^3(k))\ &\cong\ \left[H^{3-q}(\PP(V),\Omega^{0}(-k))\right]^\vee~.
\end{aligned}
\end{equation}

To compute \eqref{eq:Cohom2}, let us consider the Euler sequence \eqref{eq:EulerSequence}. We can dualise this sequence and twist by $\CO_{\PP(V)}(k)$ to obtain
\begin{equation}
 0\ \longrightarrow\ \Omega^1(k)\ \longrightarrow\ \Omega^0(k-1)\otimes V^\vee\ \longrightarrow\ \Omega^0(k)
 \ \longrightarrow\ 0~.
 \end{equation}
From this sequence and Bott's rule, we derive the long exact sequences of cohomology groups
\begin{subequations}
\begin{equation}\label{eq:Seq1}
\begin{aligned}
  0\ \longrightarrow\ H^0(\PP(V),\Omega^1(k))\ \longrightarrow\ H^0(\PP(V),\Omega^0(k-1)\otimes V^\vee)\ \overset{\kappa}{\longrightarrow}\ \kern1cm\\
\overset{\kappa}{\longrightarrow}\ H^0(\PP(V),\Omega^0(k))\ \longrightarrow\ H^1(\PP(V),\Omega^1(k))\ \longrightarrow\ 0~,
\end{aligned}
\end{equation}
and
\begin{equation}\label{eq:Seq2}
\begin{aligned}
  0\ \longrightarrow\ H^3(\PP(V),\Omega^1(k))\ \longrightarrow\ H^3(\PP(V),\Omega^0(k-1)\otimes V^\vee)\ \longrightarrow\ \kern1cm\\
\longrightarrow\ H^3(\PP(V),\Omega^0(k))\ \longrightarrow\ 0~,
\end{aligned}
\end{equation}
\end{subequations}
where we used $H^2(\PP(V),\Omega^1(k))=0$.

Let us start with $H^q(\PP(V),\Omega^1(k))$ for $q=0,1$. For $k<0$, the sequence \eqref{eq:Seq1} together with Bott's rule yield that $H^0(\PP(V),\Omega^1(k))=0=H^1(\PP(V),\Omega^1(k))$ while for $k=0$ we find $H^0(\PP(V),\Omega^1(0))=0$ and $H^1(\PP(V),\Omega^1(0))\cong H^0(\PP(V),\Omega^0(0))\cong\FC$. For $k=1$,  \eqref{eq:Seq1} also shows that  $H^0(\PP(V),\Omega^1(1))=0=H^1(\PP(V),\Omega^1(1))$ while for $k\geq2$ we find $H^1(\PP(V),\Omega^1(k))=0$ since $\kappa$ is surjective. The rest of $H^0(\PP(V),\Omega^1(k))$ then follows from the short exact sequence
\begin{equation}
  0\ \longrightarrow\ H^0(\PP(V),\Omega^1(k))\ \longrightarrow\ \odot^{k-1}V^\vee\otimes V^\vee\ \longrightarrow\ \odot^k V^\vee\ \longrightarrow\ 0~.
\end{equation}
This concludes the cases $q=0,1$.

It remains to find $H^3(\PP(V),\Omega^1(k))$. The sequence \eqref{eq:Seq2} and Bott's rule show that for $k\geq-2$, $H^3(\PP(V),\Omega^1(k))=0$ while for $k=-3$, we get $H^3(\PP(V),\Omega^1(-3))\cong V^\vee\otimes \det V$. For $k<-3$,  \eqref{eq:Seq2}  reads as
\begin{equation}
  0\ \longrightarrow\ H^3(\PP(V),\Omega^1(k))\ \longrightarrow\ \odot^{-k-3}V\otimes\det V\otimes V^\vee\ \longrightarrow\ \odot^{-k-4} V\otimes\det V\ \longrightarrow\ 0~,
  \end{equation}
which gives the remaining cases for $H^3(\PP(V),\Omega^1(k))$. This completes the proof. \hfill $\Box$

\vspace{10pt}
Next, we compute the direct image sheaves $\pi_{2*}^q\Omega^p_{\pi_1}(\CO_{P^6}(k))$. Using the short-hand notation $\Omega^p_{\pi_1}(k):=\Omega^p_{\pi_1}(\CO_{P^6}(k))$, we have the following proposition:

{\prop\label{prop:DirectImage}
Let $k_p:=2p+k$. The direct image sheaves $\pi_{2*}^q\Omega^p_{\pi_1}(k)$ are given by
\begin{subequations}\label{eq:Direct}
\begin{equation}
 \pi_{2*}^q\Omega^0_{\pi_1}(k)\ \cong\ \begin{cases}
                                               \odot^{k_0} S &\efor q\ =\ 0\ \mbox{ \& }\ k_0\ \geq\ 0~,\\[8pt]
                                               (\odot^{-k_0-4}S^\vee)[1] &\efor q\ =\ 3\ \mbox{ \& }\ k_0\ \leq\ -4~,\\[8pt]
                                               0 &\quad\mbox{otherwise}~,
                                          \end{cases}
\end{equation}
\begin{equation}
\pi_{2*}^q\Omega^1_{\pi_1}(k)\ \cong\ \begin{cases}
                                               \left(\displaystyle{\frac{\odot^{k_1-1} S^\vee\otimes_{\CO_{M^6}} S^\vee}{\odot^{k_1} S^\vee}}\right)^\vee[1] &\efor q\ =\ 0\ \mbox{ \& }\  k_1\ \geq\ 2~,\\[8pt]
                                               [1] &\efor q\ =\ 1\ \mbox{ \& }\ k_1\ =\ 0~,\\[8pt]
                                               (\odot^{-k_1-3} S^\vee\otimes_{\CO_{M^6}} S)_0[2] & \efor q\ =\ 3\ \mbox{ \& }\ k_1\ \leq\ -3~,\\[8pt]
                                               0 &\quad\mbox{otherwise}~,
                                          \end{cases}
\end{equation}
\begin{equation}\label{eq:DirectImage3}
\pi_{2*}^q\Omega^2_{\pi_1}(k)\ \cong\ \begin{cases}
                                               (\odot^{k_2-3} S\otimes_{\CO_{M^6}} S^\vee)_0[1] &\efor q\ =\ 0\ \mbox{ \& }\ k_2\ \geq\ 3~,\\[8pt]
                                               [2] &\efor q\ =\ 2\ \mbox{ \& }\ k_2\ =\ 0~,\\[8pt]
                                                \left(\displaystyle{\frac{\odot^{-k_2-1} S^\vee\otimes_{\CO_{M^6}} S^\vee}{\odot^{-k_2} S^\vee}}\right)\![2] &\efor q\ =\ 3\ \mbox{ \& }\ k_2\ \leq\ -2~,\\[8pt]
                                               0 &\quad\mbox{otherwise}~,
                                          \end{cases}
\end{equation}
and
\begin{equation}\label{eq:DirectImage4}
\pi_{2*}^q\Omega^3_{\pi_1}(k)\ \cong\ \begin{cases}
                                              (\odot^{k_3-4}S)[2] &\efor q\ =\ 0\ \mbox{ \& }\ k_3\ \geq\ 4~,\\[8pt]
                                               (\odot^{-k_3} S^\vee)[3] &\efor q\ =\ 3\ \mbox{ \& }\ k_3\ \leq\ 0~,\\[8pt]
                                               0 &\quad\mbox{otherwise}~,
                                          \end{cases}
\end{equation}
\end{subequations}
where $(\odot^{l}S^\vee\otimes_{\CO_{M^6}} S)_0$ is the totally trace-less part of $\odot^{l}S^\vee\otimes_{\CO_{M^6}} S$ which is
\begin{equation}
 (\odot^{l}S^\vee\otimes_{\CO_{M^6}} S)_0\ \cong\
 \begin{cases}
  S & \efor l\ =\ 0~,\\[8pt]
  \displaystyle{ \frac{\odot^{l}S^\vee\otimes_{\CO_{M^6}} S}{\odot^{l-1}S^\vee}} & \efor l\ \geq\ 1~.
 \end{cases}
\end{equation}
}

\vspace{10pt}
\noindent
{\it Proof:}
By definition of direct image sheaves, our task is to compute the cohomology groups $H^q(\pi_2^{-1}(U),\Omega_{\pi_1}^p(k))$ for open sets $U\subset M^6$; see also footnote \ref{foot:DirectImage}.  Notice that it suffices to work with Stein open sets $U$  so that $U':=\pi_2^{-1}(U)\cong U\times\PP^3\subset F^9$ since there are arbitrarily small Stein open sets on $M^6$. We could now apply the direct image functor to the short exact sequences of Proposition \ref{prop:RelFormsPullBackSheaves} to obtain the direct images. There is, however, a quicker way of computing these.

Consider the case  when $p=0$. It is rather straightforward to see that in this case, we have the identification
\begin{equation}
 H^q(U',\Omega_{\pi_1}^0(k))\ \cong\ \{\mbox{holomorphic functions}\,:\,U\to H^q(\PP^3,\CO_{\PP^3}(k))  \}~,
\end{equation}
and we can directly apply the results of Lemma \ref{lem:Dolbeault}. The other cohomology groups can be characterised analogously. We first recall our discussion of the relative one-forms, $\Omega_{\pi_1}^1(0)=\Omega_{\pi_1}^1$ that led to the sequence \eqref{eq:Rel1FormsPullBackSheaves}. Let $(x,\lambda)=(x^{AB},\lambda_A)$ be local coordinates on $F^9$, as before. Then the components $\omega^A$ of a relative one-form $\omega$ are of weight one in $\lambda$ and obey $\omega^A\lambda_A=0$. This essentially implies that $\omega^A=\frac12\varepsilon^{ABCD}\omega_{BC}\lambda_D$, where $\omega_{AB}=-\omega_{BA}$ depends (holomorphically) on $x$. Together with our results for the twistor space $P^6$ presented at the end of Section \ref{sec:TwistorSpace}, we may conclude that
\begin{equation}
 H^q(U',\Omega_{\pi_1}^1(0))\ \cong\ \{\mbox{holomorphic functions}\,:\,U\to H^q(\PP^3,\Omega^1(2))[1] \}~.
\end{equation}
This argument generalises to the remaining cohomology groups $H^q(U',\Omega_{\pi_1}^p)$ for $p=2,3$, and we have
\begin{equation}
 H^q(U',\Omega_{\pi_1}^p(0))\ \cong\ \{\mbox{holomorphic functions}\,:\,U\to H^q(\PP^3,\Omega^p(2p))[p]  \}~.
\end{equation}
Therefore, if we let $k_p:=2p+k$, we obtain
\begin{equation}
 H^q(U',\Omega_{\pi_1}^p(k))\ \cong\ \{\mbox{holomorphic functions}\,:\,U\to H^q(\PP^3,\Omega^p(k_p))[p]  \}~.
\end{equation}

In summary, all the cohomology groups $H^q(\pi_2^{-1}(U),\Omega_{\pi_1}^p(k))$ are characterised in terms of the cohomology groups appearing in Lemma \ref{lem:Dolbeault} for $V= S^\vee$, which yields \eqref{eq:Direct}.  \hfill $\Box$

\vspace{10pt}
So far, we have computed the direct images of the sheaves $\Omega^p_{\pi_1}(k)$. The resolutions \eqref{eq:RelDeRhamCom} and \eqref{eq:RelDeRhamComVect}  also contain the topological inverse sheaves $\pi_1^{-1}\CO_{P^6}$ and $\pi_1^{-1}\CO_{P^6}(\CO_{P^6}(k))$, respectively. The direct images of these sheaves are computed using spectral sequences. In the following, we shall merely recall a few facts about spectral sequences and we refer to \cite{Ward:1990vs} for a more detailed account.

For us, a spectral sequence is basically a sequence of two-dimensional arrays of Abelian groups $E_r=(E_r^{p,q})$ for $r=1,2,\ldots$ which are labelled by $p,q=0,1,2,\dots$ together with differential operators $\dd_r:E_r^{p,q}\to E_r^{p+r,q-r+1}$ that obey $\dd_r\circ\dd_r=0$. In addition, the arrays are linked cohomologically from one order to the next. Specifically, we have
\begin{equation}\label{eq:SpectralCohom}
 E_{r+1}^{p,q}\ \cong\ H^{p,q}(E_r)\ :=\ \frac{\ker\dd_r\,:\,E_r^{p,q}\to E_r^{p+r,q-r+1}}{\im\, \dd_r\,:\,E_r^{p-r,q+r-1}\to E_r^{p,q}}~.
\end{equation}
There also is a well-defined limit of the spectral sequence in terms of the inductive limit
\begin{equation}
 E_\infty^{p,q}\ =\ \underset{r\to\infty}{\lim\mbox{ind}}\, E_r^{p,q}~.
\end{equation}

If $U\subset M^6$ is open and $U':=\pi_2^{-1}(U)$, the resolution \eqref{eq:RelDeRhamComVect} yields a spectral sequence with initial terms $E_1^{p,q}\cong H^q(U',\Omega_{\pi_1}^p(E))$ and differential operators $\dd_1:E_1^{p,q}\to E_1^{p+1,q}$ induced by $\dd_{\pi_1}:\Omega^p_{\pi_1}(E)\ \to\ \Omega^{p+1}_{\pi_1}(E)$. This spectral sequence converges to the cohomology group $E_\infty^{p,q}\cong H^{p+q}(U',\pi_1^{-1}\CO_{P^6}(E))$, which is mnemonically written as $H^q(U',\Omega_{\pi_1}^p(E))\Rightarrow H^{p+q}(U',\pi_1^{-1}\CO_{P^6}(E))$. Altogether, we have the following proposition:

{\prop\label{prop:SpecSequence}
Let $U$ be an open set in $M^6$ and let $U':=\pi_2^{-1}(U)\subset F^9$. Then there is a spectral sequence
\begin{equation}
 E_1^{p,q}\ \cong\ H^q(U',\Omega_{\pi_1}^p(k))\ \Longrightarrow\ H^{p+q}(U',\pi_1^{-1}\CO_{P^6}(k))~,
\end{equation}
where the differential operators $\dd_1:E_1^{p,q}\to E_1^{p+1,q}$ are induced by the relative exterior derivative $\dd_{\pi_1}:\Omega^p_{\pi_1}(k)\ \to\ \Omega^{p+1}_{\pi_1}(k)$.
}

\vspace{10pt}
\noindent
Hence, we have an explicit way of computing $H^{q}(U',\pi_1^{-1}\CO_{P^6}(k))$ in terms of the cohomology groups $H^q(U',\Omega_{\pi_1}^p(k))$.

\paragraph{Cohomology groups of topological inverse sheaves.}
The final ingredient we need is a result due to Buchdahl \cite{Buchdahl:1985aa}. Above we have computed the direct images of sheaves on the correspondence space $F^9$ along the fibration $\pi_2:F^9\to M^6$ to obtain certain sheaves on space-time $M^6$. In the Penrose transform, these sheaves on $F^9$ originate from sheaves on twistor space. To connect the cohomology groups of both kinds of sheaves, we can use the following proposition:

{\prop\label{prop:Buchdahl}
(Buchdahl \cite{Buchdahl:1985aa})
Let $X$ and $Y$ be complex manifolds and $\pi:X\to Y$ a surjective holomorphic mapping of maximal rank with connected fibres. Furthermore, let $\CS$ be an Abelian sheaf on $Y$. If there is an $n_0>0$ such that $H^q(\pi^{-1}(p), \FC)= 0$ for $q=1,\ldots, n_0$ and for all $p\in Y$, then
\begin{equation}
  \pi^*\,:\, H^q(Y,\CS)\ \to\ H^q(X,\pi^{-1}\CS)~
\end{equation}
is an isomorphism for $q=0,\ldots,n_0$ and a monomorphism for $q=n_0+1$.
}

\vspace{10pt}
\noindent
The requirements of this proposition for the projection $\pi_1:F^9\rightarrow P^6$ are always satisfied in our setting. Because we always work with convex subsets $U\subset M^6$, we always have the isomorphism $H^q(\hat U,\CS)\cong H^q(U',\pi^{-1}_1\CS)$, where $U':=\pi_2^{-1}(U)\subset F^9$ and $\hat U:=\pi_1(\pi_2^{-1}(U))\subset P^6$. In a compactified version of the twistor correspondence, one has to supplement Theorem \ref{thm:6DPenrose} by the above requirements.

\subsection{Proof}\label{sec:Proof6DPW}

We are now ready to prove Theorem \ref{thm:6DPenrose}. We shall first proof the case $h>0$, that is $-2h-4<-4$, and then come to the case $h=0$, which is slightly more complicated.

\paragraph{\mathversion{bold}Case $h>0$.} Recall that sections $\psi$ of the sheaf $\CZ_h$ defined in \eqref{eq:DefZRM} obey the free field equation
\begin{equation}\label{eq:FFEh>0}
 \partial^{AB}\psi_{BA_1\cdots A_{2h-1}}\ =\ 0~.
\end{equation}
We thus have to prove that,
\begin{equation}\label{eq:eq1proof1}
 \CCP\,:\,H^3(\hat U,\CO_{\hat U}(-2h-4))\ \to\ H^0(U,\CZ_h)
\end{equation}
is an isomorphism. We already know from Proposition \ref{prop:Buchdahl} that
\begin{equation}\label{eq:AnIso}
 H^3(\hat U,\CO_{\hat U}(-2h-4))\ \cong\ H^3(U',\pi_1^{-1}\CO_{\hat U}(-2h-4))~,
\end{equation}
which reduces \eqref{eq:eq1proof1} to
\begin{equation}
 H^3(U',\pi_1^{-1}\CO_{\hat U}(-2h-4))\ \cong\ H^0(U,\CZ_h)~.
\end{equation}

Firstly, we notice that there is a particular spectral sequence, the {\em Leray spectral sequence} $L_r=(L_r^{p,q})$, which gives\footnote{In general, if $\CS$ is an Abelian sheaf on $X$ and $\pi:X\to Y$, the Leray spectral sequence $L_r=(L_r^{p,q})$ relates the cohomology of $\CS$ to that of its direct images (see e.g.~Godement \cite{Godement:1964aa}) according to $L_2^{p,q}\cong H^p(Y,\pi^q_*\CS)\ \Rightarrow\ H^{p+q}(X,\CS)$.}
\begin{equation}\label{eq:LeraySeq}
  L_2^{p,q}\ \cong\ H^p(U,\pi_{2*}^q\Omega^l_{\pi_1}(-2h-4))\ \Longrightarrow\
  H^{p+q}(U',\Omega^l_{\pi_1}(-2h-4))~.
\end{equation}
For fixed $l$, Proposition \ref{prop:DirectImage} for $h>0$ tells us that $\pi_{2*}^q\Omega^l_{\pi_1}(-2h-4)=0$ if $q\neq3$. Thus, the Leray spectral sequence $L_r^{p,q}$ is degenerate at the second level. Therefore, we have
\begin{equation}
 L_\infty^{p,q}\ \cong\ L_2^{p,q}\efor p,q\ \geq\ 0~,
\end{equation}
cf.\ \eqref{eq:SpectralCohom}. Recall that if a spectral sequence $(E_r^{p,q})$ has the property that for some $r_0$, $E_{r_0}^{p,q}=0$ for $q\neq q_0$, then $E_{r_0}^{p,q_0}\cong H^{p+q_0}$. This property together with \eqref{eq:LeraySeq} then imply
\begin{equation}\label{eq:LerayIso}
  H^p(U',\Omega^l_{\pi_1}(-2h-4))\ \cong\ \begin{cases}
                                           H^{p-3}(U,\pi_{2*}^3\Omega^l_{\pi_1}(-2h-4)) &\efor p\ \geq\ 3~,\\
					   0&\efor p\ <\ 3~.
                                          \end{cases}
\end{equation}

Secondly, Proposition \ref{prop:SpecSequence} yields another spectral sequence $E_r=(E_r^{p,q})$ with
\begin{equation}\label{eq:ParticularSpecSeq}
 E_1^{p,q}\ \cong\ H^q(U',\Omega_{\pi_1}^p(-2h-4))\ \Longrightarrow\ H^{p+q}(U',\pi_1^{-1}\CO_{P^6}(-2h-4))~.
\end{equation}
Explicitly, the $r=1$ array in this sequence reads as ($k=-2h-4$)
\begin{equation}
\begin{aligned}
 H^0(U',\Omega_{\pi_1}^0(k))\ \overset{\dd_{\pi_1}}{\longrightarrow}\ H^0(U',\Omega_{\pi_1}^1(k))\ \overset{\dd_{\pi_1}}{\longrightarrow}\ H^0(U',\Omega_{\pi_1}^2(k))\ \overset{\dd_{\pi_1}}{\longrightarrow}\ H^0(U',\Omega_{\pi_1}^3(k))\\
H^1(U',\Omega_{\pi_1}^0(k))\ \overset{\dd_{\pi_1}}{\longrightarrow}\ H^1(U',\Omega_{\pi_1}^1(k))\ \overset{\dd_{\pi_1}}{\longrightarrow}\ H^1(U',\Omega_{\pi_1}^2(k))\ \overset{\dd_{\pi_1}}{\longrightarrow}\ H^1(U',\Omega_{\pi_1}^3(k))\\
H^2(U',\Omega_{\pi_1}^0(k))\ \overset{\dd_{\pi_1}}{\longrightarrow}\ H^2(U',\Omega_{\pi_1}^1(k))\ \overset{\dd_{\pi_1}}{\longrightarrow}\ H^2(U',\Omega_{\pi_1}^2(k))\ \overset{\dd_{\pi_1}}{\longrightarrow}\ H^2(U',\Omega_{\pi_1}^3(k))\\
H^3(U',\Omega_{\pi_1}^0(k))\ \overset{\dd_{\pi_1}}{\longrightarrow}\ H^3(U',\Omega_{\pi_1}^1(k))\ \overset{\dd_{\pi_1}}{\longrightarrow}\ H^3(U',\Omega_{\pi_1}^2(k))\ \overset{\dd_{\pi_1}}{\longrightarrow}\ H^3(U',\Omega_{\pi_1}^3(k))\\
H^4(U',\Omega_{\pi_1}^0(k))\ \overset{\dd_{\pi_1}}{\longrightarrow}\ H^4(U',\Omega_{\pi_1}^1(k))\ \overset{\dd_{\pi_1}}{\longrightarrow}\ H^4(U',\Omega_{\pi_1}^2(k))\ \overset{\dd_{\pi_1}}{\longrightarrow}\ H^4(U',\Omega_{\pi_1}^3(k))\\
\vdots\kern3.5cm\vdots\kern3.5cm\vdots\kern3.55cm\vdots\kern1.1cm
\end{aligned}
\end{equation}
We may now replace these cohomology groups by $H^q(U',\Omega^p_{\pi_1}(k))$ using \eqref{eq:LerayIso} to obtain
\begin{equation}
\begin{aligned}
 0\kern3.8cm 0\kern5.35cm 0\kern1.4cm\\
 0\kern3.8cm 0\kern5.35cm 0\kern1.4cm\\
 0\kern3.8cm 0\kern5.35cm 0\kern1.4cm\\
 H^0(U,\pi_{2*}^3\Omega_{\pi_1}^0(k))\ \longrightarrow\ H^0(U,\pi_{2*}^3\Omega_{\pi_1}^1(k))\ \longrightarrow\ \cdots\ \longrightarrow\ H^0(U,\pi_{2*}^3\Omega_{\pi_1}^3(k))\\
  H^1(U,\pi_{2*}^3\Omega_{\pi_1}^0(k))\ \longrightarrow\ H^1(U,\pi_{2*}^3\Omega_{\pi_1}^1(k))\ \longrightarrow\ \cdots\ \longrightarrow\ H^1(U,\pi_{2*}^3\Omega_{\pi_1}^3(k))\\
   H^2(U,\pi_{2*}^3\Omega_{\pi_1}^0(k))\ \longrightarrow\ H^2(U,\pi_{2*}^3\Omega_{\pi_1}^1(k))\ \longrightarrow\ \cdots\ \longrightarrow\ H^2(U,\pi_{2*}^3\Omega_{\pi_1}^3(k))\\
    H^3(U,\pi_{2*}^3\Omega_{\pi_1}^0(k))\ \longrightarrow\ H^3(U,\pi_{2*}^3\Omega_{\pi_1}^1(k))\ \longrightarrow\ \cdots\ \longrightarrow\ H^3(U,\pi_{2*}^3\Omega_{\pi_1}^3(k))\\
 \vdots\kern3.95cm\vdots\kern5.45cm\vdots\kern1.4cm
\end{aligned}
\end{equation}
This diagram together with \eqref{eq:SpectralCohom} then yield the following identification:
\begin{equation}
 E_2^{0,3}\ \cong\ \ker\left\{H^0(U,\pi_{2*}^3\Omega_{\pi_1}^0(-2h-4))\ \to\ H^0(U,\pi_{2*}^3\Omega_{\pi_1}^1(-2h-4))   \right\}.
\end{equation}
Furthermore, all $E_r^{p,q}=0$ for  $p+q=3$ with $q\neq3$, and $E_2^{0,3}\cong  E_3^{0,3}\cong \cdots\cong  E_\infty^{0,3}$. From Proposition \ref{prop:DirectImage}, it follows that $\pi_{2*}^3\Omega_{\pi_1}^0(-2h-4)\cong(\odot^{2h}S^\vee)[1]$ and $\pi_{2*}^3\Omega_{\pi_1}^1(-2h-4)\cong(\odot^{2h-1}S^\vee\otimes_{\CO_U} S)_0[2]$. In addition, the relative exterior derivative $\dd_{\pi_1}:H^3(U',\Omega_{\pi_1}^0(k))\to H^3(U',\Omega_{\pi_1}^1(k))$ induces the differential operator
\begin{equation}
\dpar^{AB}\,:\,H^0(U,\pi_{2*}^3\Omega_{\pi_1}^0(-2h-4))\ \to\ H^0(U,\pi_{2*}^3\Omega_{\pi_1}^1(-2h-4))~.
\end{equation}
In summary, from \eqref{eq:AnIso} and \eqref{eq:ParticularSpecSeq} we may therefore conclude that
\begin{equation}
 H^3(\hat U,\CO_{\hat U}(-2h-4))\ \cong\ H^{3}(U',\pi_1^{-1}\CO_{\hat U}(-2h-4))\ \cong\ E_2^{0,3}\ \cong\ H^0(U,\CZ_h)~.
\end{equation}

\paragraph{\mathversion{bold}Case $h=0$.}
The proof for $h=0$ is similar to the one presented above albeit somewhat more difficult. Firstly, we shall be dealing with a second-order partial differential operator and secondly, on a more technical level, the appropriate spectral sequence will degenerate differently.

Recall that $\CZ_0$ is the sheaf of solutions to the Klein--Gordon equation. That is, its sections describe scalar fields on space-time forming the trivial representation $(\mathbf{1},\mathbf{1})$ under the little group. We wish to prove that
\begin{equation}
 \CCP\,:\,H^3(\hat U,\CO_{\hat U}(-4))\ \to\ H^0(U,\CZ_0)
\end{equation}
is an isomorphism. Again, by virtue of Proposition \ref{prop:Buchdahl}, we only need to show that
\begin{equation}
 H^3(U',\pi_1^{-1}\CO_{\hat U}(-4))\ \cong\ H^0(U,\CZ_0)~.
\end{equation}

From Proposition \ref{prop:DirectImage}, we see that
\begin{equation}\label{eq:DirectZeroHeli}
 \pi_{2*}^q\Omega^l_{\pi_1}(-4)\ \cong\ \begin{cases}
                                              [1] &\efor (q,l)=(3,0)~,\\
                                              [2] &\efor (q,l)=(2,2)~,\\
                                               0 &\quad\mbox{otherwise}~.
                                          \end{cases}
\end{equation}
When $(q,l)=(3,0)$, the corresponding Leray spectral sequence \eqref{eq:LeraySeq} yields
\begin{equation}\label{eq:LerayIso2}
  H^p(U',\Omega^0_{\pi_1}(-4))\ \cong\ \begin{cases}
H^{p-3}(U,\pi_{2*}^3\Omega^0_{\pi_1}(-4))\ \cong\ H^{p-3}(U,[1])
  &\efor p\ \geq\ 3~,\\
0 &\efor p\ <\ 3~.
                                       \end{cases}
\end{equation}
Moreover, with \eqref{eq:DirectZeroHeli} the Leray spectral sequence \eqref{eq:LeraySeq}  also gives
\begin{equation}\label{eq:LerayIso3}
 H^{p}(U,\pi_{2*}^q\Omega^l_{\pi_1}(-4))\ =\ 0
  \efor p,q\ \geq\ 0\eand l\ =\ 1,3~.
\end{equation}
When $(q,l)=(2,2)$, we derive
\begin{equation}\label{eq:LerayIso4}
  H^p(U',\Omega^2_{\pi_1}(-4))\ \cong\ \begin{cases}
H^{p-2}(U,\pi_{2*}^2\Omega^2_{\pi_1}(-4))\ \cong\ H^{p-2}(U,[2])
  &\efor p\ \geq\ 2~,\\
0 &\efor p\ <\ 2~.
\end{cases}
\end{equation}

Next, the $r=1$ part of the spectral sequence \eqref{eq:ParticularSpecSeq} for $h=0$ is given by
\begin{equation}
\begin{aligned}
 H^0(U',\Omega_{\pi_1}^0(-4))\ \overset{\dd_{\pi_1}}{\longrightarrow}\ H^0(U',\Omega_{\pi_1}^1(-4))\ \overset{\dd_{\pi_1}}{\longrightarrow}\ H^0(U',\Omega_{\pi_1}^2(-4))\ \overset{\dd_{\pi_1}}{\longrightarrow}\ H^0(U',\Omega_{\pi_1}^3(-4))\\
H^1(U',\Omega_{\pi_1}^0(-4))\ \overset{\dd_{\pi_1}}{\longrightarrow}\ H^1(U',\Omega_{\pi_1}^1(-4))\ \overset{\dd_{\pi_1}}{\longrightarrow}\ H^1(U',\Omega_{\pi_1}^2(-4))\ \overset{\dd_{\pi_1}}{\longrightarrow}\ H^1(U',\Omega_{\pi_1}^3(-4))\\
H^2(U',\Omega_{\pi_1}^0(-4))\ \overset{\dd_{\pi_1}}{\longrightarrow}\ H^2(U',\Omega_{\pi_1}^1(-4))\ \overset{\dd_{\pi_1}}{\longrightarrow}\ H^2(U',\Omega_{\pi_1}^2(-4))\ \overset{\dd_{\pi_1}}{\longrightarrow}\ H^2(U',\Omega_{\pi_1}^3(-4))\\
H^3(U',\Omega_{\pi_1}^0(-4))\ \overset{\dd_{\pi_1}}{\longrightarrow}\ H^3(U',\Omega_{\pi_1}^1(-4))\ \overset{\dd_{\pi_1}}{\longrightarrow}\ H^3(U',\Omega_{\pi_1}^2(-4))\ \overset{\dd_{\pi_1}}{\longrightarrow}\ H^3(U',\Omega_{\pi_1}^3(-4))\\
H^4(U',\Omega_{\pi_1}^0(-4))\ \overset{\dd_{\pi_1}}{\longrightarrow}\ H^4(U',\Omega_{\pi_1}^1(-4))\ \overset{\dd_{\pi_1}}{\longrightarrow}\ H^4(U',\Omega_{\pi_1}^2(-4))\ \overset{\dd_{\pi_1}}{\longrightarrow}\ H^4(U',\Omega_{\pi_1}^3(-4))\\
\vdots\kern3.8cm\vdots\kern3.8cm\vdots\kern3.8cm\vdots\kern1.4cm
\end{aligned}
\end{equation}
Our above calculations show that the second and fourth columns of this diagram are zero, while the first and third ones are non-zero in general. Hence, the differential operator  $\dd_1$ on $E_1^{p,q}$ vanishes identically and therefore, we have the identification $E_1^{p,q}\cong E_2^{p,q}$. Substituting \eqref{eq:LerayIso2}--\eqref{eq:LerayIso4} into this diagram, we eventually find
\begin{equation}
\begin{aligned}
0\kern.8cm \longrightarrow\ 0\ \longrightarrow\kern.93cm 0\kern.8cm \longrightarrow\ 0\\
0\kern.8cm \longrightarrow\ 0\ \longrightarrow\kern.93cm 0\kern.8cm \longrightarrow\ 0\\
0\kern.8cm \longrightarrow\ 0\ \longrightarrow\ H^0(U,[2])\ \longrightarrow\ 0\\
H^0(U,[1])\ \longrightarrow\ 0\ \longrightarrow\ H^1(U,[2])\ \longrightarrow\ 0\\
H^1(U,[1])\ \longrightarrow\ 0\ \longrightarrow\ H^2(U,[2])\ \longrightarrow\ 0\\
H^2(U,[1])\ \longrightarrow\ 0\ \longrightarrow\ H^3(U,[2])\ \longrightarrow\ 0\\
H^3(U,[1])\ \longrightarrow\ 0\ \longrightarrow\ H^4(U,[2])\ \longrightarrow\ 0\\
\vdots\kern3.95cm\vdots\kern2cm
\end{aligned}
\end{equation}
Furthermore, the differential operator $\dd_2$ on $E_2^{0,3}$ maps $E_2^{0,3}$ to $E_2^{2,2}$ and since $E_1^{p,q}\cong E_2^{p,q}$ and thus, $E_2^{0,3}\cong H^0(U,[1])$ and $E_2^{2,2}\cong H^0(U,[2])$, respectively, we have a map $\Box:H^0(U,[1])\to H^0(U,[2])$ which is induced by $\dd_2$. One can see that this map is a composition of first-order differential operators and it is indeed the one we defined in \eqref{eq:DefZRM}.

Finally, we note that
\begin{equation}
 E_3^{0,3}\ \cong\ \ker\left\{\Box\,:\,H^0(U,[1])\ \to\ H^0(U,[2]) \right\},
\end{equation}
together with $E_3^{0,3}\cong\cdots\cong  E_\infty^{0,3}$. Altogether,
\begin{equation}
 H^3(\hat U,\CO_{\hat U}(-4))\ \cong\ H^{3}(U',\pi_1^{-1}\CO_{\hat U}(-4))\ \cong\ E_3^{0,3}\ \cong\ H^0(U,\CZ_0)~,
\end{equation}
which completes the proof for $h=0$.

\subsection{Integral formul{\ae}}\label{sec:IntegralFormulae}

Similarly to four dimensions \cite{Penrose:1969aa}, we can write down certain contour integral formul{\ae} yielding solutions to the zero-rest-mass field equations in six dimensions. As already indicated, such formul{\ae} appeared first in works by Hughston \cite{Hughston:1979TN,Hughston:1982TN,Hughston:1984TN,Hughston:1987aa}.

\paragraph{Integral formul{\ae} on twistor space.}
Let us choose a sufficiently fine open Stein covering $\hat{\mathfrak{U}}=\{\hat U_{a}\}$ of $\hat U$. We shall make use of the abbreviations $\hat U_{ab}:=\hat U_{a}\cap\hat U_{b}$, $\hat U_{abc}:=\hat U_{a}\cap\hat U_{b}\cap\hat U_{c}$, etc. The simplest choice for $\hat{\mathfrak{U}}$ is a lift of the standard cover of $\PP^3$ to $\hat{U}$ requiring four patches $\hat U_{a}$, $a=1,\ldots,4$. In this case, there is only one quadruple overlap of four patches, and a holomorphic function $\hat f_{-2h-4}= \hat f_{-2h-4}(z,\lambda)$ on $\hat{U}_{1234}\subset\hat U$ of homogeneity $-2h-4$ represents an element of $H^3(\hat{U},\CO_{\hat U}(-2h-4))$. For simplicity, we shall assume a \v Cech cocycle $\hat f_{-2h-4}$ of this form in the following. Note that this is not the most general way of representing elements of $H^3(\hat{U},\CO_{\hat{U}})$. This, however, requires merely a technical extension of our discussion below using branched contour integrals, cf.~Penrose \& Rindler \cite{Penrose:1986ca}.

Let us now restrict to $h\geq 0$ and construct zero-rest-mass fields $\psi\in H^0(U,\CZ_h)$. That is, $\psi$ forms the representation $(\mathbf{2h+1},\mathbf{1})$ of the little group $\sSL(2,\FC)\times\widetilde{\sSL(2,\FC)}$, cf.\ \eqref{eq:DefZRM}. We start from a \v Cech cocycle $\hat f_{-2h-4}$, which we restrict to $\hat x\cong \PP^3$ to obtain $\hat f_{-2h-4}= \hat f_{-2h-4}(x\cdot\lambda,\lambda)$ on the intersection $\hat U_{1234}\cap\hat x$. Using the holomorphic $\sSL(4,\FC)$-invariant measure on $\PP^3$ given by
\begin{equation}
 \Omega^{(3,0)}\ :=\ \frac{1}{4!}\eps^{ABCD}\lambda_A\dd\lambda_B\wedge\dd\lambda_C\wedge \dd \lambda_D~,
\end{equation}
we can write down the contour integral
\begin{equation}\label{eq:intform1}
 \psi_{A_1\cdots A_{2h}}(x)\ =\ \oint_\CCC \Omega^{(3,0)}~ \lambda_{A_1}\cdots\lambda_{A_{2h}} \hat f_{-2h-4}(x\cdot\lambda,\lambda)~,
\end{equation}
where the contour $\CCC$ is topologically a three-torus contained in $\hat{U}_{1234}$. Clearly
\begin{equation}
 \partial^{AB}\psi_{BA_1\cdots A_{2h-1}}\ =\ 0\efor h\ >\ 0\eand\Box\psi\ =\ 0\efor h\ =\ 0~,
\end{equation}
as follows from straightforward differentiation under the integral.

\paragraph{Integral formul{\ae} on thickened twistor space.}
More recently, similar integral formul{\ae} were discussed by Berkovits \& Cherkis \cite{Berkovits:2004bw} and Chern \cite{Chern:2009nt} also for the cohomology groups $H^3(\hat{U},\CO_{\hat U}(2h-4))$ with $h>0$. However, these cohomology groups yield trivial space-time fields as we shall discuss in Remark \ref{rem:zeroH3} (see also \cite{Baston:1989}). Therefore, their integral formul{\ae} make only sense if one {\it thickens} (via {\it infinitesimal neighbourhoods}) $P^6$ into its ambient space $\PP^7_\circ \cong \CO_{\PP^3}(1)\otimes\FC^4$.\footnote{Recall that $P^6$ is a hypersurface in $\PP^7_\circ$  as follows from \eqref{eq:NormalSequence2}.}  Thickenings of manifolds occur in various twistor geometric contexts. The most prominent examples appear in the twistor descriptions of Yang--Mills theory and Einstein gravity
\cite{Witten:1978xx,Isenberg:1978kk,Isenberg:1978qd,Eastwood:1986aa,Lebrun:1986it,Manin:1988ds,springerlink:10.1007/BF00419314,LeBrun:1991jh,0264-9381-2-4-020,Merkulov:1992qa} in four space-time dimensions (see also Section \ref{ss:Ambitwistors}).

To thicken our twistor space $P^6$, consider $\CO_{\PP^7_\circ}$, the sheaf of holomorphic functions on $\PP^7_\circ$, and $\CI$, the ideal subsheaf of $\CO_{\PP^7_\circ}$ consisting of those functions that vanish on $P^6\hookrightarrow \PP^7_\circ$. The $\ell$-th order thickening (or $\ell$-th infinitesimal neighbourhood) of $P^6$ inside $\PP^7_\circ$ is the scheme $P^6_{[\ell]}$ defined by
\begin{equation}
 P^6_{[\ell]}\ :=\ (P^6,\CO_{\PP^7_\circ}/\CI^{\ell+1})~.
\end{equation}
Notice that we recover the twistor space as the zeroth order thickening, i.e.~$P_{[0]}^6=P^6$. Moreover, a cover of $P^6$ will also form a cover of $P^6_{[\ell]}$. The spaces $P^6_{[\ell]}$ can be thought of as the jets of the embedding of $P^6$ into the larger manifold $\PP^7_\circ$. In local coordinates $(z^A,\lambda_A)$ on $\PP^7_\circ$, we have $(z^A\lambda_A)^{i+1}=0$ for $i\geq \ell$ but $(z^A\lambda_A)^{i}\neq 0$ for $0< i\leq \ell$ on $P^6_{[\ell]}$. This implies that on the first order thickening $P^6_{[1]}$, the four vector fields $\der{z^A}$ are linearly independent and act freely on functions on $P^6_{[1]}$. Differential operators of order $\ell$ constructed out of these four vector fields act freely on functions on $P^6_{[\ell]}$. As we shall see momentarily, this fact is the essential ingredient for writing down a contour integral leading to zero-rest-mass fields.

Proceeding analogously to four dimensions, we shall now construct a second contour integral by replacing $\lambda_A$ in \eqref{eq:intform1} by the derivatives $\der{z^A}$ and adjusting the homogeneity of $\hat f_{2h-4}$ for $h>0$ accordingly. The resulting $2h$ derivatives in the contour integral should act freely, and therefore we have to consider a thickening of $\hat U\subset P^6$ to $2h$-th order, that is, $\hat U_{[2h]}\subset P^6_{[2h]}$. Let $\hat f_{2h-4}^{[2h]}= \hat f_{2h-4}^{[2h]}(z,\lambda)$ be a representative of the cohomology group $H^3(\hat U_{[2h]},\CO_{\hat U_{[2h]}}(2h-4))$ for $h>0$. It is expanded as
\begin{equation}\label{eq:NormalExpThick1}
  \hat f_{2h-4}^{[2h]}(z,\lambda)\ =\ \hat g(\lambda)+\sum_{l\geq1} \frac{1}{l!} z^{A_1}\cdots z^{A_l}\hat g_{A_1\cdots A_l}(\lambda)~,
\end{equation}
where the coefficients $\hat g_{A_1\cdots A_l}$ for $l\leq 2h$ are uniquely defined for $0< l\leq 2h$. We may rewrite the above expansion as
\begin{equation}\label{eq:NormalExpThick2}
  \hat f_{2h-4}^{[2h]}(z,\lambda)\ =\ \frac{1}{(2h)!} z^{A_1}\cdots z^{A_{2h}} \hat f_{A_1\cdots A_{2h}}(z,\lambda)+\cdots~,
\end{equation}
where the ellipsis denotes terms that contain at most $2h-1$ factors of $z^A$. As the coefficients $\hat f_{A_1\cdots A_{2h}}$ are uniquely fixed, they can be extracted from $\hat f_{2h-4}^{[2h]}$. Upon restriction to $\hat x\cong\PP^3$, we may write
\begin{equation}
  \hat f_{A_1\cdots A_{2h}}(x\cdot\lambda,\lambda)\ =\ \left. \der{z^{A_1}}\cdots \der{z^{A_{2h}}} \hat f^{[2h]}_{2h-4}(z,\lambda)\right|_{z=x\cdot\lambda}.
\end{equation}
The latter relation can then be used to construct the contour integral formula
\begin{equation}\label{eq:intform2}
\begin{aligned}
 {\psi}_{A_1\cdots A_{2h}}(x)\ &=\ \oint_\CCC \Omega^{(3,0)}\, \hat f_{A_1\cdots A_{2h}}(x\cdot\lambda,\lambda)\\
  &=\ \left.\oint_\CCC \Omega^{(3,0)}\, \der{z^{A_1}}\cdots \der{z^{A_{2s}}} \hat f^{[2h]}_{2h-4}(z,\lambda)\right|_{z=x\cdot\lambda}~,
\end{aligned}
 \end{equation}
where the contour is again a three-torus. By differentiation under the integral, one may check that this is indeed a zero-rest-mass field, i.e.
\begin{equation}
 \partial^{AB}\psi_{BA_1\cdots A_{2h-1}}\ =\ 0~,
\end{equation}
since $\der{x^{AB}}=\lambda_{[A}\der{z^{B]}}$ under the integral.

More generally, we can write down the following contour integral, which interpolates between the above two formul\ae{} \eqref{eq:intform1} and \eqref{eq:intform2}:
\begin{equation}
\begin{aligned}
 {\psi}_{A_1\cdots A_{2h}}(x)\ &=\ \left.\oint_\CCC \Omega^{(3,0)}\, \lambda_{(A_1}\cdots\lambda_{A_{j+h}}\der{z^{A_{j+h+1}}}\cdots \der{z^{A_{2h})}} \hat f^{[h-j]}_{-2j-4}(z,\lambda)\right|_{z=x\cdot\lambda}~.
\end{aligned}
\end{equation}
Here, $j=-h,\ldots,h$ and the indices $A_1,\ldots,A_{2h}$ are symmetrised in the integrand. Again, it is straightforward to check that these fields satisfy the field equation $\dpar^{AB}\psi_{BA_1\cdots A_{2h-1}}=0$.

\section{Penrose--Ward transform in six dimensions}\label{sec:PWTransform}

In the previous section, we have seen how the Penrose transform relates  $H^3(\hat{U},\CO_{\hat U}(-2h-4))$ for  $h\geq 0$ to spinor fields on space-time subject to certain field equations. Recalling now the four-dimensional case, one would expect a Penrose transform for all $h\in\frac12\RZ$. However, the situation in six dimensions is rather different since  the cohomology group $H^3(\hat{U},\CO_{\hat U}(-2h-4))$ yields trivial space-time fields for $h<0$.  In fact, what replaces $H^3(\hat{U},\CO_{\hat U}(-2h-4))$ in this case is  the cohomology group $H^2(\hat{U},\CO_{\hat U}(2h-2))$ with $h> 0$. Notice that apart from these two cohomology groups all other (higher) cohomology groups (for $\CO_{\hat U}(k)$ with $k\in\RZ$) appear to give trivial fields on space-time; see Remark \ref{rem:zeroH3} below for more details.\footnote{See also \cite{Baston:1989} for an abstract discussion with different arguments.}

In this section, we wish to establish the following theorem:

{\theorem\label{thm:6DPWH2}
Let $U\subset M^6$ be open and convex and set  $U':=\pi_2^{-1}(U)\subset F^9$ and $\hat U:=\pi_1(\pi_2^{-1}(U))\subset P^6$, respectively. For $h\in\frac12\NN_0$, there is a canonical isomorphism
\begin{equation}
\CCP\,:\,H^2(\hat U,\CO_{\hat U}(2h-2))\ \to\ H^0(U,\CZ_h)~,
\end{equation}
where $\CZ_h$ is the sheaf of chiral zero-rest-mass fields defined in \eqref{eq:DefZRM}.  This transformation is called Penrose--Ward transform.
}

\vspace{10pt}
\noindent
We shall prove this theorem for $h>0$ in a very elementary way via so-called Riemann--Hilbert problems and gauge potentials. This derivation forms a direct generalisation of the potential formulation that was given in the four-dimensional setting by the authors of \cite{Penrose:1977aa,Ward:1977aa,Ward:1977ta,Eastwood:1981jy}. The case $h=0$ is somewhat exceptional and we shall treat it differently by using spectral sequence arguments similar to our previous discussion of the Penrose transform.

\subsection{Proof}

\paragraph{\mathversion{bold}Case $h>0$.}
We begin our considerations with elements of the cohomology group\footnote{We shall again consider convex open subsets $U\subset M^6$ of space-time together with the corresponding subsets $U'=\pi_2^{-1}(U)\subset F^9$ of the correspondence space and $\hat U=\pi_1(\pi_2^{-1}(U))\subset P^6$ of twistor space.} $H^2(\hat U,\CO_{\hat U}(0))\equiv H^2(\hat U,\CO_{\hat U})$, that is, $h=1$. This case is particularly interesting since this cohomology group encodes holomorphic {\it one-gerbes}\footnote{Some basic facts on $n$-gerbes are collected in Appendix \ref{app:gerbes}.} on $\hat U$: consider the exponential sheaf sequence on $\hat U$,
\begin{equation}\label{eq:ExpSheafSeq}
 0\ \longrightarrow\ \RZ\ \longrightarrow\ \CO_{\hat U}\ \stackrel{\exp}{\longrightarrow}\ \CO_{\hat U}^*\ \longrightarrow \ 0~.
\end{equation}
Here, $\CO_{\hat U}^*$ is the sheaf of non-vanishing holomorphic functions on $\hat U$ and $\exp:\CO_{\hat U}\to\CO_{\hat U}^*$ is the exponential map $\exp(f):=\mbox{e}^{2\pi\sqrt{-1}f}$. The induced long exact sequence of cohomology groups on $\hat U$ then yields
\begin{equation}
 H^1(\hat U,\CO_{\hat U}^*)\ \stackrel{c_1}{\longrightarrow}\ H^2(\hat U,\RZ)\ \longrightarrow\ H^2(\hat U,\CO_{\hat U})\
 \longrightarrow H^2(\hat U,\CO_{\hat U}^*)\ \stackrel{\rm DD}{\longrightarrow}\ H^3(\hat U,\RZ)~.
\end{equation}
Notice that the cohomology group $H^1(\hat U,\CO_{\hat U}^*)$ is the moduli space of holomorphic line bundles (zero-gerbes) over $\hat U$ while $H^2(\hat U,\CO_{\hat U}^*)$ is the moduli space of holomorphic one-gerbes over $\hat U$. In addition, $c_1$ is the first Chern class map which gives the characteristic class of line bundles while DD the Dixmier--Douady class map which, correspondingly, gives the characteristic class of one-gerbes. One can check\footnote{There are no odd-dimensional cells in the cell decomposition of $\hat U$.} that $H^3(\hat U,\RZ)=0$. Moreover, $c_1:H^1(\hat U,\CO_{\hat U}^*)\to H^2(\hat U,\RZ)$ is surjective. Therefore, the above cohomology sequence reduces to
\begin{equation}
 H^2(\hat U,\CO_{\hat U})\ \cong\ H^2(\hat U,\CO_{\hat U}^*)~.
\end{equation}
Thus, holomorphic one-gerbes over $\hat U$, which we denote by $\hat\Gamma$ in the following, are characterised by elements of $H^2(\hat U,\CO_{\hat U})$.  Our goal is now to establish a Penrose--Ward transform to find  the corresponding space-time interpretation of such gerbes. To slenderise the discussion, we shall consider the more general case of $H^2(\hat{U},\CO_{\hat U}(2h-2))$ with $h>0$ and loosely speak of holomorphic one-gerbes also in this case. Note that  the restriction of  $H^2(\hat{U},\CO_{\hat U}(2h-2))$ to $\hat x=\pi_1(\pi_2^{-1}(x))\hookrightarrow\hat U$ for any $x\in U$ vanishes by Lemma \ref{lem:Dolbeault}  (remember that $\hat x\cong\PP^3$), i.e.~a holomorphic one-gerbe $\hat\Gamma$ described by $H^2(\hat U,\CO_{\hat U}(2h-2))$ becomes holomorphically trivial upon restriction to any $\hat x$.

To make our constructions explicit, let us choose an open Stein cover $\hat{\mathfrak U}=\{\hat U_{a}\}$ of $\hat U$ and a (smooth) partition of unity $\hat\theta=\{\hat\theta_{a}\}$ subordinate to $\hat{\mathfrak U}$. As before, we shall write $\hat U_{ab}:=\hat U_{a}\cap\hat U_{b}$, $\hat U_{abc}:=\hat U_{a}\cap\hat U_{b}\cap\hat U_{c}$, etc. Consider a holomorphic one-gerbe $\hat\Gamma$ which is described by a  \v Cech cocycle $[\hat f]=[\{\hat f_{abc}\}]\in H^2(\hat U,\CO_{\hat U}(2h-2))$. The Dolbeault isomorphism allows us to identify the \v Cech cohomology group $H^2(\hat U,\CO_{\hat U}(2h-2))$ with the Dolbeault cohomology group $H^{(0,2)}_{\bar\partial}(\hat U,\CO_{\hat U}(2h-2))$ of $\bar\partial$-closed $(0,2)$-forms on $\hat U$ with values in the holomorphic line bundle $\CO_{\hat U}(2h-2)$.\footnote{Strictly speaking, they take values in some Abelian Lie algebra $\mathfrak{g}$.} Explicitly, this is done by using the partition of unity $\hat\theta=\{\hat\theta_{a}\}$: we may
introduce a smooth \v Cech one-cochain $\hat s$ by setting
\begin{equation}
 \hat s_{ab}\ :=\ \sum_c \hat f_{abc}\hat \theta_{c}~~~\mbox{on}~~~\hat U_{ab}~.
\end{equation}
This cochain gives rise to a smooth splitting of $\hat f$,
\begin{equation}
 \hat f_{abc}\ =\ \hat s_{ab}+\hat s_{bc}+\hat s_{ca}\quad\mbox{on}\quad \hat U_{abc}~.
\end{equation}
From this splitting, we can now define $(0,q)$-forms with $q=1,2$ by
\begin{equation}\label{eq:FormsT}
\begin{aligned}
 \hat A_{ab}\ &:=\ \bar\partial\hat s_{ab}\ =\ \sum_c \hat f_{abc} \bar\partial\hat\theta_{c}\quad\mbox{on}\quad \hat U_{ab}~,\\
 \hat{B}_{a}\ &:=\ \sum_{b,c}\hat f_{abc}\, \bar\partial\hat\theta_{b}\wedge\dparb\hat\theta_{c}
 \quad\mbox{on}\quad \hat U_{a}~.
 \end{aligned}
\end{equation}
These $(0,q)$-forms define a so-called {\em holomorphic connective structure} on $\hat\Gamma$. Clearly, they are all $\bar\partial$-closed on the respective intersections of the coordinate patches $\hat U_{a}$. Furthermore, the $(0,2)$-forms $\hat{B}_{a}$ yield a globally defined $\bar\partial$-closed $(0,2)$-form $\hat{B}^{(0,2)}$ with $\hat{B}_{a}=\hat{B}^{(0,2)}|_{\hat U_{a}}$ since $\hat B_{a}=\sum_{b,c}\hat s_{bc} \bar\partial\hat\theta_{b}\wedge\dparb\hat\theta_{c}$. This is the desired Dolbeault representative. We therefore have $\hat{H}^{(0,3)}:=\dparb \hat{B}^{(0,2)}=0$, which is the one-gerbe analogue of the equations of motion of (Abelian) holomorphic Chern--Simons theory for holomorphic vector bundles. Here, $\hat{H}^{(0,3)}$ is understood as the $(0,3)$-part of the three-form curvature.

Because $\hat\Gamma$ is holomorphically trivial on any $\hat x\hookrightarrow\hat U$, we have a holomorphic splitting of $\hat f$ on $\hat x$,
\begin{equation}
 \hat f_{abc}\ =\ \hat h_{ab}+\hat h_{bc}+\hat h_{ca}\quad\mbox{on}\quad \hat U_{abc}\ \cap\ \hat x~.
\end{equation}
Here, the \v Cech one-cochain $\hat h=\{\hat h_{ab}\}$ is holomorphic, i.e.\ $\hat h=\hat h(x,\lambda)$ depends holomorphically on $(x,\lambda)$. Finding such a splitting is known as {\em Riemann--Hilbert problem}. Notice that such splittings are not unique, as we can always shift $\hat h_{ab}\mapsto \hat\varphi_{a}-\hat\varphi_{b}$ for holomorphic $\hat\varphi=\{\hat\varphi_{a}\}$.

Consider now the cover $\mathfrak{U}':=\{U'_{a}\}$ of $U'\subset F^9$ which is induced by $\hat{\mathfrak{U}}$, i.e.\ $U'_{a}=\pi_1^{-1}(\hat U_{a})$, and a (smooth) partition of unity $\theta'=\{\theta'_{a}\}$ subordinate to $\mathfrak{U}'$. The pull-back $f'=\pi_1^*\hat f$ defines the pull-back one-gerbe $\Gamma'=\pi_1^*\hat\Gamma$ on $U'$. This one-gerbe is holomorphically trivial on $U'$ (since $\hat\Gamma$ was so on any $\hat x$), and we therefore have a holomorphic splitting
\begin{equation}\label{eq:HoloSplitCS}
 f'_{abc}\ =\ \pi_1^*\hat f_{abc}\ =\ h'_{ab}+h'_{bc}+h'_{ca}\quad\mbox{on}\quad U'_{abc}
\end{equation}
with $h'_{ab}=h'_{ab}(x,\lambda)$ holomorphic on $U'_{ab}\subset U'$. Notice that $h'=\hat h$. Moreover, we have another, smooth splitting $s'_{ab}:=\sum_c f'_{abc}\theta'_{c}$ obtained from the partition of unity. Evidently $h'_{ab}\neq s'_{ab}$, but both splittings are related by a gauge transformation $s'_{ab}=h'_{ab}-\varphi'_{a}+\varphi'_{b}$ with
\begin{equation}\label{eq:GaugePara}
 \varphi'_{a}\ :=\ \sum_b h'_{ab}\theta'_{b}~.
\end{equation}
We shall come back to this point below. In Appendix \ref{sec:splittings} we present an explicit derivation of the holomorphic splitting.

By the definition of a pull-back, $f'=\pi_1^*f$ is constant along the fibres of the fibration $\pi_1$. That is, $\dd_{\pi_1}  f'=0$ $\Leftrightarrow$ $V^A f'_{abc}=0$, where $\dd_{\pi_1}$ is  the relative exterior derivative introduced in \eqref{eq:RED1}. We now have the following natural differential forms on correspondence space:
\begin{equation}\label{eq:PBForms}
\begin{aligned}
A'_{ab}\ :=\ \dd_{\pi_1} h'_{ab} \quad\mbox{on}\quad U'_{ab}\eand
 B'_{a}\ :=\ \sum_b \dd_{\pi_1} h'_{ab}\wedge \bar\partial \theta'_{b}
 \quad\mbox{on}\quad U'_{a}~.
 \end{aligned}
\end{equation}
Notice that the differential forms $\pi_1^*\hat A_{ab}$ and $\pi_1^*\hat B_{a}$ obtained from \eqref{eq:FormsT} are gauge equivalent to \eqref{eq:PBForms} via the gauge transformation that is mediated by the gauge parameter \eqref{eq:GaugePara}.  Furthermore as one may check, $A'_{ab}$ is \v Cech-closed, thus representing an element of $H^1(U',\Omega^1_{\pi_1}(2h-2))$. However, by Proposition \ref{prop:DirectImage}, this cohomology group vanishes and therefore, $A'_{ab}$ can be split holomorphically as $A'_{ab}=A'_{a}-A'_{b}$. As above, such a splitting is not unique, and this fact will turn out to correspond to space-time gauge transformations. The splitting of $A'$ leads naturally to relative two-forms $\tilde B'_{a}:=\dd_{\pi_1} A'_{a}$ which are gauge equivalent to $B'_{a}$ in \eqref{eq:PBForms} by construction. They  now define a relative two-form $\tilde B'$ via $\tilde B'_{a}=\tilde B'|_{U'_{a}}$, which is defined globally (i.e.~on $U'$) and which is relatively closed, i.e.~the relative three-form
curvature $H':=\dd_{\pi_1}B'=0$ vanishes. Altogether, we have obtained a {\it flat} relative connective structure on the pull-back one-gerbe $\Gamma'$.

Using the Leray sequence and \eqref{eq:DirectImage3} of Proposition \ref{prop:DirectImage}, we find the identification
\begin{equation}
 H^0(U',\Omega^2_{\pi_1}(2h-2))\ \cong\ H^0(U,(\odot^{2h-1}S\otimes_{\CO_U}S^\vee)_0[1])~.
\end{equation}
We may express this explicitly as
\begin{equation}
 \tilde B'\ =\ e_A\wedge e_B\lambda_C\, \varepsilon^{ABCD} B_D{}^{A_1\cdots A_{2h-1}}\lambda_{A_1}\cdots\lambda_{A_{2h-1}}~,
\end{equation}
where $B_A{}^{A_1\cdots A_{2h-1}}=B_A{}^{(A_1\cdots A_{2h-1})}$ depends only on space-time and is totally trace-less.  In the above field expansion, we used relative one-forms $e_A$ of homogeneity $-1$, which combine with the tangent vectors $V^A$ to give the relative exterior derivative $\dd_{\pi_1}=e_A V^A$. The $e_A$ are not unique since $\lambda_A V^A=0$ implies that we can shift the $e_A$ by terms proportional to $\lambda_A$. Notice that the above expansion of $\tilde B'$ reflects this property. It is then a straightforward exercise to verify that $\dd_{\pi_1}\tilde B'=0$ if and only if
\begin{equation}\label{eq:PotEqB}
 \partial^{A (A_1}B_A{}^{A_2\cdots A_{2h})}\ =\ 0~.
\end{equation}

The choice one has in the splittings of the \v Cech cocycles involved in the above construction result in
$\tilde B'$s that differ by $\dd_{\pi_1}$-exact relative three-forms, i.e.\ $ \tilde B' \mapsto \tilde B'+\dd_{\pi_1}\tilde\Lambda'$, where $\tilde\Lambda'\in H^0(U',\Omega_{\pi_1}^1)$. Since
\begin{equation}
 H^0(U',\Omega^1_{\pi_1}(2h-2))\ \cong\ H^0\left(U,\left(\frac{\odot^{2h-1}S^\vee\otimes_{\CO_U} S^\vee}{\odot^{2h}S^\vee}  \right)^\vee[1]\right)
\end{equation}
by virtue of Proposition \ref{prop:DirectImage}, we have
\begin{equation}
 \tilde\Lambda'\ =\ e_A\lambda_B\eps^{ABCD}\,\Lambda_{CD}{}^{A_1\cdots A_{2h-2}}\lambda_{A_1}\cdots \lambda_{A_{2h-2}}~,
\end{equation}
where $\Lambda_{AB}{}^{A_1\cdots A_{2h-2}}=\Lambda_{[AB]}{}^{(A_1\cdots A_{2h-2})}$ depends only on space-time and is totally trace-less. Therefore, $\tilde B'\mapsto\tilde B'+\dd_{\pi_1}\tilde\Lambda'$ corresponds on space-time to
\begin{equation}\label{eq:GaugeTrafePotB}
 B_B{}^{A A_1\cdots A_{2h-2}}\ \mapsto\  B_B{}^{A A_1\cdots A_{2h-2}}+\left[\partial_{CB}\Lambda^{C(AA _1\cdots A_{2h-2})}-\partial^{C(A}\Lambda_{CB}{}^{A_1\cdots A_{2h-2})}\right]_0~,
\end{equation}
where the subscript zero refers to the totally trace-less part. Note that the trace-part of $\big[\partial_{CB}\Lambda^{C(AA _1\cdots A_{2h-2})}-\partial^{C(A}\Lambda_{CB}{}^{A_1\cdots A_{2h-2})}\big]$ does not enter in \eqref{eq:PotEqB} because the partial derivative is anti-symmetric in its indices. These are precisely the space-time gauge transformations displayed in \eqref{eq:GaugeTrafePotB-Intro}.

From the potentials $B_A{}^{A_1\cdots A_{2h-1}}$, one can derive fields $H_{A_1\cdots A_{2h}}\in H^0(U,(\odot^{2h}S^\vee)[1])$ by setting
\begin{equation}
 H_{A_1\cdots A_{2h}}\ :=\ \partial_{(A_1B_1}\cdots\partial_{A_{2h-1}B_{2h-1}} B_{A_{2h})}{}^{B_1\cdots B_{2h-1}}~.
\end{equation}
Because of \eqref{eq:PotEqB}, $H_{A_1\cdots A_{2h}}$ is a chiral zero-rest-mass field
\begin{equation}
 \partial^{AA_1}H_{A_1\cdots A_{2h}}\ =\ 0~.
\end{equation}
As an example, consider the case $h=1$. Here, $B_A{}^B$ represents a two-form potential of a self-dual three-form field since from \eqref{eq:PotEqB} we have $H^{AB}=\partial^{C(A}B_C{}^{B)}=0$ and $H_{AB}$ obeys the Dirac equation. In this case, the gauge transformation \eqref{eq:GaugeTrafePotB} reduces to the familiar one for two-form potentials, that is, $B_B{}^A\mapsto B_B{}^A+\partial^{AC}\Lambda_{CB}-\partial_{BC}\Lambda^{CA}$.

{\rem\label{rem:zeroH3}
Notice that the above discussion can be used to verify that $H^2(\hat{U},\CO_{\hat U}(-2h-2))$ with $h<0$ yields trivial fields on space-time since by virtue of Proposition \ref{prop:DirectImage}, the appropriate direct images vanish. Likewise, the above discussion can be straightforwardly adapted to the case $H^3(\hat{U},\CO_{\hat U}(-2h-4))$ with $h<0$. In this case, a short derivation reveals that $H^3(\hat{U},\CO_{\hat U}(-2h-4))$ does not give anything non-trivial on space-time either. Finally, one can show that the same is true for $H^1(\hat{U},\CO_{\hat U}(k))$, $k\in\RZ$. }

\paragraph{\mathversion{bold}Case $h=0$.}
The case $h=0$ is somewhat exceptional since in this case $H^1(U',\Omega^1_{\pi_1}(-2))$ does not vanish as follows from  Proposition \ref{prop:DirectImage}. Thus, $A'_{ab}$ in \eqref{eq:PBForms} cannot be split a priori. Let us therefore treat this case differently by following the same procedure we used when proving that $H^3(\hat U,\CO_{\hat U}(-4))$ yields Klein--Gordon fields on space-time. We can be rather brief, however, as the discussion is very similar.

In particular, we find from Proposition \ref{prop:DirectImage}
\begin{equation}\label{eq:DirectZeroHeli-PW}
 \pi_{2*}^q\Omega^l_{\pi_1}(-2)\ \cong\ \begin{cases}
                                              [1] &\efor (q,l)=(1,1)~,\\
                                              [2] &\efor (q,l)=(0,3)~,\\
                                               0 &\quad\mbox{otherwise}~.
                                          \end{cases}
\end{equation}
When $(q,l)=(1,1)$, the corresponding Leray spectral sequence \eqref{eq:LeraySeq} yields
\begin{equation}\label{eq:LerayIso2-PW}
  H^p(U',\Omega^1_{\pi_1}(-2))\ \cong\ \begin{cases}
H^{p-1}(U,\pi_{2*}^3\Omega^1_{\pi_1}(-2))\ \cong\ H^{p-1}(U,[1])
  &\efor p\ \geq\ 1~,\\
0 &\efor p\ <\ 1~.
                                       \end{cases}
\end{equation}
Moreover, with \eqref{eq:DirectZeroHeli-PW} the Leray spectral sequence \eqref{eq:LeraySeq}  also gives
\begin{equation}\label{eq:LerayIso3.2}
 H^{p}(U,\pi_{2*}^q\Omega^l_{\pi_1}(-2))\ =\ 0
  \efor p,q\ \geq\ 0\eand l\ =\ 0,2~.
\end{equation}
When $(q,l)=(0,3)$, we obtain
\begin{equation}\label{eq:LerayIso4-PW}
  H^p(U',\Omega^3_{\pi_1}(-2))\ \cong\
H^{p}(U,\pi_{2*}^0\Omega^3_{\pi_1}(-2))\ \cong\ H^{p}(U,[2])
  \efor p\ \geq\ 0~.
\end{equation}

Using these ingredients,  the $r=1$ part of the spectral sequence $E_r^{p,q}$ given in \eqref{eq:ParticularSpecSeq}  reads as
\begin{equation}
\begin{aligned}
0\ \longrightarrow\ \kern.81cm 0\kern.8cm \longrightarrow\ 0\ \longrightarrow\ H^0(U,[2])\\
0\ \longrightarrow\ H^0(U,[1])\ \longrightarrow\ 0\ \longrightarrow\ H^1(U,[2])\\
0\ \longrightarrow\ H^1(U,[1])\ \longrightarrow\ 0\ \longrightarrow\ H^2(U,[2])\\
0\ \longrightarrow\ H^2(U,[1])\ \longrightarrow\ 0\ \longrightarrow\ H^3(U,[2])\\
0\ \longrightarrow\ H^3(U,[1])\ \longrightarrow\ 0\ \longrightarrow\ H^4(U,[2])\\
\vdots\kern3.95cm\vdots\kern.7cm
\end{aligned}
\end{equation}
Since the first and third columns of this diagram are zero, while the second and fourth ones are non-zero in general, the differential operator  $\dd_1$ on $E_1^{p,q}$ vanishes identically and therefore, we have the identification $E_1^{p,q}\cong E_2^{p,q}$. The differential operator $\dd_2$ on $E_2^{1,1}$ maps $E_2^{1,1}$ to $E_2^{3,0}$ and since $E_1^{p,q}\cong E_2^{p,q}$ and thus, $E_2^{1,1}\cong H^0(U,[1])$ and $E_2^{3,0}\cong H^0(U,[2])$, respectively, we have a map $\Box:H^0(U,[1])\to H^0(U,[2])$ which is induced by $\dd_2$. One can check that this map is indeed the Klein--Gordon operator  defined in \eqref{eq:DefZRM}. We then have
\begin{equation}
 E_3^{1,1}\ \cong\ \ker\left\{\Box\,:\,H^0(U,[1])\ \to\ H^0(U,[2]) \right\},
\end{equation}
together with  $E_3^{1,1}\cong\cdots\cong  E_\infty^{1,1}$ such that
\begin{equation}
 H^2(\hat U,\CO_{\hat U}(-2))\ \cong\ H^{2}(U',\pi_1^{-1}\CO_{\hat U}(-2))\ \cong\ E_3^{1,1}\ \cong\ H^0(U,\CZ_0)~.
\end{equation}

\subsection{Twistor space action}

We have seen so far that elements of both $H^2(\hat U,\CO_{\hat U}(2h-2)$  and $H^3(\hat U,\CO_{\hat U}(-2h-4))$ correspond to chiral zero-rest-mass fields of spin $h\geq0$. This is rather different when compared to four dimensions, where only a first cohomology group appears. Curiously, this difference allows us to write down a twistor space action principle for chiral fields (even without supersymmetry).

To demonstrate this, we first note that  the holomorphic measure on $P^6$ is a (6,0)-form of homogeneity $+6$ given by
\begin{equation}
 \Omega^{(6,0)}\ :=\ \oint_\CCC\frac{\Omega^{(4,0)}(z)\wedge \Omega^{(3,0)}(\lambda)}{z^A\lambda_A}~,
\end{equation}
where $\CCC$ is any contour encircling $P^6\hookrightarrow\PP^7_\circ$. We have again used holomorphic volume forms $\Omega^{(4,0)}(z):=\frac{1}{4!}\eps_{ABCD}\dd z^A\wedge\dd z^B\wedge\dd z^C\wedge\dd z^D$ and
$\Omega^{(3,0)}(\lambda):=\frac{1}{4!}\eps^{ABCD}\lambda_A\dd\lambda_B\wedge\dd\lambda_C\wedge\dd\lambda_D$.
Next we switch to the Dolbeault picture and represent elements of the cohomology groups $H^2(\hat U,\CO_{\hat U}(2h-2)$  and $H^3(\hat U,\CO_{\hat U}(-2h-4))$  in terms of their Dolbeault representatives. We shall denote them by $\hat B^{(0,2)}_{2h-2}$ and $\hat C^{(0,3)}_{-2h-4}$, respectively, where the subscript indicates the respective homogeneity. Now we extend these fields to off-shell fields by assuming that they are not holomorphic and introduce the following twistor space action functional:
\begin{equation}\label{eq:TwistorAction}
 S\ =\ \int\Omega^{(6,0)} \wedge \hat B^{(0,2)}_{2h-2}\wedge\bar\partial \hat C^{(0,3)}_{-2h-4}~.
\end{equation}
Note that this action is well-defined as the respective weights cancel out. Clearly, the equations of motion resulting from the action \eqref{eq:TwistorAction} are $\bar\partial\hat B^{(0,2)}_{2h-2}=0=\bar\partial\hat C^{(0,3)}_{-2h-4}$. On-shell, $\hat B^{(0,2)}_{2h-2}$ and $\hat C^{(0,3)}_{-2h-4}$ therefore correspond indeed to representatives of the \v Cech cohomology groups $H^2(\hat U,\CO_{\hat U}(2h-2))$ and $H^3(\hat U,\CO_{\hat U}(-2h-4))$.

The case of $h=1$ is of particular interest since in that case we obtain self-dual three-form fields on space-time. In that sense, \eqref{eq:TwistorAction} can be understood as a twistor space action for self-dual three-form fields. It would be interesting to see if, after imposing appropriate reality conditions (and partially fixing gauge), the action \eqref{eq:TwistorAction} is related to one of the chiral space-time actions of Pasti, Sorokin \& Tonin \cite{Pasti:1995ii,Pasti:1995tn,Pasti:1996vs,Pasti:1997gx}. We shall return to this issue in the near future.

\section{Reduction to lower dimensions}\label{sec:RedToLowerDim}

In this section, we shall present various dimensional reductions of the twistor space $P^6$  to lower dimensions, specifically to twistor spaces of three- and four-dimensional space-times. Concretely, we shall focus on the reductions to the ambitwistor space $P^5$ \cite{Witten:1978xx,Isenberg:1978kk}, to a new twistor space $P^3$ and to Hitchin's minitwistor space $P^2$ \cite{Hitchin:1982gh}. We shall also discuss the corresponding Penrose and Penrose--Ward transforms. In particular, as is well-known, the ambitwistor space underlies a Penrose--Ward transform  for the Maxwell equation in four dimensions, while the minitwistor space gives rise to a Penrose--Ward transform for the Abelian Bogomolny monopole equation in three dimensions.\footnote{Both ambitwistor space and minitwistor space can be used also in the non-Abelian setting, but in this paper, we are only interested in the Abelian setting.} As we shall show below, the new twistor space $P^3$ underlies a Penrose--Ward transform for the Abelian self-dual
string equation in four dimensions. We shall refer to $P^3$ as the {\em hyperplane twistor space}---the reason for this name becomes transparent shortly.

Note that a dimensional reduction on space-time induces a dimensional reduction of the subspaces of space-time that are parametrised by twistor space. In particular, the three-dimensional null-planes in $M^6$ parametrised by $P^6$ split under dimensional reduction from $M^6$ to $M^4$ into three classes of null spaces, as we shall argue. As there are no three-dimensional null vector spaces in $M^4$, we are left with the so-called $\alpha$- and $\beta$-planes in $M^4$ and the null-lines given by intersections of both types of null-planes. These null-spaces should be in one-to-one correspondence with points on the reduced twistor spaces. Therefore, the dimensional reduction on space-time naturally induces further reductions on twistor space, which depend on the kind of null-set we choose to work with.

Starting from the double fibration \eqref{eq:DoubleFibration}, we shall dimensionally reduce space-time $M^6$, which induces reductions of the incidence relation and, correspondingly, of the associated twistor space $P^6$ and the correspondence space $F^9$. We thus arrive at the following chain of double fibrations:
\begin{eqnarray}\label{diag:reductions}
  \begin{picture}(50,150)(0,0)
  \put(-14.0,120.0){\makebox(0,0)[c]{$P^6$}}
  \put(78.0,120.0){\makebox(0,0)[c]{$M^6$}}
  \put(34.0,148.0){\makebox(0,0)[c]{$F^9$}}
  \put(7.0,138.0){\makebox(0,0)[c]{$\pi_1$}}
  \put(55.0,138.0){\makebox(0,0)[c]{$\pi_2$}}
  \put(25.0,140.0){\vector(-2,-1){28}}
  \put(37.0,140.0){\vector(2,-1){28}}
  \put(-14.0,80.0){\makebox(0,0)[c]{$P^5$}}
  \put(78.0,80.0){\makebox(0,0)[c]{$M^4$}}
  \put(34.0,108.0){\makebox(0,0)[c]{$F^6$}}
  \put(7.0,98.0){\makebox(0,0)[c]{$\pi_3$}}
  \put(55.0,98.0){\makebox(0,0)[c]{$\pi_4$}}
  \put(25.0,100.0){\vector(-2,-1){28}}
  \put(37.0,100.0){\vector(2,-1){28}}
  \put(-14.0,40.0){\makebox(0,0)[c]{$P^3$}}
  \put(78.0,40.0){\makebox(0,0)[c]{$M^4$}}
  \put(34.0,68.0){\makebox(0,0)[c]{$F^6$}}
  \put(7.0,58.0){\makebox(0,0)[c]{$\pi_5$}}
  \put(55.0,58.0){\makebox(0,0)[c]{$\pi_6$}}
  \put(25.0,60.0){\vector(-2,-1){28}}
  \put(37.0,60.0){\vector(2,-1){28}}
  \put(-14.0,0.0){\makebox(0,0)[c]{$P^2$}}
  \put(78.0,0.0){\makebox(0,0)[c]{$M^3$}}
  \put(34.0,28.0){\makebox(0,0)[c]{$F^4$}}
  \put(7.0,18.0){\makebox(0,0)[c]{$\pi_7$}}
  \put(55.0,18.0){\makebox(0,0)[c]{$\pi_8$}}
  \put(25.0,20.0){\vector(-2,-1){28}}
  \put(37.0,20.0){\vector(2,-1){28}}
  \put(76.0,32.0){\vector(0,-1){23}}
  \put(-16.0,32.0){\vector(0,-1){23}}
  \put(32.0,60.0){\vector(0,-1){23}}
  \put(76.0,72.0){\vector(0,-1){23}}
  \put(-16.0,72.0){\vector(0,-1){23}}
  \put(32.0,100.0){\vector(0,-1){23}}
  \put(76.0,112.0){\vector(0,-1){23}}
  \put(-16.0,112.0){\vector(0,-1){23}}
  \put(32.0,140.0){\vector(0,-1){23}}
  \put(83.0,60.0){\makebox(0,0)[c]{$\cong$}}
  \put(39.0,88.0){\makebox(0,0)[c]{$\cong$}}
 \end{picture}\notag\\[-3cm]
\end{eqnarray}

\vspace{2.2cm}
\noindent where $M^n:=\FC^n$. To jump ahead of our story a bit, we shall find:
\begin{equation*}
\begin{aligned}
  P^6\ &\cong\ \Omega^1_{\PP^3}\otimes\CO_{\PP^3}(2)\ \ \hat=\ \ \mbox{twistor space of six-dimensional space-time}~M^6~,\\
  P^5\ &\cong\ {\rm Jet}^1\CO_{\PP^1\times\PP^1}(1,1) \ \ \hat=\ \ \mbox{ambitwistor space of four-dimensional space-time}~M^4~,\\
  P^3\ &\cong\ \CO_{\PP^1\times\PP^1}(1,1) \ \ \hat=\ \ \mbox{hyperplane twistor space of four-dimensional space-time}~M^4~,\\
  P^2\ &\cong\ \CO_{\PP^1}(2) \ \ \hat=\ \ \mbox{minitwistor space of three-dimensional space-time}~M^3~.
\end{aligned}
\end{equation*}
The above terminology shall be clarified when constructing the respective twistor spaces.

\subsection{Field equations in lower dimensions}\label{sec:FEqnsLowerDim}

Before presenting these reductions, we explain how the self-dual string equation, the Maxwell equation and the Bogomolny equation arise via dimensional reductions  of the equations of motion of the six-dimensional self-dual three-form field strength $H=\dd B$. As we have already discussed in Section \ref{sec:6dZRMfields}, a general three-form $H=\dd B$ in six dimensions is given by a pair of symmetric bi-spinors $H_{AB}=\partial_{C(A}B_{B)}{}^{C}$ and $H^{AB}=\partial^{C(A}B_C{}^{B)}$ via a (trace-less) two-form potential $B_B{}^A$. Imposing self-duality onto $H$ is equivalent to saying that $H^{AB}=0$.

\paragraph{\mathversion{bold}Field equations in four dimensions.} To dimensionally reduce $M^6$ to $M^4$, we split the $\mathbf{6}\cong\mathbf{4}\wedge\mathbf{4}$ representation of $\aso(6,\FC)\cong \asl(4,\FC)$ into the bi-fundamental representation $(\mathbf{2},\mathbf{2})$ of $\asl(2,\FC)\oplus \asl(2,\FC)\cong \aso(4,\FC)$ plus twice the trivial representation to obtain
\begin{equation}\label{eq:4dsplitx}
 x^{AB}\ \to\ (x^{\alpha\dot\beta},\varepsilon^{\alpha\beta}x^+,\varepsilon^{\dot\alpha\dot\beta}x^-)~,
\end{equation}
where $\alpha,\beta,\ldots,\dot\alpha,\dot\beta,\ldots=1,2$. The symplectic forms $\varepsilon^{\alpha\beta}$ and $\varepsilon^{\dot\alpha\dot\beta}$  of $\asl(2,\FC)\oplus \asl(2,\FC)$ can be used to raise and lower $\asl(2,\FC)$ spinor indices. Four-dimensional space-time $M^4$ is then given by the quotient $M^4:= M^6/D_{M^6}$ with the distribution $D_{M^6}:=\langle \partial_\pm\rangle$.

A two-form potential in six dimensions $B_B{}^A$ reduces then to
\begin{equation}
 B_B{}^A\ \to\ (A_{\alpha\dot\alpha}^+,A_{\alpha\dot\alpha}^-,B_{\alpha\beta}=B_{(\alpha\beta)},B_{\dot\alpha\dot\beta}=B_{(\dot\alpha\dot\beta)},\phi)~
 \end{equation}
 and represents in four dimensions two one-form potentials $A_{\alpha\dot\alpha}^\pm$, a two-form potential $(B_{\alpha\beta},B_{\dot\alpha\dot\beta})$ and a scalar field $\phi$. Notice that we used the symplectic forms $\varepsilon_{\alpha\beta}$ and $\varepsilon_{\dot\alpha\dot\beta}$ to rise and lower spinor indices. Correspondingly, gauge transformations of $B_B{}^A$,
 \begin{equation}
 B_B{}^A\ \mapsto\ B_B{}^A+\partial^{AC}\Lambda_{CB}-\partial_{BC}\Lambda^{CA}~,
\end{equation}
where $\Lambda_{AB}=-\Lambda_{BA}$, reduce in four dimensions to $\Lambda_{AB}\to (\Lambda_{\alpha\dot\alpha},\Lambda^+,\Lambda^-)$ with
\begin{equation}\label{eq:GaugeTrafosB}
\begin{aligned}
 A_{\alpha\dot\alpha}^\pm\ &\mapsto\ A_{\alpha\ald}^\pm+\dpar_{\alpha\ald}\Lambda^\pm~,\\
 B_{\alpha\beta}\ &\mapsto\ B_{\alpha\beta}+\eps^{\ald\bed}\dpar_{(\alpha\ald}\Lambda_{\beta)\dot\beta}~,\\
B_{\ald\bed}\ &\mapsto\  B_{\ald\bed}+\eps^{\alpha\beta}\dpar_{\alpha(\ald}\Lambda_{\beta\dot\beta)}~,\\
\phi\ &\mapsto\ \phi~.
\end{aligned}
\end{equation}
Furthermore, the (first-order) self-duality equation $H^{AB}=\partial^{C(A}B_C{}^{B)}=0$ reduces to
\begin{subequations}
\begin{equation}\label{eq:SDSSpinorNotation}
  H_{\alpha\dot\alpha}\ :=\ \varepsilon^{\bed\dot\gamma}\dpar_{\alpha\bed}B_{\ald\dot\gamma}-
\varepsilon^{\beta\gamma}\dpar_{\beta\ald}B_{\alpha\gamma}\ =\ \dpar_{\alpha\ald}\phi~
\end{equation}
and
\begin{equation}\label{eq:SDASDYM}
  f_{\dot\alpha\dot\beta}(A^+)\ =\ 0\eand f_{\alpha\beta}(A^-)\ =\ 0~,
\end{equation}
where $f_{\alpha\beta}$ and $f_{\ald\bed}$ are the self-dual and anti-self-dual parts of the curvature of a potential $A_{\alpha\dot\alpha}$, i.e.
\begin{equation}
  f_{\alpha\beta}(A)\ :=\ \eps^{\ald\bed}\dpar_{(\alpha\ald}A_{\beta)\dot\beta}\eand
 f_{\ald\bed}(A)\ :=\ \eps^{\alpha\beta}\dpar_{\alpha(\ald}A_{\beta\dot\beta)}~.
\end{equation}
\end{subequations}
Note that under this decomposition, $H_{AB}\to(f_{\alpha\beta}(A^+),f_{\dot\alpha\dot\beta}(A^-),\dpar_{\alpha\ald}\phi)$. Equation \eqref{eq:SDSSpinorNotation} is the self-dual string equation $H=\star_4\dd \phi$ in spinor notation, while \eqref{eq:SDASDYM} says that the curvature of $A^+$ (respectively, $A^-$) is self-dual (respectively, anti-self-dual). Recall that in four dimensions, the Maxwell equation and the Bianchi identity for a gauge potential $A_{\alpha\dot\alpha}$ read in spinor notation as
\begin{subequations}
\begin{equation}\label{eq:4dMaxwell}
 \varepsilon^{\bed\dot\gamma}\dpar_{\alpha\bed}f_{\ald\dot\gamma}+
\varepsilon^{\beta\gamma}\dpar_{\beta\ald}f_{\alpha\gamma}\ =\ 0~,
\end{equation}
and
\begin{equation}\label{eq:BianchiAndMaxwell}
 \varepsilon^{\bed\dot\gamma}\dpar_{\alpha\bed}f_{\ald\dot\gamma}-
\varepsilon^{\beta\gamma}\dpar_{\beta\ald}f_{\alpha\gamma}\ =\ 0~,
\end{equation}
\end{subequations}
so that the equations for $f_{\alpha\beta}$ and $f_{\ald\bed}$ decouple, i.e.\ $\varepsilon^{\bed\dot\gamma}\dpar_{\alpha\bed}f_{\ald\dot\gamma}=0=\varepsilon^{\beta\gamma}\dpar_{\beta\ald}f_{\alpha\gamma}$. Therefore, $A_{\alpha\dot\alpha}^\pm$ constitute the degrees of freedom of a general Maxwell field $A_{\alpha\dot\alpha}$, and we may write $A_{\alpha\dot\alpha}=A_{\alpha\dot\alpha}^++A_{\alpha\dot\alpha}^-$.

\paragraph{\mathversion{bold}Field equations in three dimensions.} To further reduce to three dimensions, we split the $(\mathbf{2},\mathbf{2})$ of $\asl(2,\FC)\oplus \asl(2,\FC)$ into the $\mathbf{3}\oplus \mathbf{1}$ of $\asl(2,\FC)$: $x^{\alpha\ald}\rightarrow (x^{\alpha\beta}=x^{(\alpha\beta)},x^{[12]})$. We then have $M^3:=M^4/D_{M^4}$ with $D_{M^4}:=\left\langle \der{x^{[12]}}\right\rangle$. Here, the field strength of a gauge potential reduces directly according to
\begin{equation}\label{eq:3dFS}
 \dpar_{\alpha\beta} A_{\gamma\delta}-\dpar_{\gamma\delta}A_{\alpha\beta}\ =\ \tfrac12(\eps_{\alpha\gamma}f_{\beta\delta}+\eps_{\beta\delta}f_{\alpha\gamma}+(\alpha\leftrightarrow\beta))~,
\end{equation}
and the BPS subsector of the reduced Maxwell equation is described by the Abelian Bogomolny equation\footnote{or Dirac monopole equation} $F=\star_3 \dd \phi$, which reads in spinor notation as
\begin{equation}\label{eq:Bogomolny}
 f_{\alpha\beta}\ =\ \dpar_{\alpha\beta}\phi~.
\end{equation}
Note that this equation can be obtained from the self-dual string equation by defining $F=\der{x^{[14]}}\lrcorner H$. In spinor notation, this amounts to defining $f_{\alpha\beta}:=H_{(\alpha\beta)}$ and \eqref{eq:SDSSpinorNotation} reduces to \eqref{eq:Bogomolny}.

\subsection{Ambitwistors and Maxwell fields}\label{ss:Ambitwistors}

\paragraph{\mathversion{bold}Ambitwistor space.}
The first reduction in the sequence \eqref{diag:reductions} is that of \eqref{eq:DoubleFibration} to the double fibration $(\pi_3,\pi_4)$ containing the ambitwistor space $P^5$. Recall from \eqref{eq:IncidenceSolution} that three-dimensional totally null-planes in $M^6$, which are in one-to-one correspondence with points in $P^6$, are given by
\begin{equation}\label{eq:null3plane}
 x^{AB}\ =\ x^{AB}_0+\eps^{ABCD}\mu_C\lambda_D~,
\end{equation}
where $\mu_A$ is defined modulo terms proportional to $\lambda_A$. In view of the splitting $\asl(4,\FC)$ into $\asl(2,\FC)\oplus \asl(2,\FC)$ used in reducing $M^6$ to $M^4$, we now  split the spinors $\lambda_A$ and $\kappa_A$ according to
\begin{equation}
 \lambda_A\ \rightarrow\ (\mu_\alpha,\lambda_\ald)\eand \mu_A\ \rightarrow\ (\kappa_\alpha,\nu_\ald)~.
\end{equation}
Correspondingly, upon using \eqref{eq:4dsplitx}, the equation \eqref{eq:null3plane} decomposes as
\begin{equation}\label{eq:reducinci}
\begin{aligned}
 x^{\alpha\bed}\ = \ x^{\alpha\bed}_0+\mu^\alpha\nu^\ald-\kappa^\alpha\lambda^\ald~,\quad
 x^+\ =\ x_0^++\nu_\ald\lambda^\ald~,\eand
 x^-\ =\ x_0^-+\kappa_\alpha\mu^\alpha~,
\end{aligned}
\end{equation}
where we have raised spinor indices with $\varepsilon^{\alpha\beta}$ and $\varepsilon^{\ald\bed}$, respectively. Next we impose $x^\pm=0=x^\pm_0$ to dimensionally reduce to four dimensions. Hence, $\nu_\ald\lambda^\ald=0$ and $\kappa_\alpha\mu^\alpha=0$. There are essentially three cases emerging from these equations: firstly, $\mu_\alpha=0$ but $\lambda_\ald\neq0$ and $\kappa_\alpha$ arbitrary and $\nu_\ald\propto\lambda_\ald$ so that the first equation \eqref{eq:reducinci} reduces to
\begin{equation}
 x^{\alpha\ald}\ =\ x^{\alpha\ald}_0-\kappa^\alpha\lambda^\ald~.
\end{equation}
This equation parametrises anti-self-dual two-planes in four dimensions (so-called $\alpha$-planes). Secondly, $\mu_\alpha\neq0$ but $\lambda_\ald=0$ and $\kappa_\alpha\propto\mu_\alpha$ and $\nu_\ald$ arbitrary so that
\begin{equation}
 x^{\alpha\ald}\ =\ x^{\alpha\ald}_0+\mu^\alpha\nu^\ald~,
\end{equation}
which  parametrises self-dual two-planes in four dimensions (so-called $\beta$-planes). Thirdly, we can have both $\mu_\alpha\neq0$ and $\lambda_\ald\neq0$ together with $\kappa_\alpha\propto\mu_\alpha$ and $\nu_\ald\propto\lambda_\ald$. This gives
\begin{equation}
 x^{\alpha\ald}\ =\ x^{\alpha\ald}_0+\varrho\mu^\alpha\lambda^\ald\efor \varrho\in\FC~,
\end{equation}
which are the null-lines in four dimensions arising from intersecting $\alpha$-planes and $\beta$-planes.

At this point, one has to make a choice which null-spaces one would like to work with. Choosing the $\alpha$- or $\beta$-planes will lead to either Penrose's twistor space or Penrose's dual twistor space which are the twistor spaces parametrising such null-planes. In the following, we shall work with the null-lines obtained from intersections of $\alpha$- and $\beta$-planes. This yields the ambitwistor space $P^5$ which is the twistor space parametrising such null-lines. As we shall show momentarily, the ambitwistor space $P^5$ as well as the correspondence space $F^6$ can be obtained by factoring out certain distributions. The quadric equation \eqref{eq:quadric} defining $P^6$ in the open subset $\PP_\circ^7$ of $\PP^7$ will reduce to the quadric equation defining $P^5$ in the open subset $\PP^3_\circ\times \PP^3_\circ$ of $\PP^3\times \PP^3$.

To proceed, we now decompose all the twistor coordinates $(z^A,\lambda_A)$ on $P^6$ according to the splitting of $\asl(4,\FC)$ into $\asl(2,\FC)\oplus \asl(2,\FC)$ to obtain
\begin{equation}
(z^A,\lambda_A)\ \rightarrow\ (z^\alpha,-w^\ald,\mu_\alpha,\lambda_\ald)~.
\end{equation}
Let us first consider the base space $\PP^3$ of the fibration $P^6\to\PP^3$. To reduce $\PP^3$ with homogeneous coordinates $\lambda_A$ to $\PP^1\times \PP^1$ with homogeneous coordinates $(\mu_\alpha,\lambda_\ald)$, we consider the corresponding reduction of the structure sheaf $\CO_{\PP^3}$ of $\PP^3$. Local sections $f$ of $\CO_{\PP^3}$ fulfil the equation $\Upsilon f=0$, where $\Upsilon:=\lambda_A\der{\lambda_A}$ is the Euler vector field on $\PP^3$. This reflects the invariance under re-scalings $\lambda_A\mapsto t\lambda_A$, with $t\in \FC^*$. Local sections $f$ of $\CO_{\PP^1\times \PP^1}$ fulfil $\mu_\alpha\der{\mu_\alpha}f=0$ and $\lambda_\ald\der{\lambda_\ald}f=0$. Therefore, the quotient of $\PP^3$ by the distribution
\begin{equation}
 D_{\PP^3}\ :=\ \left\langle\mu_\alpha\frac{\partial}{\partial\mu_\alpha}-\lambda_{\dot\alpha}\frac{\partial}{\partial  \lambda_{\dot\alpha}} \right\rangle
\end{equation}
can be identified with $\PP^1\times \PP^1$.

Analogously, one reduces $\PP^7$ with homogeneous coordinates $(z^A,\lambda_A)$ to $\PP^3\times \PP^3$. Since we are interested in non-compact versions, let us directly remove the $\PP^3$ defined by $z^A\neq 0$ and $\lambda_A=0$ from $\PP^7$ to obtain the ambient space $\PP^7_\circ=\PP^7\setminus\PP^3\cong\CO_{\PP^3}(1)\otimes\FC^4$ of the twistor space $P^6$ we encountered before. The quotient of $\PP^7_\circ$  by the distribution
\begin{equation}
 D_{\PP^7_\circ}\ :=\ \left\langle z^\alpha\der{z^\alpha}+\lambda_\ald\der{\lambda_\ald}-w^\ald\der{w^\ald}-\mu^\alpha\der{\mu^\alpha}\right\rangle~
\end{equation}
can be identified with $\PP^3_\circ\times \tilde\PP^3_\circ$, where $\PP^3_\circ$ and $\tilde \PP^3_\circ$ are each bi-holomorphic to the total space of the bundle $\CO_{\PP^1}(1)\otimes\FC^2$. The quadric condition $z^A\lambda_A=0$, which defines $P^6\hookrightarrow \PP^7_\circ$, descends to the quadric equation
\begin{equation}
 z^\alpha\mu_\alpha-w^\ald\lambda_\ald\ =\ 0~,
\end{equation}
which defines the ambitwistor space $P^5\hookrightarrow\PP^3_\circ\times \tilde\PP^3_\circ$ as a quadric hypersurface of $\PP^3_\circ\times \tilde\PP^3_\circ$.  Note that $\PP^3_\circ$ is Penrose's twistor-space of four-dimensional space time while $ \tilde\PP^3_\circ$ is the dual twistor space.

The correspondence space $F^6$ is obtained as the quotient of  $F^9\cong\FC^6\times\PP^3$ by the distribution
\begin{equation}
 D_{F^9}\ :=\ \left\langle\frac{\partial}{\partial x^\pm}, \mu_\alpha\frac{\partial}{\partial\mu_\alpha}-\lambda_{\dot\alpha}\frac{\partial}{\partial  \lambda_{\dot\alpha}} \right\rangle~,
\end{equation}
and we have $F^6:=F^9/D_{F^9}\cong\FC^4\times \PP^1\times \PP^1$. Altogether, we arrive at the following double fibration:
\begin{equation}\label{eq:doubleFibrationATwistorSpace}
\begin{picture}(50,40)
  \put(0.0,0.0){\makebox(0,0)[c]{$P^5$}}
                 \put(64.0,0.0){\makebox(0,0)[c]{$M^4$}}
                 \put(34.0,33.0){\makebox(0,0)[c]{$F^6$}}
                 \put(7.0,18.0){\makebox(0,0)[c]{$\pi_3$}}
                 \put(55.0,18.0){\makebox(0,0)[c]{$\pi_4$}}
                 \put(25.0,25.0){\vector(-1,-1){18}}
                 \put(37.0,25.0){\vector(1,-1){18}}
\end{picture}
\end{equation}
where $\pi_4$ is the trivial projection and
\begin{equation}
 \pi_3\,:\, (x^{\alpha\ald},\lambda_\ald,\mu_\alpha)\ \mapsto\ (z^\alpha,w^\ald,\mu_\alpha,\lambda_\ald)\ =\ (x^{\alpha\ald}\lambda_\ald,x^{\alpha\ald}\mu_\alpha,\mu_\alpha,\lambda_\ald)~.
\end{equation}
Note that the twistor distribution in this case is of rank one and generated by the vector field $\mu_\alpha\lambda_\ald\partial^{\alpha\ald}$, i.e.~$P^5\cong F^6/\langle \mu_\alpha\lambda_\ald\partial^{\alpha\ald}\rangle$ with $\dpar^{\alpha\ald}:=\eps^{\alpha\beta}\eps^{\ald\dot\beta}\der{x^{\beta\dot\beta}}$.

Geometrically, a point $x$ in four-dimensional space-time $M^4$ corresponds to a holomorphic embedding of  $\hat x:=\pi_3(\pi_4^{-1}(x))\cong \PP^1\times \PP^1\hookrightarrow P^5$. On the other hand, a point $p$ in ambitwistor space $P^5$ corresponds to a null line $\pi_4(\pi_3^{-1}(x))\hookrightarrow M^4$ given by
\begin{equation}
  x^{\alpha\ald}\ =\  x^{\alpha\ald}_0+\varrho\mu^\alpha\lambda^\ald~,\ewith\varrho\ \in\ \FC~,
\end{equation}
in agreement with our initial choice of null-space.

Moreover, if we introduce the two projections $\mbox{pr}_{1,2}\,:\,\PP^1\times\PP^1\to\PP^1$ to the first and second copy of $\PP^1$, respectively, and in addition
\begin{equation}\label{eq:DefO(k,l)}
\begin{aligned}
  \CO_{\PP^1\times\PP^1}(k,l)\ &:=\ \mbox{pr}_1^*\CO_{\PP^1}(k)\otimes\mbox{pr}_2^*\CO_{\PP^1}(l)~, \\ \Omega^p_{\PP^1\times\PP^1}(k,l)\ &:=\  \Omega^p_{\PP^1\times\PP^1}( \mbox{pr}_1^*\CO_{\PP^1}(k)\otimes\mbox{pr}_2^*\CO_{\PP^1}(l))
  \end{aligned}
\end{equation}
for $k,l\in\RZ$, then the sequence \eqref{eq:NormalSequence2} naturally reduces to a corresponding sequence for the ambitwistor space
\begin{equation}
 0\ \longrightarrow\ P^5\ \longrightarrow\ \big(\CO_{\PP^1\times\PP^1}(1,0)\oplus \CO_{\PP^1\times\PP^1}(0,1)\big)\otimes \FC^2\ \stackrel{\kappa}{\longrightarrow}\ \CO_{\PP^1\times\PP^1}(1,1) \ \longrightarrow \ 0~.
\end{equation}
Here, $\kappa:(z^\alpha,w^\ald,\mu_\alpha,\lambda_\ald)\mapsto z^\alpha\mu_\alpha- w^\ald\lambda_\ald$. Upon dualising and twisting by $\CO_{\PP^1\times\PP^1}(1,1)$ the Euler sequence for $\PP^1\times \PP^1$, we find
\begin{equation}
 0\ \longrightarrow\ \Omega^1_{\PP^1\times \PP^1}(1,1)\ \longrightarrow\ P^5\ \longrightarrow\ \CO_{\PP^1\times\PP^1}(1,1) \ \longrightarrow \ 0~.
\end{equation}
This implies that $P^5$  can be identified with the bundle of first-order jets ${\rm Jet}^1\CO_{\PP^1\times\PP^1}(1,1)$ of $\CO_{\PP^1\times\PP^1}(1,1)$ as a consequence of the jet-sequence
\begin{equation}
 0\ \longrightarrow\ \Omega^1_X(\CS)\ \longrightarrow\ {\rm Jet}^1\CS\ \longrightarrow\ \CS\ \longrightarrow\ 0
\end{equation}
for an Abelian sheaf $\CS$ on a complex manifold $X$ (see~e.g.~\cite{Manin:1988ds}).

{\rem\label{rem:NotationYM}
The above constructions show that we have a factorisation of the tangent bundle $T_{M^4}$ into the two bundles of undotted and dotted chiral spinors. We shall denote these bundles by $S$ and $\tilde S$ and therefore, $T_{M^4}\cong S\otimes_{\CO_{M^4}}\tilde S$, which is the reduction of the corresponding factorisation \eqref{eq:TangentBundle} in six dimensions. Note that such a factorisation amounts to choosing a holomorphic conformal structure. Furthermore, we shall make use of the following notation (for $k,l\in\RZ$):
\begin{equation}
 [k,l]\ :=\ \begin{cases}
 \otimes^k\det S^\vee\otimes_{\CO_{M^4}} \otimes^l\det\tilde S^\vee &\efor k,l>0~,\\
  \otimes^k\det S^\vee\otimes_{\CO_{M^4}} \otimes^{|l|}\det\tilde S &\efor k>0~,~~l<0~,\\
   \otimes^{|k|}\det S\otimes_{\CO_{M^4}} \otimes^l\det\tilde S^\vee &\efor k<0~,~~l>0~,\\
    \otimes^{|k|}\det S\otimes_{\CO_{M^4}} \otimes^{|l|}\det\tilde S &\efor k,l<0~,
 \end{cases}
\end{equation}
and we shall write $\CS[k,l]:=\CS\otimes_{\CO_{M^4}}[k,l]$ for an Abelian sheaf $\CS$ on $M^4$. In addition, we introduce
\begin{equation}
\CO_{P^5}(k,l)\ :=\ {\rm pr}^*\CO_{\PP^1\times\PP^1}(k,l)\efor k,l\ \in\ \RZ~,
\end{equation}
where ${\rm pr}$ is the bundle projection ${\rm pr}:P^5\to\PP^1\times\PP^1$ and $\CO_{\PP^1\times\PP^1}(k,l)$ was defined in \eqref{eq:DefO(k,l)}.
}

\paragraph{\mathversion{bold}Penrose--Ward transform.}
Let us consider an open set $U\subset M^4$ and define $\hat U:=\pi_3(\pi_4^{-1}(U))\subset P^5$ with covering $\mathfrak{\hat U}=\{\hat U_{a}\}$. We start from holomorphic line bundles over $\hat U$ which are holomorphically trivial on any $\hat x\cong \PP^1\times\PP^1\hookrightarrow \hat U$. Such line bundles are characterised by \v Cech one-cocycles $\hat f=\{\hat f_{ab}\}\in H^1(\hat{U},\CO_{\hat U})$. The pull-back of $\hat f$ to the correspondence space can be split holomorphically, $f'_{ab}=\pi_3^*\hat f_{ab}=h'_{a}-h'_{b}$. Since $f'_{ab}$ gets annihilated by the twistor distribution, we find $A':=\mu_\alpha\lambda_\ald\partial^{\alpha\ald}h'_{a}$ which is globally defined. Hence $A'$ must be of the form $A':=\mu_\alpha\lambda_\ald A^{\alpha\ald}$, where $A_{\alpha\ald}$ depends only on space-time. Since the twistor distribution is one-dimensional, we do not obtain any space-time field equations for $A_{\alpha\ald}$. Moreover, since the splitting $f'_{ab}=h'_{a}-h'_{b}$ is not
unique, we can always consider $h'_{a}\mapsto h'_{a}+\varphi'$, where $\varphi'$ is defined globally on $\hat U':=\pi_4^{-1}(U)\subset F^6$. Therefore, $\varphi'$ can only depend on space-time (since the $\PP^1$s are compact) and thus, it corresponds to transformations of the form $A_{\alpha\ald}\mapsto A_{\alpha\ald}+\partial_{\alpha\ald}\varphi'$. In summary, this shows that $H^1(\hat{U},\CO_{\hat U})$ can be identified with the Maxwell potentials on $U$ modulo gauge transformations. Notice that this construction also applies to the non-Abelian setting, that is, to Yang--Mills potentials.

In order to find a twistorial description of Maxwell/Yang--Mills fields which do satisfy the corresponding field equations, one has to do more work. Such descriptions (including their supersymmetric extensions) have been found a long time ago by Witten \cite{Witten:1978xx}, Isenberg, Green \& Yasskin \cite{Isenberg:1978kk,Isenberg:1978qd} and Manin \cite{Manin:1988ds}; see \cite{Buchdahl:1985aa,Pool:1981aa} for a cohomological analysis, Mason \& Skinner \cite{Mason:2005kn} for an action principle for Yang--Mills theory on ambitwistor space, and \cite{Popov:2004rb} for a recent review in conventions similar to ours. For the sake of completeness, we shall recall these constructions but keep the discussion very brief.

As is well-known, in order to construct self-dual (or anti-self-dual) solutions to the Maxwell/Yang--Mills equations, one employs Ward's construction \cite{Ward:1977ta} starting from holomorphic vector bundles over Penrose's twistor space $\PP^3_\circ$ (or the dual twistor space $\tilde \PP^3_\circ$) subject to certain triviality conditions. Because ambitwistor space incorporates both twistors and dual twistors, it can be used to give a twistor interpretation of the Maxwell/Yang--Mills equations. As we have seen above, however, the ambitwistor space itself is not quite sufficient to recover these equations. To resolve this problem, one needs to thicken the ambitwistor space into its ambient space $\PP^3_\circ\times \tilde \PP^3_\circ$ to a certain order. This is fully analogous to the thickening of $P^6$ in $\PP^7_\circ$ as encountered in Section \ref{sec:IntegralFormulae}. The {\em $\ell$-th order thickening} (or the $\ell$-th infinitesimal neighbourhood) is defined by
$P^5_{[\ell]}:=(P^5,\CO_{\PP^3_\circ\times \tilde \PP^3_\circ}/\CI^{\ell+1})$. Here, $\CO_{\PP^3_\circ\times \tilde \PP^3_\circ}$ is the sheaf of holomorphic functions on $\PP^3_\circ\times \tilde \PP^3_\circ$ and $\CI$ is the ideal subsheaf of $\CO_{\PP^3_\circ\times \tilde \PP^3_\circ}$ consisting of those functions that vanish on $P^5$. Now we have the following theorem:

\pagebreak[4]

{\theorem (\!\!\cite{Witten:1978xx,Isenberg:1978kk})
Let $U$ be an open subset of $M^{4}$ such that any null line intersects $U$ in a convex set. Then there is a one-to-one correspondence between gauge equivalence classes of complex holomorphic solutions to the Yang--Mills equations on $U$ and equivalence classes of holomorphic vector
bundles which are holomorphically trivial on any $\hat x\cong \PP^1\times\PP^1\hookrightarrow P^5$ for all $x\in U$ and which admit an extension to a 3-rd order thickening $P^5_{[3]} $ of $P^5$ in $\PP^3_\circ\times \tilde \PP^3_\circ$.
}

\vspace{10pt}
\noindent
Note that if the holomorphic vector bundle can be extended to a finite neighbourhood within the ambient space $\PP^3_\circ\times \tilde \PP^3_\circ$, then the space-time gauge field constructed from this vector bundle is either self-dual, anti-self-dual or Abelian \cite{Isenberg:1978kk}. Thus, if one is only interested in the Maxwell equation (as we are in the present case) one may work with holomorphic line bundles on the ambient space $\PP^3_\circ\times \tilde \PP^3_\circ$ which are holomorphically trivial on $\PP^1\times\PP^1\hookrightarrow \PP^3_\circ\times \tilde \PP^3_\circ$. Since $\PP^3_\circ$ is Penrose's twistor space while $\tilde \PP^3_\circ$ its dual, one finds a self-dual and an anti-self dual field strength. Both can be linearly superposed to obtain a solution to the Maxwell equation. This is possible as the equations for the two helicities decouple, as discussed in Section \ref{sec:FEqnsLowerDim}.

\paragraph{Penrose transform.}
Besides the Penrose--Ward transform and Maxwell fields, one may also consider other spinor fields  on space-time which can be obtained from certain cohomology groups on ambitwistor space. In fact, we have the following theorems due to Pool \cite{Pool:1981aa} and Eastwood \cite{JSTOR:2990349}.

{\theorem\label{thm:4DPenrose-P} (Pool \cite{Pool:1981aa})
Consider the double fibration \eqref{eq:doubleFibrationATwistorSpace}. Let $U\subset M^4$ be open and convex and set $U':=\pi_4^{-1}(U)\subset F^6$ and $\hat U:=\pi_3(\pi_4^{-1}(U))\subset P^3$. For $h_{1,2}\in\frac12\NN$, there is a canonical isomorphism
\begin{equation}
\begin{aligned}
 \CCP\,:\,H^1(\hat U,\CO_{\hat U}(2h_1-2,2h_2-2))\ \to\kern4cm \\ \to\
 \left\{\frac{H^0(U,(\odot^{2h_1-1}S\otimes_{\CO_U}\odot^{2h_2-1}\tilde S)[1,1])}{\partial^{\alpha\ald}H^0(U,\odot^{2h_1-2}S\otimes_{\CO_U}\odot^{2h_2-2}\tilde S)}  \right\} .
 \end{aligned}
\end{equation}
}

\vspace{10pt}
\noindent
In particular, for $h_1=h_2=1$ we recover the identification of $H^1(\hat U,\CO_{\hat U})$ with Maxwell potentials on $U\subset M^4$ modulo gauge transformations.

{\theorem\label{thm:4DPenrose-E} (Eastwood \cite{JSTOR:2990349})
Let $U\subset M^4$ be open and convex and set $U':=\pi_4^{-1}(U)\subset F^6$ and $\hat U:=\pi_3(\pi_4^{-1}(U))\subset P^3$, where the maps $\pi_{3,4}$ are those appearing in the double fibration \eqref{eq:doubleFibrationATwistorSpace}. For $h_{1,2}\in\frac12\NN_0$, there is a canonical isomorphism
\begin{equation}
\begin{aligned}
 \CCP\,:\,H^2(\hat U,\CO_{\hat U}(-2h_1-2,-2h_2-2))\ \to\kern4cm \\ \to\
 \left\{\begin{array}{cc} \psi_{\alpha_1\cdots\alpha_{2h_1}\ald_1\cdots\ald_{2h_2}}\in H^0(U,(\odot^{2h_1}S\otimes_{\CO_U}\odot^{2h_2}\tilde S)[1,1])  \\ \mbox{such that}~~\partial^{\alpha_1\ald_1}\psi_{\alpha_1\cdots\alpha_{2h_1}\ald_1\cdots\ald_{2h_2}}=0 \end{array}\right\} .
 \end{aligned}
\end{equation}
}

\vspace{10pt}
\noindent
Eastwood's result is particularly interesting for the case when $h_1=h_2=\frac12$ since then, we have an identification of $H^2(\hat U,\CO_{\hat U}(-3,-3))$ with all conserved currents on $U\subset M^4$. Notice that Eastwood also gives a twistorial interpretation of {\em massive} space-time fields in \cite{JSTOR:2990349}.

\subsection{Hyperplane twistors and self-dual strings}\label{sec:HPT-SDS}

In this section, we introduce a new twistor space $P^3$ which, to our knowledge, has not been considered before. For that reason, we shall present a more detailed discussion in the following. As we shall see below, this twistor space underlies a Penrose--Ward transform mapping a certain cohomology group on $P^3$ to solutions to the self-dual string equation on $M^4$ in a bijective manner.

\paragraph{\mathversion{bold}Hyperplane twistor space.} While ambitwistor space describes the intersection of $\alpha$- and $\beta$-planes, the {\em hyperplane twistor space} will describe the span of the union of both types of intersecting null-planes. Because these two kinds of planes intersect along a null-line, the span describes a three-dimensional hyperplane in $M^4$, hence the name hyperplane twistor space.

In the corresponding double fibration, space-time obviously remains the same. Recall that the two spheres in the correspondence space $F^6\cong \FC^4\times \PP^1\times \PP^1$ specify the choice of an $\alpha$- and a $\beta$-plane. Because we need the same data in the definition of a hyperplane twistor, the correspondence space remains the same, too. The equivalence relation between points, however, is different: while two points in the correspondence space are equivalent if they correspond to the same null-line in the case of ambitwistors, in the case of hyperplane twistors,  two points in the correspondence space are considered equivalent if they correspond to the same hyperplane. Therefore the twistor distribution for the hyperplane twistor space contains that of the ambitwistor space, but it is strictly larger, and the hyperplane twistor space is a subspace of the ambitwistor space.

Explicitly, the hyperplane twistor space $P^3$ can be obtained by quotenting $P^5$ by the distribution
\begin{equation}
 D_{P^5}\ :=\ \left\langle \mu^\alpha\der{z^\alpha},\lambda^\ald\der{w^\ald} \right\rangle.
\end{equation}
It is rather straightforward to see that $P^3:=P^5/D_{P^5}$ is bi-holomorphic to the total space of the holomorphic line bundle $\CO_{\PP^1\times\PP^1}(1,1)\to \PP^1\times \PP^1$. Altogether, we may write down the following double fibration:
\begin{equation}\label{eq:DFforSDS}
\begin{picture}(50,40)
  \put(0.0,0.0){\makebox(0,0)[c]{$P^3$}}
                 \put(64.0,0.0){\makebox(0,0)[c]{$M^4$}}
                 \put(34.0,33.0){\makebox(0,0)[c]{$F^6$}}
                 \put(7.0,18.0){\makebox(0,0)[c]{$\pi_5$}}
                 \put(55.0,18.0){\makebox(0,0)[c]{$\pi_6$}}
                 \put(25.0,25.0){\vector(-1,-1){18}}
                 \put(37.0,25.0){\vector(1,-1){18}}
\end{picture}
\end{equation}
where $\pi_6$ is the trivial projection and
\begin{equation}
 \pi_5\,:\, (x^{\alpha\ald},\mu_\alpha,\lambda_\ald)\ \mapsto\ (z,\mu_\alpha,\lambda_\ald)\ =\ (x^{\alpha\ald}\mu_\alpha\lambda_\ald,\mu_\alpha,\lambda_\ald)~.
\end{equation}
Note that the twistor distribution is of rank three and generated by the vector fields $\mu_\alpha\partial^{\alpha\ald}$ and $\lambda_\ald\partial^{\alpha\ald}$, i.e.~$P^3\cong F^6/\langle \mu_\alpha\partial^{\alpha\ald},\lambda_\ald\partial^{\alpha\ald}\rangle$, with $\dpar^{\alpha\ald}:=\eps^{\alpha\beta}\eps^{\ald\dot\beta}\der{x^{\beta\dot\beta}}$ as before.

The geometric twistor correspondence here is as follows. By virtue of the incidence relation $z=x^{\alpha\ald}\mu_\alpha\lambda_\ald$, a point $x\in M^4$ corresponds to a holomorphic embedding of $\hat x:=\pi_5(\pi_6^{-1}(x))\cong \PP^1\times \PP^1\hookrightarrow P^3$, while a point $p\in P^3$ corresponds to a hyperplane  $\pi_6(\pi_5^{-1}(p))\hookrightarrow M^4$ in space-time. To see this, note that the incidence relation $z=x^{\alpha\ald}\mu_\alpha\lambda_\ald$ can be solved for fixed $p=(z,\mu,\lambda)\in P^3$ by
\begin{equation}\label{eq:SolIncSD}
 x^{\alpha\ald}\ =\ x^{\alpha\ald}_0+\mu^\alpha \nu^\ald-\kappa^\alpha\lambda^\ald~.
\end{equation}
Here, $x^{\alpha\ald}_0$ is a particular solution and $\kappa_\alpha$ and $\nu_\ald$ are arbitrary, which parametrise translations of $x^{\alpha\ald}_0$ along totally null two-planes (the $\alpha$- and $\beta$-planes). The apparent four parameters in the spinors $\nu^\ald$ and $\kappa^\alpha$ are reduced to three, because the shifts
\begin{equation}
\kappa_\alpha\ \mapsto\ \kappa_\alpha+\varrho\mu_\alpha\eand \nu_\ald\ \mapsto\ \nu_\ald+\varrho\lambda_\ald\efor \varrho\ \in\ \FC
\end{equation}
leave the solution \eqref{eq:SolIncSD} invariant.\footnote{Upon imposing reality conditions corresponding to Euclidean signature, the hyperplane twistor space parametrises oriented lines in $\FR^4$.}

{\rem\label{rem:NotationSDS}
The above constructions show again that we have a factorisation of the tangent bundle $T_{M^4}$ into the two bundles of undotted and dotted chiral spinors. As before, we shall denote these bundles by $S$ and $\tilde S$ and therefore, $T_{M^4}\cong S\otimes_{\CO_{M^4}}\tilde S$. Similarly to Remark \ref{rem:NotationYM}, we introduce
\begin{equation}
\CO_{P^3}(k,l)\ :=\ {\rm pr}^*\CO_{\PP^1\times\PP^1}(k,l)\efor k,l\ \in\ \RZ~,
\end{equation}
where ${\rm pr}$ is the bundle projection ${\rm pr}:P^3\to\PP^1\times\PP^1$ and $\CO_{\PP^1\times\PP^1}(k,l)$ was defined in \eqref{eq:DefO(k,l)}.
}

\paragraph{\mathversion{bold}Penrose--Ward transform.}
Let us fix a Stein cover $\hat{\frU}=\{\Uh_{a}\}$ of $P^3$ together with a partition of unity $\thetah=\{\thetah_{a}\}$ subordinate to $\hat\frU$. As before, we shall use the abbreviations $\Uh_{ab}=\Uh_{a}\cap\Uh_{b}$, etc.~for intersections of patches. Instead of working with all of $P^3$, we again allow for the restriction to open neighbourhoods $\Uh$  corresponding to open sets $U\subset M^4$ via $\Uh=\pi_5(\pi_6^{-1}(U))$.

Consider a representative $\hat f$ of the cohomology group $H^2(\Uh,\CO_{\hat U})$. The exponential sheaf sequence yields again a long exact sequence of cohomology groups containing
\begin{equation}
H^1(\hat U,\CO_{\hat U}^*)\ \stackrel{c_1}{\longrightarrow}\ H^2(\hat U,\RZ)\ \longrightarrow\ H^2(\hat U,\CO_{\hat U})\ \longrightarrow\ H^2(\hat U,\CO_{\hat U}^*)\ \stackrel{\rm DD}{\longrightarrow}\ H^3(\hat U,\RZ)~.
\end{equation}
Now the group $H^3(\hat U,\RZ)$ vanishes, which implies that all one-gerbes on the hyperplane twistor space have vanishing Dixmier--Douady class and therefore become holomorphically trivial on any $\hat x$. Moreover, the first map in this sequence is surjective and therefore elements of $H^2(\Uh,\CO_{\hat U})$ specify holomorphic one-gerbes on the hyperplane twistor space.

Using the partition of unity, we have a smooth splitting of $\fh$,
\begin{equation}
 \fh_{abc}\ =\ \sh_{ab}+\sh_{bc}+\sh_{ca}~
\end{equation}
on $\Uh_{abc}$, where $\sh_{ab}:=\sum_c \fh_{abc}\thetah_{c}$.  As before, we perform the transition to the Dolbeault picture by introducing certain differential $(0,1)$- and $(0,2)$-forms on $\Uh_{ab}$ and $\Uh_{a}$, respectively:
\begin{equation}
  \Ah_{ab}\ :=\ \dparb \sh_{ab}\ =\ \sum_c \fh_{abc}\dparb\thetah_{c}\eand
  \Bh_{a}\ :=\ \sum_{b,c}\fh_{abc}\dparb\thetah_{b}\wedge\dparb\thetah_{c}~.
\end{equation}
Altogether, we have obtained a holomorphic connective structure. Since $H^2(\Uh,\CO_{\hat U})$ vanishes when restricted to any $\hat x\cong\PP^1\times\PP^1$ for any $x\in U$, we can find a holomorphic splitting of $\fh$ on $\Uh_{abc}\cap \hat x$.

The next step in the Penrose--Ward transform is to pull-back $\hat f$ to the correspondence space $F^6$, which yields  $f'=\pi_5^*\hat f$. Analogously to the six-dimensional setting, we shall make use of the relative $p$-forms along $\pi_5:F^6\rightarrow P^3$. We denote them by $\Omega^p_{\pi_5}$ and they are given by a short exact sequence
\begin{equation}
 0\ \longrightarrow\ \pi_5^*\Omega^1_{P^3}\wedge\Omega^{p-1}_{F^6}\ \longrightarrow\ \Omega^p_{F^6}\ \longrightarrow\ \Omega^p_{\pi_5}\ \longrightarrow\ 0~.
\end{equation}
This sequence also yields the projection ${\rm pr}_{\pi_5}:\Omega^p\rightarrow \Omega^p_{\pi_5}$ via the quotient mapping, which allows us to introduce the relative differential $\dd_{\pi_5}:={\rm pr}_{\pi_5}\circ \dd:\Omega^p_{\pi_5}\rightarrow \Omega^{p+1}_{\pi_5}$, cf.\ \eqref{eq:RED1}. On $U'=\pi_5^{-1}(\hat U)=\pi_6^{-1}(U)$, we choose the cover $\frU':=\{U'_{a}\}$ that is induced by the cover $\hat\frU$.

Next we observe that the  \v Cech cocycle $f'=\pi_5^*\hat f$ can be split holomorphically on $U'_{abc}$. Therefore, we have
\begin{equation}\label{eq:HoloSplitSDS}
 f'_{abc}\ =\ h'_{ab}+h'_{bc}+h'_{ca}~,
\end{equation}
where $h'=\{h'_{ab}\}$ is holomorphic and, as before, its choice is not unique. Since $f'$ is the pull-back of $\hat f$ via $\pi_5$, it is annihilated by the relative exterior derivative $\dd_{\pi_5}$. This allows us to introduce the following differential forms:
\begin{equation}
A'_{ab}\ :=\ \dd_{\pi_5} h'_{ab} \quad\mbox{on}\quad U'_{ab}\eand
 B'_{a}\ :=\ \sum_b \dd_{\pi_5} h'_{ab}\wedge \bar\partial \theta'_{b}~
 \quad\mbox{on}\quad U'_{a}~.
 \end{equation}

Similarly to the six-dimensional construction, one may verify that the relative differential one-form $A'_{ab}$ is \v Cech-closed and thus represents an element of $H^1(U',\Omega^1_{\pi_5})$. To compute this cohomology group, we point out that we have a similar representation of the relative $p$-forms in terms of certain pull-back sheaves as in the six-dimensional setting, cf.~Proposition \ref{prop:RelFormsPullBackSheaves}. If we let $\Omega^{p}_{\pi_5}(k,l):=\Omega^{p}_{\pi_5}\otimes_{\CO_{F^6}}\pi_5^*\CO_{P^3}(k,l)$, then one can check that $\Omega^p_{\pi_5}$ is characterised by the following short exact sequence:
\begin{equation}\label{eq:RelFormsSDS}
\begin{aligned}
 0\ \longrightarrow\ \Omega^p_{\pi_5}
 \ \longrightarrow\ \Lambda^p\big[\pi_6^*(S[1,1])\otimes_{\CO_{F^6}}\pi_5^*\CO_{P^3}(0,1)\,\oplus\kern3cm\\ \oplus\, \pi_6^*(\tilde S[1,1])\otimes_{\CO_{F^6}}\pi_5^*\CO_{P^3}(1,0)\big]\ \longrightarrow\kern2cm \\
 \longrightarrow\ \pi_6^*[1,1]\otimes_{\CO_{F^6}}\Omega^{p-1}_{\pi_5}(1,1)\ \longrightarrow\ 0~,
 \end{aligned}
\end{equation}
where we have used the notation introduced in Remarks \ref{rem:NotationYM} and \ref{rem:NotationSDS}. The induced long exact sequence of cohomology groups yields that $H^1(U',\Omega^1_{\pi_5})=0$ since the map
\begin{equation}
\begin{aligned}
 H^0(U',\pi_6^*(S[1,0])\otimes_{\CO_{U'}}\pi_5^*\CO_{P^3}(0,1)\oplus&\pi_6^*(\tilde S[0,1])\otimes_{\CO_{U'}}\pi_5^*\CO_{P^3}(1,0))\\&\ \to\  H^0(U', \pi_6^*[1,1]\otimes_{\CO_{U'}}\Omega^{0}_{\pi_5}(1,1))
\end{aligned}
\end{equation}
is surjective. Therefore, we have a splitting $A'_{ab}=A'_{a}-A'_{b}$, which allows us to define a relative two-form by setting $\tilde B'_{a}:=\dd_{\pi_5}A'_{a}$. By construction, the two-forms $B'_{a}$ and $\tilde B'_{a}$ are gauge equivalent. Moreover, $\tilde B'_{a}$ defines a global (i.e.~on $U'$) relative two-form $\tilde B'$, which is relatively closed, i.e.~$\dd_{\pi_5}\tilde B'=0$.  Altogether, we have obtained a flat relative connective structure on the correspondence space.

The final step in the construction is to push-down $\tilde B'\in H^0(U',\Omega^2_{\pi_5})$ to space-time. To this end, we first point out that the sequence \eqref{eq:RelFormsSDS} also yields the isomorphism
\begin{equation}\label{eq:Rel2FonM}
  H^0(U',\Omega^2_{\pi_5})\ \cong\ H^0(U,(\odot^2S\otimes_{\CO_{U}}\Lambda^2\tilde S~\oplus~\Lambda^2 S\otimes_{\CO_{U}}\odot^2\tilde S~\oplus~\Lambda^2S\otimes_{\CO_{U}}\Lambda^2\tilde S)[2,2])~,
\end{equation}
which implies that a relative two-form corresponds to the space-time fields  $B_{\alpha\beta}=B_{(\alpha\beta)}$, $B_{\ald\bed}=B_{(\ald\bed)}$ and $\phi$ representing the field content of the self-dual string. To show that these fields indeed obey the self-dual string equation \eqref{eq:SDSSpinorNotation}, let us work in local coordinates $(x^{\alpha\ald},\mu_{\alpha},\lambda_{\dot\alpha})$ on the correspondence space. Recall that the relative tangent bundle of the fibration $\pi_5:F^6\to P^3$ (respectively, the twistor distribution) is spanned by the vector fields $V^\alpha:=\lambda_\ald\dpar^{\alpha\ald}$ and $V^\ald:=\mu_\alpha\dpar^{\alpha\ald}$, which satisfy $\mu_\alpha V^\alpha=\lambda_\ald V^\ald$. Correspondingly, the relative cotangent bundle, i.e.\ the bundle of relative one-forms, will be spanned by one-forms $e_\alpha$ and $e_\ald$ subject to the equivalence relation
\begin{equation}\label{eq:ERSDS}
(e_\alpha,e_\ald)\sim (e_\alpha+\mu_\alpha e,e_\ald-\lambda_\ald e)
 \end{equation}
for some $e$ to accommodate $\mu_\alpha V^\alpha=\lambda_\ald V^\ald$. The relative exterior derivative is then given by $\dd_{\pi_5}=e_\alpha V^\alpha+e_\ald V^\ald$.
In terms of the $e_\alpha$ and $e_{\dot\alpha}$, the most general expansion of a relative two-form then reads as
\begin{equation}\label{eq:FieldExpBSDS}
\begin{aligned}
 \tilde B'\ =\ e^\alpha\wedge e_\alpha \lambda^\ald\lambda^\bed B_{\ald\bed}+e^\ald\wedge e_\ald \mu^\alpha\mu^\beta B_{\alpha\beta}\,+\kern2cm\\
 +\, 2e^\alpha\wedge e^\ald\left(\mu_\alpha \lambda^\bed B_{\ald\bed}-\lambda_\ald\mu^\beta B_{\alpha\beta}-\mu_\alpha\lambda_\ald\phi\right),
 \end{aligned}
\end{equation}
where all the fields $B_{\alpha\beta}$, $B_{\ald\bed}$ and $\phi$ depend only on $x^{\alpha\ald}$ and represent the isomorphism \eqref{eq:Rel2FonM}. Note that the expression \eqref{eq:FieldExpBSDS} is invariant under \eqref{eq:ERSDS} as required for consistency. Using this expansion together with the expression for the relative exterior derivative, we obtain after some algebra that $\dd_{\pi_5}\tilde B'=0$ is equivalent to
\begin{equation}
  \varepsilon^{\bed\dot\gamma}\dpar_{\alpha\bed}B_{\ald\dot\gamma}-
\varepsilon^{\beta\gamma}\dpar_{\beta\ald}B_{\alpha\gamma}\ =\ \dpar_{\alpha\ald}\phi~
\end{equation}
on space-time. This, however, is precisely the self-dual string equation \eqref{eq:SDSSpinorNotation}. Finally, we would like to point out that the above splittings are, as always, not unique, which results in gauge transformations in $\tilde B'$ of the form $\tilde B'\mapsto \tilde B'+\dd_{\pi_5}\tilde\Lambda'$ where $\tilde\Lambda'\in H^0(U',\Omega^1_{\pi_5})$. Such $\tilde\Lambda'$ are of the form $\tilde\Lambda'=(e_\alpha\lambda_\ald+e_\ald\mu_\alpha)\Lambda^{\alpha\ald}$, where $\Lambda^{\alpha\ald}=\Lambda^{\alpha\ald}(x)$ depends only on space-time. Then $\tilde B'\mapsto \tilde B'+\dd_{\pi_5}\tilde\Lambda'$ induces the space-time gauge transformations displayed in \eqref{eq:GaugeTrafosB}.

In summary, we have established the following theorem:
{\theorem
Consider the double fibration \eqref{eq:DFforSDS}. Let $U\subset M^4$ be open and convex and set $U':=\pi_6^{-1}(U)\subset F^6$ and $\hat U:=\pi_5(\pi_6^{-1}(U))\subset P^3$, respectively. Then there is a canonical isomorphism:
\begin{equation}
 H^2(\hat U,\CO_{\hat U})\ \cong\
 \left\{\begin{array}{cc} \mbox{gauge equivalence classes of (complex holomorphic) solutions}  \\ \mbox{to the self-dual string equation}\  \mbox{on}\ U \end{array}\right\}.
\end{equation}
}

\vspace{10pt}
\noindent
Note that similarly to the discussion presented in Appendix \ref{sec:splittings}, also here one may construct the holomorphic splitting \eqref{eq:HoloSplitSDS} explicitly by using the Green function of the Dolbeault operator on $\PP^1\times\PP^1$.

{\rem One may also consider the case of $H^2(\hat U,\CO_{\hat U}(-2h_1-2,-2h_2-2)$ for $h_{1,2}\leq-1$ using the above constructions. Specifically, one would obtain $B_{\ald_1\cdots\ald_{-2h_2-2}\beta_1\cdots\beta_{-2h_1}}$, $B_{\alpha_1\cdots\alpha_{-2h_1-2}\bed_1\cdots\bed_{-2h_2}}$ and $\phi_{\alpha_1\cdots\alpha_{-2h_1-2}\bed_1\cdots\bed_{-2h_2-2}}$ as space-time fields, which are totally symmetric in all of their spinor indices. Likewise, the self-dual string equation generalises straightforwardly to
\begin{equation}
\begin{aligned}
  \varepsilon^{\bed\dot\gamma}\dpar_{(\alpha_1\bed}B_{\alpha_2\cdots\alpha_{-2h_1-1})\dot\gamma\bed_1\cdots\bed_{-2h_2-1}}-
\varepsilon^{\beta\gamma}\dpar_{\beta(\bed_1}B_{\bed_2\cdots\bed_{-2h_2-1})\gamma\alpha_1\cdots\alpha_{-2h_1-1}}\kern1.5cm\\
\ =\ \dpar_{(\alpha_1(\bed_1}\phi_{\alpha_2\cdots\alpha_{-2h_1-1})\bed_2\cdots\bed_{-2h_2-1})}~.
\end{aligned}
\end{equation}
The cases with either $h_1=-\frac12$ or $h_2=-\frac12$ seem not to give anything non-trivial. The cases with $h_{1,2}\geq0$ will be discussed below.
}

{\rem
One might wonder what the Penrose--Ward transform would yield for holomorphic line bundles over $P^3$ which are holomorphically trivial on any $\hat x$. Such line bundles are characterised by \v Cech one-cocycles $\hat f=\{\hat f_{ab}\}\in H^1(\hat{U},\CO_{\hat U})$. The pull-back of $\hat f$ can be split holomorphically, $f'_{ab}=\pi_5^*\hat f_{ab}=h'_{a}-h'_{b}$. This allows us to introduce a (global) relative one-form $A'=e_\alpha A^\alpha+e_\ald A^\ald$ with components $A^\alpha:=V^\alpha h'_{a} =: \lambda_\ald A^{\alpha\ald}$ and $A^\ald:=V^\ald h'_{a}=:\mu_\alpha A^{\alpha\ald}$, as a consequence of $\lambda_\ald A^\ald=\mu_\alpha A^\alpha$. Here, $A_{\alpha\ald}$ depends only on space-time. From the flatness condition on the corresponding curvature, we obtain the equation $V_\alpha A_\ald-V_\ald A_\alpha= \lambda^\ald\mu^\beta(\dpar_{\alpha\ald}A_{\beta\ald}-\dpar_{\beta\ald}A_{\alpha\ald})=0$. Hence, $A_{\alpha\ald}$ has to be pure gauge. Notice that this even holds true  in the non-Abelian case for rank-$r$ holomorphic vector bundle over $P^3$ with $r>1$.
}

\paragraph{Penrose transform.}
So far, we have discussed the Penrose--Ward transform yielding the identification of $H^2(\hat U,\CO_{\hat U})$ with the moduli space of solutions (obtained from the solution space as a quotient with respect to the group of gauge transformations) of the self-dual string equation. In this paragraph we would like to demonstrate that there is a natural extension to other field equations in four dimensions. Specifically, we are interested in space-time fields of the form $\psi_{\alpha_1\cdots\alpha_{2h_1}\ald_1\cdots\ald_{2h_2}}=\psi_{(\alpha_1\cdots\alpha_{2h_1})(\ald_1\cdots\ald_{2h_2})}$ with $h_{1,2}\in\frac12\NN_0$ which obey
\begin{subequations}\label{eq:EoM4dSDS}
\begin{equation}
 \partial^{\alpha_1\dot\beta}\psi_{\alpha_1\cdots\alpha_{2h_1}\ald_1\cdots\ald_{2h_2}}\ =\ 0\ =\ \partial^{\beta\ald_1}\psi_{\alpha_1\cdots\alpha_{2h_1}\ald_1\cdots\ald_{2h_2}}~.
\end{equation}
We shall refer to such fields as zero-rest-mass fields of helicity $(h_1,h_2)$. When either $h_1$ or $h_2$ vanishes then we have chiral spinors, $\psi_{\alpha_1\cdots\alpha_{2h_1}}$, or anti-chiral spinors, $\psi_{\ald_1\cdots\ald_{2h_2}}$. In the special case $h_1=h_2=0$, we have a scalar field (denoted by $\phi$) and, as always, a second-order field equation
\begin{equation}
 \Box\phi\ =\ 0~.
\end{equation}
\end{subequations}
Such zero-rest-mass fields can be constructed from representatives of cohomology groups on twistor space $P^3$ via the following theorem:

{\theorem\label{thm:4DPenroseSD}
Consider the double fibration \eqref{eq:DFforSDS}. Let $U\subset M^4$ be open and convex and set $U':=\pi_6^{-1}(U)\subset F^6$ and $\hat U:=\pi_5(\pi_6^{-1}(U))\subset P^3$. For $h_{1,2}\in\frac12\NN_0$, there is a canonical isomorphism\footnote{Notice that for either $h_1=0$ or $h_2=0$ one has a reduction of $H^2(\hat U,\CO_{\hat U}(-2h_1-2,-2h_2-2))$ to $H^1(\hat U_1,\CO_{\hat U_1}(-2h_1-2))$ or $H^1(\hat U,\CO_{\hat U}(-2h_2-2))$, where $\hat U_1$ is an open subset of Penrose's twistor space $\PP^3_\circ$ and $\hat U_2$ of its dual $\tilde \PP^3_\circ$.
}
\begin{equation}
 \CCP\,:\,H^2(\hat U,\CO_{\hat U}(-2h_1-2,-2h_2-2))\ \to\
 \left\{\begin{array}{cc} \mbox{zero-rest-mass fields}  \\ \mbox{of helicity}\ (h_1,h_2)\, \mbox{on}\ U \end{array}\right\} .
\end{equation}
}

\vspace{10pt}
\noindent
The proof of this theorem is essentially the same as the one presented in Section \ref{sec:Proof6DPW}. One can use again the machinery of spectral sequences and the direct image sheaves computed from the short exact sequences \eqref{eq:RelFormsSDS} of the relative differential forms after twisting by the pull-back of $\CO_{P^3}(k,l)$ for appropriate $k$ and $l$. Therefore, we shall refrain from repeating analogous arguments for the present situation, but concern ourselves with the corresponding contour integral formul{\ae}, which will make the theorem more transparent.

To this end, let us consider the canonical four-patch covering of $P^3$. An element of the cohomology group  $H^2(\hat U,\CO_{\hat U}(-2h_1-2,-2h_2-2))$ can be represented by a collection of four functions $\hat f=\{\hat f_{abc}\}$, each of which is of homogeneity $(-2h_1-2,-2h_2-2)$. For our choice of cover, all four triple overlaps $U_{abc}$ are of the same topology and moreover, also two double overlaps have the same topology as the triple overlaps. Therefore, two of the four functions defining the cocycle $\hat f$ can be fixed using equivalence relations, i.e.\ adding an appropriate \v Cech one-cochain. One of the remaining two functions is then fixed by the cocycle condition for $f$, leaving us with just one function defining the cocycle $\hat f$. Let us denote this function by $\hat f_{-2h_1-2,-2h_2-2}=\hat f_{-2h_1-2,-2h_2-2}(z,\mu,\lambda)$.  Using now the holomorphic measure on $\PP^1\times\PP^1$ given by
\begin{equation}
 \Omega^{(2,0)}\ :=\ \frac14\eps^{\alpha\beta}\eps^{\ald\bed}\,\mu_\alpha\dd\mu_\beta\wedge\lambda_\ald\dd\lambda_\bed~,
\end{equation}
we can define spinor fields via the following integral:
\begin{equation}
  \psi_{\alpha_1\cdots\alpha_{2h_1}\ald_1\cdots\ald_{2h_2}}(x)\ =\ \oint_\CCC\Omega^{(2,0)}\mu_{\alpha_1}\cdots\mu_{\alpha_{2h_1}}\lambda_{\ald_1}\cdots\lambda_{\ald_{2h_2}}\hat f_{-2h_1-2,-2h_2-2}(x^{\beta\bed}\mu_\beta\lambda_\bed,\mu,\lambda)~.
\end{equation}
It is trivial to check that these spinor fields obey the field equations \eqref{eq:EoM4dSDS}. Notice that one can write this covariantly using branched contour integrals (see e.g.~Penrose \& Rindler \cite{Penrose:1986ca} for a discussion in the ordinary twistor setting). Notice also that the above integral is somewhat similar to Eastwood's integral in the ambitwistor setting \cite{JSTOR:2990349}.

\subsection{Minitwistors and monopoles}

\paragraph{\mathversion{bold}Minitwistor space.}
The twistor space used to describe monopoles on three-dimensional space-time $M^3:=\FC^3$ is Hitchin's minitwistor space $P^2$ \cite{Hitchin:1982gh}. It can be regarded as the tangent space of $\PP^1$ or, equivalently, the total space of the holomorphic line bundle $\CO_{\PP^1}(2)\to\PP^1$.

In the twistor picture, the restriction of the moduli space of sections from $M^4$ to $M^3$ amounts to restricting the line bundle $P^3$ to the diagonal $\PP^1$ with $\mu_\alpha=\lambda_\alpha$ in the base $\PP^1\times \PP^1$ of $P^3$. We can achieve this by quotenting by the distribution
\begin{equation}
 D_{P^3}\ :=\ \left\langle\mu^\beta\lambda_\beta\left(\lambda_\alpha\der{\mu_\alpha}- \mu_\alpha\der{\lambda_\alpha}\right)\right\rangle.
\end{equation}
That is, $P^2:= P^3/D_{P^3}$, and the holomorphic line bundle $\CO_{\PP^1\times\PP^1}(1,1)\to \PP^1\times \PP^1$ reduces to the line bundle $\CO_{\PP^1}(2)\rightarrow \PP^1$. The correspondence space is obtained by taking the quotient of $F^6$ by the distribution
\begin{equation}
 D_{F^6}=\left\langle \der{x^{[12]}}, \mu^\beta\lambda_\beta\left(\lambda_\alpha\der{\mu_\alpha}- \mu_\alpha\der{\lambda_\alpha}\right)\right\rangle~,
\end{equation}
so that $F^4:=F^6/D_{F^6}\cong \FC^3\times\PP^1$. Here, we have the double fibration
\begin{equation}
\begin{picture}(50,40)
  \put(0.0,0.0){\makebox(0,0)[c]{$P^2$}}
                 \put(64.0,0.0){\makebox(0,0)[c]{$M^3$}}
                 \put(34.0,33.0){\makebox(0,0)[c]{$F^4$}}
                 \put(7.0,18.0){\makebox(0,0)[c]{$\pi_7$}}
                 \put(55.0,18.0){\makebox(0,0)[c]{$\pi_8$}}
                 \put(25.0,25.0){\vector(-1,-1){18}}
                 \put(37.0,25.0){\vector(1,-1){18}}
\end{picture}
\end{equation}
with $\pi_7:(x^{\alpha\beta},\mu_\alpha)\mapsto(z,\mu_\alpha)=(x^{\alpha\beta}\mu_\alpha\mu_\beta,\mu_\alpha)$ and $\pi_8$ being the trivial projection. In the case of $P^2$, we have a geometric twistor correspondence between points in $M^3$ and holomorphic embeddings $\PP^1\embd P^2$, as well as between points in $P^2$ and two-planes in $M^3$.\footnote{Upon imposing reality conditions corresponding to Euclidean signature, the minitwistor space parametrises oriented lines in $\FR^3$.} Notice that the twistor distribution here is of rank two and it is generated by the vector fields $\mu_\alpha\partial^{\alpha\beta}$, i.e.~$P^2\cong F^4/\langle \mu_\alpha\partial^{\alpha\beta}\rangle$ with $\dpar^{\alpha\beta}:=\eps^{\alpha\gamma}\eps^{\beta\delta}\der{x^{\gamma\delta}}$.

{\rem
There is an alternative way of obtaining the minitwistor space from the ambitwistor space in the non-Abelian setting. Firstly, one reduces to the miniambitwistor space \cite{Saemann:2005ji} underlying a Penrose--Ward transform for solutions to the three-dimensional Yang--Mills--Higgs theory. Restricting to BPS solutions then amounts to restricting the miniambitwistor space to the minitwistor space.
}

\paragraph{\mathversion{bold}Penrose--Ward transform.}
The construction of the Abelian monopole equations in the twistor context has been discussed extensively in the literature e.g.~in \cite{Hitchin:1982gh,Ward:1989aa}; see also \cite{Atiyah:1988jp,Popov:2004rt} and \cite{Popov:2005uv} for a review in conventions similar to ours. Let us therefore just make a few comments in the following.

The Penrose--Ward transform works here in the familiar way. A holomorphic vector bundle over $P^2$ which becomes holomorphically trivial upon restriction to the submanifolds $\hat x\cong\PP^1\embd P^2$ can be pulled back to $F^4$. Specifically, we have $[\hat f]=[\{\hat f_{ab}\}]\in H^1(\hat{U},\CO_{\hat U})$ for $\hat U\subset P^2$. The pull-back of $\hat f$ can be split holomorphically, $f'_{ab}=\pi_7^*\hat f_{ab}=h'_{a}-h'_{b}$. Using the Liouville theorem, this allows us to introduce a global relative one-form $A'$ with components
\begin{equation}
 A'^\alpha\ :=\ \mu_\beta\partial^{\alpha\beta} h'_{a}\ =:\ \mu_\beta (A^{\alpha\beta}-\eps^{\alpha\beta}\phi)~,
 \end{equation}
 where the fields on the right-hand-side depend only on space-time. From the flatness condition on the corresponding curvature, we obtain
\begin{equation}
 f_{\alpha\beta}\ =\ \partial_{\alpha\beta}\phi~,
\end{equation}
where $f_{\alpha\beta}$ is the curvature of $A_{\alpha\beta}$ as given in \eqref{eq:3dFS}. This is the spinorial form of the Bogomolny monopole equation $F:=\dd A=\star_3 \dd\phi$ in three dimensions.

\section{Conclusions and outlook}\label{sec:ConclusionsAndOutlook}

We have discussed the Penrose  transform on the twistor space $P^6$ associated with flat six-dimensional space-time. We have proved that this transform yields canonical isomorphisms between certain cohomological data on $P^6$ and solutions to the chiral zero-rest-mass field equations on space-time. We have also reviewed various contour integral formul\ae{} yielding chiral zero-rest-mass-fields. We have pointed out that some of them implicitly require an infinitesimal extension of the twistor space $P^6$ into its ambient space $\PP^7_\circ=\PP^7\setminus\PP^3\cong\CO_{\PP^3}(1)\otimes\FC^4$. Interestingly, it is also possible to establish a Penrose--Ward transform for \v Cech two-cocycles with values in various sheaves on twistor space (one-gerbes). The \v Cech two-cocycles lead to potentials giving rise to zero-rest-mass fields on space-time, which satisfy the appropriate field equations. After lifting the Dolbeault representatives of the \v Cech cocycles used in both the Penrose- and the Penrose-Ward transform to off-shell fields on twistor space, we were able to give a twistor space action principle for these fields. This action is to be seen analogous to holomorphic Chern--Simons theory in the twistor description of self-dual fields on four-dimensional space-time.

We found that dimensional reductions of chiral spinor-fields from six-dimensional space-time to four and three dimensions yield corresponding reductions of twistor space $P^6$ and the cohomological data in question. In particular, we have shown that the quadric $P^6$ in the open subset $\PP^7_\circ$ of $\PP^7$ reduces to the ambitwistor space $P^5$, which is a quadric in the open subset $\PP^3_\circ\times \tilde \PP^3_\circ$ of $\PP^3\times \tilde \PP^3$. Moreover, the infinitesimal extension of $P^6$ to $\PP^7_\circ$ that we encountered in the integral formul\ae{} is mapped to the familiar infinitesimal extension of the ambitwistor space $P^5$ inside $\PP^3_\circ\times\tilde \PP^3_\circ$. We have also found the
novel hyperplane twistor space $P^3\cong \CO_{\PP^1\times\PP^1}(1,1)$. The Penrose--Ward transform for this case maps very naturally one-gerbes over the hyperplane twistor space to solutions to the self-dual string equation. This result is particularly interesting, as the self-dual string has been very rarely discussed in the context of integrable field theories. Moreover, it fills a gap by providing, at least in the Abelian case, the twistor picture underlying the generalised Atiyah--Drinfeld--Hitchin--Manin--Nahm construction of \cite{Gustavsson:2008dy,Saemann:2010cp,Palmer:2011vx}. We have been able to complement this by a Penrose transform as well as corresponding contour integral formul\ae{} for other fields.

Our results lead to a number of questions that are beyond the scope of this paper, but which we shall address in forthcoming publications (cf.\ e.g.\ \cite{Saemann:2012uq}): 

An issue to be resolved is that of thickenings of the twistor space $P^6$. It could be preferable to work directly with a supertwistor space containing $P^6$ as its body, instead of thickenings of $P^6$. This is suggested by our experience with the corresponding four-dimensional discussions. A supersymmetric extension of the twistor space $P^6$ which has some of the desired properties has been proposed by Chern \cite{Chern:2009nt}.\footnote{The dimensional reduction to superambitwistor space, however, suggests to use a slightly different supertwistor space.} On the field theory side, it then remains to perform an analysis of the supersymmetric field equations in terms of constraint equations of a certain superconnection along the lines of \cite{Sohnius:1978wk,Harnad:1984vk,Harnad:1985bc}. Such an
analysis has recently been worked out for the M2-brane models by Samtleben \& Wimmer \cite{Samtleben:2009ts,Samtleben:2010eu}.

An in-depth analysis of the twistor space action \eqref{eq:TwistorAction} that appeared very naturally from the Dolbeault representatives of the cohomology groups of interest in the Penrose--Ward transform is also necessary. It would be exciting if it was possible to establish a connection to the actions describing self-dual three-forms in six dimensions as given by Pasti, Sorokin \& Tonin \cite{Pasti:1995ii,Pasti:1995tn,Pasti:1996vs,Pasti:1997gx}. As indicated, this would require imposing reality conditions and partially fixing gauges.

Moreover, it is certainly of major importance to find non-Abelian extensions of the Penrose--Ward transforms we presented here, possibly based on the non-Abelian generalisations of gerbes as introduced e.g.\ in \cite{Breen:math0106083} or \cite{Aschieri:2003mw,Jurco:2009px}. The fact that a non-Abelian version of the self-dual string equation can be found on loop space \cite{Gustavsson:2008dy,Saemann:2010cp,Palmer:2011vx} indeed suggests that such non-Abelian generalisations should exist. Amongst other things, an advantage of this approach is that, given a non-Abelian cohomology on twistor space, all space-time structures such as gauge transformations, field equations, etc.\ will follow automatically via Riemann--Hilbert problems.

Another interesting issue to address concerns the hidden (non-local)  symmetries of chiral six-dimensional field theories using twistor constructions as the ones presented here.  Specifically, one would be interested in the deformation theory of the corresponding cohomological data on twistor space to obtain hidden symmetries and integrable hierarchies for the equations of motion on space-time. Similar constructions were performed in four dimensions e.g.\ in \cite{Pohlmeyer:1979ya,Dolan:1983bp,Popov:1995qb,Ivanova:9702144,Popov:1998pc,Popov:1998fb,Wolf:2004hp,Wolf:2005sd,Popov:2006qu}. 

All of the above issues, i.e.\ the supersymmetric and non-Abelian extensions as well as the action principle and hidden symmetries and integrable hierarchies, should also be studied in the case of the hyperplane twistor space. In this case, we expect some of these problems to have more straightforward solutions than in the case of $P^6$.

Finally, one might also want to develop a twistor description of the (non-chiral) $\CN=(1,1)$ supersymmetric Yang--Mills theory in six dimensions.  In fact, Samtleben \& Wimmer \cite{Samtleben:2010eu} (see Devchand \cite{Devchand:1985au} for an earlier account) gave a Lax formulation for the constraint equations of the superconnection of this theory (see Harnad \& Shnider \cite{Harnad:1985bc})  with a six-dimensional null-vector as spectral parameter. This implies that one should be considering the space of all (super-)null-rays in six-dimensional (super-)space-time for a twistorial formulation of this theory \cite{Saemann:2012rr}.

\paragraph{Acknowledgements.}
We are very grateful to M.~Dunajski, J.~Figueroa-O'Farrill, \linebreak J.~Grant, L.~Hughston, L.~Mason, J.~McOrist, S.~Palmer, N.~Rink, A.~Torrielli, and R.~Wimmer for helpful discussions and comments. C.S. was supported by an EPSRC Career Acceleration Fellowship. M.W. was in part supported by an STFC Postdoctoral Fellowship at DAMTP, University of Cambridge and by a Senior Research Fellowship at Wolfson College Cambridge.

\appendices

\subsection{Spinor notation and reality conditions in six and four dimensions}\label{app:spinors}

\paragraph{Six dimensions.} We wish to describe a six-dimensional chiral field theory containing a self-dual three-form field strength. Similarly to four dimensions, where self-duality of a two-form field strength is only possible in Euclidean and split signature (Kleinian signature), we also have a limited choice of signatures in six dimensions: The Hodge star on three-forms is only idempotent for Minkowski and split signature. That is, we have to restrict ourselves to the spaces $\FR^{p,6-p}:=(\FR^6,\eta)$ with $p$ odd and metric $\eta=(\eta_{MN})=\diag(-\unit_p,\unit_{6-p})$ for $M,N,\ldots=1,\ldots,6$.

Consider now the generators $\gamma_M$ of the Clifford algebra $\CC\ell(\FR^{p,6-p})$ of $\FR^{p,6-p}$ with $p$ odd. We thus have $\{\gamma_M,\gamma_N\}=-2\eta_{MN}$. The spinor representation ${\bf 8}_s$ of the corresponding generators $\gamma_{MN}$ of $\sSpin(p,6-p)$ splits into the direct sum $S\oplus \tilde{S}$ of the subspaces of chiral and anti-chiral spinors. There is a natural isomorphism between $\tilde{S}$ and $S^\vee$ \cite{Penrose:1986ca}, so that we can exclusively work with elements of $S$ and $S^\vee$. We label the corresponding spinors by upper and lower indices, e.g.\ $(\psi^A)\in S$ and $(\chi_A)\in S^\vee$ with $A,B,\ldots=1,\ldots,4$.

We can choose a representation of $\CC\ell(\FR^{p,6-p})$ such that the sigma-matrices\footnote{They are the generalisations of the sigma-matrices in four dimensions, i.e.\ they correspond to the non-vanishing off-diagonal blocks in the Clifford generators.} $\sigma^M_{AB}$ and $\tilde{\sigma}^{M\,AB}$ are antisymmetric and satisfy $\tilde{\sigma}^{M\,AB}=\tfrac{1}{2}\eps^{ABCD}\sigma^M_{CD}$. By definition, we have
\begin{equation}
 \sigma^M_{AC}\tilde{\sigma}^{N\,CB}+\sigma^N_{AC}\tilde{\sigma}^{M\,CB}\ =\ -2\eta^{MN}\unit_4~.
\end{equation}
We use the sigma-matrices to convert back and forth between vector and spinor notation:
\begin{equation}
 x^{AB}\ :=\ \tilde{\sigma}_M^{AB}x^M\eand x^M\ =\ \tfrac{1}{4}\sigma^M_{AB}x^{AB}~,
\end{equation}
so that we can use the Levi-Civita symbol $\eps_{ABCD}$ as a metric:
\begin{equation}
 x^Mx_M\ =\ \tfrac{1}{4}x_{AB}x^{AB}~,\ewith x_{AB}\ =\ x_M\sigma^M_{AB}\ =\ \tfrac{1}{2}\eps_{ABCD}x^{CD}~.
\end{equation}
Derivatives act according to
\begin{equation}
 \der{x^{AB}}x^{CD}\ =\ \tfrac{1}{4}\sigma^M_{AB}\der{x^M} x^{CD}\ =\ \tfrac{1}{2}(\delta_A^C\delta_B^D-\delta_A^D\delta_B^C)~.
\end{equation}
For signature $(p,6-p)=(1,5)$, a set of sigma-matrices is given by
\begin{subequations}
\begin{equation}
x_{AB}\ =\ x_M\sigma^M_{AB}\ =\ \left(\begin{array}{cccc}
                        0 & x_0+x_5 & -x_3-\di x_4 & -x_1+\di x_2\\
			-x_0- x_5 & 0 & -x_1-\di x_2 & x_3-\di x_4\\
			x_3+\di x_4 & x_1+\di x_2 & 0 & -x_0+x_5\\
			x_1-\di x_2 & -x_3+\di x_4 & x_0-x_5 & 0
                       \end{array}\right),
\end{equation}
while for signature $(p,6-p)=(3,3)$ we may use
\begin{equation}
x_{AB}\ =\ x_M\sigma^M_{AB}\ =\ \left(\begin{array}{cccc}
                        0 & x_0+x_5 & -x_3-\di x_4 & -x_2-\di x_1\\
			-x_0- x_5 & 0 & x_2-\di x_1 & x_3-\di x_4\\
			x_3+\di x_4 & -x_2+\di x_1 & 0 & -x_0+x_5\\
			x_2+\di x_1 & -x_3+\di x_4 & x_0-x_5 & 0
                       \end{array}\right).
\end{equation}
\end{subequations}

On spinors $\psi\in S$ and $\chi\in S^\vee$, we can introduce an antilinear map
\begin{equation}
 \tau_p(\psi,\chi)\ =\ (C_p\psi^*,C^{-1}_p\chi^*)~,\ewith C_p\ :=\ \left(\begin{array}{cccc}
0 & 1 & 0 & 0 \\
p-2 & 0 & 0 & 0 \\
0 & 0 & 0 & 1 \\
0 & 0 & p-2 & 0
\end{array}\right).
\end{equation}
Here, $\psi^*$ and $\chi^*$ denote the complex conjugate of the spinors $\psi$ and $\chi$; $p-2=\pm 1$ depending on the two cases $p=1$ and $p=3$. Notice that with our choice of sigma-matrices,  $\tau_p(\sigma^{MN}\lambda)=(\sigma^{MN})^*\tau_p(\lambda)$ and therefore $\tau_p$ yields a real structure on spinors. This induces the following reality conditions on the bi-spinors $(x^{AB})\in T_{M^6}\cong S\wedge S$:
\begin{equation}
 x^{AB}\ =\ (C_p)^A{}_C (x^{CD})^* (C^{-1}_p)_D{}^B~.
\end{equation}

We define weighted, totally antisymmetric products of sigma-matrices according to their index structure, for example:
\begin{equation}\label{eq:AntiSymSigmaMatrices}
\begin{aligned}
 \sigma^{MN\,A}{}_B\ &:=\ \tfrac{1}{2}\left(\sigma^{M\,AC}\sigma^N_{CB}-\sigma^{N\,AC}\sigma^M_{CB}\right)
 \ =\ -\sigma^{MN}{}_B{}^A,\\
 \sigma^{MNK\,AB}\ &:=\ \tfrac{1}{3!}\left(\sigma^{M\,AC}\sigma^N_{CD}\sigma^{K\,DB}\pm\mbox{permutations}\right)\ =\ \sigma^{MNK\,BA}~,\\
 \sigma^{MNK}_{AB}\ &:=\ \tfrac{1}{3!}\left(\sigma^M_{AC}\sigma^{N\,CD}\sigma^K_{DB}\pm\mbox{permutations}\right)\ =\ \sigma^{MNK}_{BA}~,\\
\sigma^{MNKL\,A}{}_B\ &:=\ \tfrac{1}{4!}\left(\sigma^{M\,AC}\sigma^N_{CD}\sigma^{K\,DE}\sigma^L_{EB}
\pm\mbox{permutations}
\right)\ =\ \sigma^{MNKL}{}_B{}^A~.
\end{aligned}
\end{equation}
These products are used in the translation between vector and spinor notation for differential forms:\footnote{In spinor notation, two- and four-forms actually correspond to pairs $(B^A{}_B,B_A{}^B)$ and $(G^A{}_B,G_A{}^B)$. Depending on the choice of sigma-matrices, i.e.\ the choice of the underlying space-time, there are relations within these pairs, which allow for the distinction between two- and four-forms on space-time. To simplify our formul\ae{}, we use the above identification of spinor fields.}
\begin{equation}
\begin{aligned}
 A_M\ &\rightarrow\ A_{AB}\ =\ \tfrac{1}{4}\sigma^M_{AB}A_M~,\\
 B_{MN}\ &\rightarrow\  B_B{}^A\ =\ \sigma^{MN}{}_B{}^AB_{MN}~,\\
 H_{MNK}\ &\rightarrow\ (H_{AB},H^{AB})\ =\ (\sigma^{MNK}_{AB}H_{MNK},\sigma^{MNK\,AB}H_{MNK})~,\\
 G_{MNKL}\ &\rightarrow\ G^A{}_B\ =\ \sigma^{MNKL\,A}{}_B G_{MNKL}~.
\end{aligned}
\end{equation}
Notice that the products $\sigma^{MNK}_{AB}$ and $\sigma_{MNK}^{AB}$ form projectors onto self-dual and anti-self-dual three-forms, respectively:
\begin{equation}
 \begin{aligned}
  \sigma^{MNK\,AB}\ =\ -\frac{1}{3!}\eps^{MNKRST}\sigma_{RST}^{AB}\eand \sigma_{MNK}^{AB}\ =\ -\frac{1}{3!}\eps_{MNKRST}\,\sigma^{RST\,AB}~,
 \end{aligned}
\end{equation}
where $\eps^{012345}=+1$ and $\eps_{012345}:=-1$.

\paragraph{Four dimensions.} The six-dimensional spaces $\FR^{p,p-6}$ allow for dimensional reductions to four dimensions with arbitrary signature. The process is always the same. For brevity, we focus here on the reduction $\FR^{p,q}$ to $\FR^{p-1,q-1}$, which is done by imposing $\der{x^0}=\der{x^5}=0$. In spinor notation, this amounts to a real form of the branching of the isometry group corresponding to $\sSL(4,\FC)\rightarrow \sSL(2,\FC)\times \widetilde{\sSL(2,\FC)}$. Correspondingly, we restrict the sigma-matrices to their off-diagonal blocks:
\begin{equation}
 \sigma^M_{AB}\ \rightarrow\ \left(\begin{array}{cc} 0 & \sigma^\mu_{\alpha\ald} \\ -\sigma^\mu_{\alpha\ald} & 0 \end{array}\right),
\end{equation}
where $\mu,\nu,\ldots=1,\ldots,4$ and $\alpha,\beta,\ldots,\ald,\bed,\ldots=1,2$. The translation between vector and spinor notation then reads as
\begin{equation}
 x^{\alpha\ald}\ :=\ \sigma^{\alpha\ald}_\mu x^\mu\quad\Longleftrightarrow\quad x^\mu\ =\ \tfrac{1}{2}\sigma^\mu_{\alpha\ald}x^{\alpha\ald}~.
\end{equation}
Recall that indices can be raised and lowered with the antisymmetric tensor of $\asl(2,\FC)$: $\sigma^{\mu\,\alpha\ald}=\eps^{\alpha\beta}\eps^{\ald\bed}\sigma^\mu_{\beta\bed}$. We use the conventions $\eps_{12}=1$ and $\eps_{\alpha\beta}\eps^{\beta\gamma}=\delta_{\alpha}^{\gamma}$. The norm-squared of $x^\mu$  is thus given by
\begin{equation}
 x^\mu x_\mu\ =\ \tfrac{1}{2}x^{\alpha\ald}x_{\alpha\ald}\ :=\ \tfrac{1}{2}x^{\alpha\ald}\eps_{\alpha\beta}\eps_{\ald\bed}x^{\beta\bed}~.
\end{equation}
The real structures we obtain from reducing the six-dimensional real structure read as
\begin{equation}
 \tau_p(\psi,\chi)\ =\ (C_p\psi^*,C^{-1}_p\chi^*)\ewith C_p\ :=\ \left(\begin{array}{cc}
0 & 1 \\
p-2 & 0
\end{array}\right),
\end{equation}
and for the  coordinate vector, we have
\begin{equation}
 x^{\alpha\ald}\ =\ (C_p)^\alpha{}_\beta (x^{\beta\bed})^* (C^{-1}_p)_\bed{}^\ald~~~\Rightarrow~~~x^{1\dot{1}}\ =\ \bar{x}^{2\dot{2}}\eand x^{1\dot{2}}\ =\ (p-2)\bar{x}^{2\dot{1}}~.
\end{equation}

\subsection{Some remarks about $n$-gerbes}\label{app:gerbes}

Recall that the transition functions of a smooth principal $\sU(1)$-bundle\footnote{One can equivalently consider the associated line bundle.} over a manifold $X$ form a \v Cech one-cocycle with values in the sheaf $\CE_X^*$ of invertible smooth functions on $X$. Thus, a principal $\sU(1)$-bundle is a geometric realisation of an element in $H^1(X,\CE^*_X)$. For a holomorphic $\sU(1)$-bundle, the relevant sheaf is that of invertible holomorphic functions on $X$, $\CO_X^*$, and the transition functions represent an element of $H^1(X,\CO_X^*)$.

The {\em $n$-gerbes} defined in \cite{0817647309}, or rather the {\it bundle $n$-gerbes} developed in \cite{Hitchin:1999fh,Chatterjee:1998,Murray:9407015,Stevenson:0004117}, form geometric realisations of the cohomology groups $H^{n+1}(X,\CE^*_X)$ for smooth $n$-gerbes and $H^{n+1}(X,\CO^*_X)$ for holomorphic $n$-gerbes, respectively. For the sake of concreteness, we now focus on smooth $n$-gerbes. We have the isomorphism
\begin{equation}\label{eq:isoH2}
 H^{n+1}(X,\CE^*_X)\ \cong\ H^{n+2}(X,\RZ)~,
\end{equation}
which follows from the exactness of the exponential sequence
\begin{equation}
0\ \longrightarrow\ \RZ\ \longrightarrow\ \CE_X\ \stackrel{\exp}{\longrightarrow}\ \CE^*_X\ \longrightarrow\ 0
\end{equation}
and the fact that $\CE_X$ is fine which, in turn, implies that $H^{n}(X,\CE_X)=0$ for $n\geq1$. Furthermore, we have a natural morphism
\begin{equation}
i\,:\, H^n(X,\RZ)\ \hookrightarrow\ H^n(X,\FR)~.
\end{equation}
The elements of $\ker(i)$ are known as the {\it torsion elements} of $H^n(X,\RZ)$. Intuitively, if $[f]\in H^n(X,\RZ)$ is a torsion element, it means that there is a non-zero $m\in\RZ$ with $m[f] = [mf] = [0]$. Hence, $ [mf]$ can be split over $\RZ$, and, consequently, $[f]$ can be split over $\FR$ implying that $i([f])=[0]$ in $H^n(X,\FR)$.\footnote{Note that if an $[f]\in H^n(X,\RZ)$ can be split over $\FR$, it can also be split over $\FQ$ and hence, it must be torsion. As a specific example, let $\{U_a\}$ be a finite open cover of $X$ (this is no real restriction as any connected, paracompact Hausdorff manifold admits such a cover) and consider an $[f]=[\{f_{ab}\}]\in H^1(X,\RZ)$ with $f_{ab}=r_a-r_b$ on $U_{ab}$, where $r_a\in\FR$. Since $f_{ab}\in\RZ$, we must have the decomposition $r_a = \frac{p_a}{q_a} + \tilde r_a$ with $p_a,q_a\in\RZ$ and $\tilde r_a$ irrational such that $\tilde r_a=\tilde r_b$ which, in turn, implies that $f_{ab}= \frac{p_a}{q_a}-\frac{p_b}{q_b}$. This shows that $[f]$ can be split over $\FQ$. Furthermore, let $m$ be the least common multiple of all the $q_a$. Then, we may write $f_{ab}= \frac{\tilde p_a}{m}-\frac{\tilde p_b}{m}$ for some $\tilde p_a\in\RZ$ and so $m[f]=[0]$ in $H^1(X,\RZ)$, that is, $[f]$ is torsion.}  Since
\begin{equation}
H^n(X,\FR)\ \cong\ H^n_{\rm dR}(X,\FR)~,
\end{equation}
where $H^n_{\rm dR}(X,\FR)$ is the $n$-th de Rham cohomology group, torsion elements cannot be represented by differential forms. Altogether, modulo torsion, smooth $n$-gerbes are classified by their characteristic classes which are integral elements in $H^{n+2}_{\rm dR}(X,\FR)$. 

To be more explicit, let us consider the examples $n=0,1$. Suppose we have an open cover $\{U_{a}\}$ of $X$ and a smooth partition of unity $\{\theta_{a}\}$ subordinate to the cover. Then, a smooth zero-gerbe $\Gamma$ is characterised by a  \v Cech two-cocyle $f=\{f_{abc}\}$ representing an element in $H^2(X,\RZ)$. Assuming that $[f]$ is not torsion and using the partition of unity, we may define a smooth \v Cech one-cochain $s_{ab}:=\sum_c f_{abc} \theta_c$ yielding a smooth splitting $f_{abc}=s_{ab}+s_{bc}+s_{ca}$. We also set $A_{a}:=\sum_b s_{ab} \dd \theta_{b}$ for which we have $A_a=A_b-\dd s_{ab}$ on non-trivial overlaps of the coordinate patches. This then gives a representative $F$ of an element of the de Rham comology group $H^{2}_{\rm dR}(X,\FR)$ by means of $F_a=\dd A_{a}$ which, in turn, corresponds to the first Chern class of $\Gamma$.  Conversely, starting with an integral $[F]\in H^{2}_{\rm dR}(X,\FR)$ representing the first Chern class, we may apply the Poincar\'e lemma to obtain a one-form potential $A_{a}$ with $\dd A_{a}=F_a$ on each patch $U_{a}$ with $A_a=A_b-\dd s_{ab}$ on intersection of patches $U_{ab}$. Here, $s=\{s_{ab}\}$ is a smooth  \v Cech one-cochain. Furthermore, we have $s_{ab}+s_{bc}+s_{ca}=f_{abc}\in 2\pi\di\RZ$ on intersections of patches $U_{abc}$. Therefore, the exponentials $t_{ab}:=\exp(s_{ab})$ yield a \v Cech one-cocycle $t=\{t_{ab}\}$ representing an element in $H^1(X,\CE^*_X)$, that is, the transition functions of $\Gamma$.

In the case of a one-gerbe $\Gamma$, we start from a three-form $H$ representing an integral element in $H^3_{\rm dR}(X,\FR)$ also called the {\em Dixmier--Douady class} of $\Gamma$. The Poincar\'e lemma leads to a two-form potential $[B]\in H^1(X,\Omega^2_X)$, a one-form potential $[A]\in H^2(X,\Omega^1_X)$, and a \v Cech two-cocycle $[t]=[\{t_{abc}\}]\in H^3(X,\CE^*_X)$ with $t_{abc}:=\exp(s_{abc})$ all of which satisfy
\begin{equation}
\begin{aligned}
 H_a\ =\ \dd B_{a}\eon U_{a}~,~~~B_{a}-B_{b}\ =\ \dd A_{ab}\eon U_{ab}~,\\
A_{ab}+A_{bc}+A_{ca}\ =\ \dd s_{abc}\eon U_{abc}~.\hspace{1.2cm}
\end{aligned}
\end{equation}
The set $\{A,B\}$ is also known as a {\it connective structure} on the one-gerbe $\Gamma$ and $H$ as its {\it curvature}. The converse construction of $H$ from an element in $H^3(X,\RZ)$ is again performed as above by using the partition of unity and again, one should keep in mind torsion.

\subsection{Formul\ae{} for splittings}\label{sec:splittings}

In this Appendix, we present explicit formul\ae{} for holomorphic splitting \eqref{eq:HoloSplitCS} used in the construction of the differential forms \eqref{eq:PBForms}. To this end, we shall follow the ideas of Helfer \cite{JSTOR:2398112}, who constructed similar holomorphic splittings in the context of self-dual Yang--Mills theory in four dimensions.

Recall that the smooth splitting $s'=\{s'_{ab}\}$ with $s'_{ab}=\sum_c f'_{abc}\theta'_{d}$ and the holomorphic splitting $h'=\{h'_{ab}\}$ are related by a gauge transformation
\begin{equation}
 s'_{ab}\ =\ h'_{ab}-\varphi'_{ab}\ewith \varphi'_{ab}\ :=\ \varphi'_{a}-\varphi'_{b}~,
\end{equation}
cf.\  \eqref{eq:GaugePara}. Hence in order to find $h'$, we need to construct $\varphi'$, which is a solution to the differential equation
\begin{equation}\label{eq:EqGP}
 \bar\partial \varphi'_{ab}\ =\ -\sum_c f'_{abc}\bar\partial\theta'_{c}\ =:\ a^{(0,1)}_{ab}~.
\end{equation}
Here, $\bar\partial$ is the Dolbeault operator on $\PP^3$ and its Green function directly yields $\varphi'$.

Fortunately, this Green function has been computed before in \cite{Mason:2008jy,Mason:2009qx,Adamo:2011cb,Boels:2006ir} (see also \cite{Adamo:2011pv,Wolf:2010av}). It is a $(0,2)$-form $G^{(0,2)}=G^{(0,2)}(\lambda_1,\lambda_2)$ on $\PP^3\times\PP^3$ satisfying $\bar\partial G^{(0,2)}=\delta^{(0,3)}$, where the projective Dolbeault--Dirac delta distribution is given by
\begin{equation}
 \delta^{(0,3)}(\lambda_1,\lambda_2)\ :=\ \int_{\FC} \frac{\dd s}{s}\, \delta^{(0,4)}(s\lambda_1+\lambda_2)~.
\end{equation}
Here, $\lambda_i$ is short-hand notation for $\lambda_{iA}$ and $\delta^{(0,4)}$ is the Dolbeault--Dirac delta distribution on $\FC^4$.\footnote{On the complex plane $\FC$ described by the coordinate $\lambda$, the Dolbeault--Dirac distribution is given by $\delta^{(0,1)}(\lambda)=\bar\partial (2\pi\di \lambda)^{-1}$. Therefore, $\delta^{(0,4)}(\lambda)=\delta^{(0,1)}(\lambda_{A=1})\wedge\cdots\wedge\delta^{(0,1)}(\lambda_{A=4})$.} We have  $\delta^{(0,3)}(t_1\lambda_1,t_2\lambda_2)=t_1^{0}t_2^{-4}\delta^{(0,3)}(t_1\lambda_1,t_2\lambda_2)$ for $t_{1,2}\in\FC^*$, and for holomorphic functions $f=f(\lambda)$ of homogeneity zero (defined on appropriate regions of $\PP^3$), we have the identity
\begin{equation}
 f(\lambda_1)\ =\ \int \Omega^{(3,0)}(\lambda_2)\wedge\delta^{(0,3)}(\lambda_1,\lambda_2) f(\lambda_2)~,
\end{equation}
where $\Omega^{(3,0)}(\lambda):=\frac{1}{4!}\eps^{ABCD}\lambda_A\dd\lambda_B\wedge\dd\lambda_C\wedge \dd \lambda_D$ is the holomorphic measure on $\PP^3$. Notice that  the Green function cannot be unique since one can always add a holomorphic $(0,2)$-form to $G^{(0,2)}$ while preserving $\bar\partial G^{(0,2)}=\delta^{(0,3)}$. By virtue of Lemma \ref{lem:Dolbeault}, however, all holomorphic $(0,2)$-forms (of any homogeneity) on $\PP^3\times \PP^3$ must be Dolbeault-exact and therefore, $\bar\partial G^{(0,2)}=\delta^{(0,3)}$ enjoys a gauge freedom of the form $G^{(0,2)}\mapsto G^{(0,2)}+\bar\partial\varphi^{(0,1)}$ for some (smooth) differential $(0,1)$-form $\varphi^{(0,1)}$ of appropriate homogeneity. A convenient choice of gauge is the Cachazo--Svr\v cek--Witten gauge \cite{Cachazo:2004kj},
\begin{equation}
 \bar\xi^A\der{\bar\lambda^A_1}\lrcorner\, G^{(0,2)}(\lambda_1,\lambda_2)\ =\ 0~,
\end{equation}
where bar denotes complex conjugation and $\xi_A$ is some fixed reference spinor. In this gauge, the Green function is given by
\begin{equation}
 G^{(0,2)}(\lambda_1,\lambda_2)\ =\ \int_{\FC} \frac{\dd s}{s}\,  \delta^{(0,3)}(s\,\xi+\lambda_1,\lambda_2)~,
\end{equation}
cf.\ e.g.\ \cite{Adamo:2011pv}. It scales as $G^{(0,2)}(t_1\lambda_1,t_2\lambda_2)=t_1^{0}t_2^{-4}G^{(0,2)}(t_1\lambda_1,t_2\lambda_2)$ for $t_{1,2}\in\FC^*$, and it is invariant under re-scalings of the reference spinor $\xi_A$.

In order to compute the gauge parameter as a solution of \eqref{eq:EqGP}, we choose a covering of $\PP^3$ coming from $U'$ and consider the respective double overlaps. We shall assume that the reference spinor $\xi_A$ is not contained in any of these double overlaps.\footnote{This assumption is needed to avoid certain `error terms', see e.g.~the review \cite{Adamo:2011pv} for details.} Altogether, we arrive at the gauge parameter $\varphi'_{ab}$, which is given by the integral
\begin{equation}
 \varphi'_{ab}\ =\ \int \Omega^{(3,0)}(\lambda_2)\wedge G^{(0,2)}(\lambda_1,\lambda_2)\wedge a^{(0,1)}_{ab}(\lambda_2)~,
\end{equation}
where we integrate over a suitable region.

\end{document}